\documentclass{aa}  
\usepackage{graphicx}
\usepackage{txfonts}
\usepackage{hyperref}
\hypersetup{
  colorlinks   = true, 
  urlcolor     = blue, 
  linkcolor    = blue, 
  citecolor    =[RGB]{255, 104, 68}
}

\usepackage{booktabs}
\usepackage[flushleft]{threeparttable}
\usepackage{upgreek}

\newcommand{\RI}{FRB~20121102A}
\newcommand{\RItwin}{FRB~20190520B}

\newcommand{\clu}{CLU-compiled~}

\newcommand\sbullet[1][.5]{\mathbin{\vcenter{\hbox{\scalebox{#1}{$\bullet$}}}}}

\begin{document}

   \title{A LOFAR sample of luminous compact sources coincident with nearby dwarf galaxies\thanks{Tables \ref{table:candidates} and \ref{table:fluxes} are available in electronic form at the CDS via anonymous ftp to \url{cdsarc.cds.unistra.fr} (\url{130.79.128.5}) or via \url{https:// cdsarc.cds.unistra.fr/viz- bin/cat/J/A+A/XXX/XXX}}}
   
   \author{D.~Vohl
          \inst{1,2}\fnmsep\thanks{Email: d.vohl@uva.nl}
          \and
          H.~K.~Vedantham\inst{2,3}
          \and
          J.~W.~T. Hessels\inst{1,2}
          \and
          C.~G. Bassa\inst{2}
          \and
          D.~O. Cook\inst{4}
          \and
          D.~L.~Kaplan\inst{5}
          \and
          T.~W.~Shimwell\inst{2,6}
          \and
          C. Zhang\inst{5}
          }

   \institute{Anton Pannekoek Institute for Astronomy, University of Amsterdam, 
              P.O. Box 94249, 1090 GE Amsterdam, Nederland
         \and
            ASTRON, Netherlands Institute for Radio Astronomy, 
            Oude Hoogeveensedijk 4, Dwingeloo, 7991\,PD, The Netherlands
         \and
            Kapteyn Astronomical Institute, University of Groningen, 
            PO\,Box 72, 97200\,AB, Groningen, The Netherlands
         \and
            Caltech/IPAC, 1200 E. California Boulevard, Pasadena, CA 91125, USA
         \and
            Center for Gravitation, Cosmology, and Astrophysics, 
            Department of Physics, University of Wisconsin-Milwaukee, 
            PO Box 413, Milwaukee, WI, 53201, USA
         \and
            Leiden Observatory, Leiden University, PO\,Box 9513, 
            2300\,RA, Leiden, The Netherlands
             }
 
  \abstract
   {The vast majority of extragalactic compact continuum radio sources are associated with star formation or jets from (super)massive black holes and, as such, are more likely to be found in association with starburst galaxies or early-type galaxies. 
   Two new populations of radio sources were recently identified: (a) compact and persistent sources (PRSs) associated with fast radio bursts (FRBs) in dwarf galaxies and (b) compact sources in dwarf galaxies that could belong to the long-sought population of intermediate-mass black holes. 
   Despite the interesting aspects of these newly found sources, the current sample size is small, limiting scrutiny of the underlying population. 
   Here, we present a search for compact radio sources coincident with dwarf galaxies. 
   We search the LOFAR Two-meter Sky Survey (LoTSS), the most sensitive low-frequency (144\,MHz central frequency) large-area survey for optically thin synchrotron emission to date.
   Exploiting the high spatial resolution ($6\arcsec$) and low astrometric uncertainty ($\sim0\,\farcs2$) of the LoTSS, we match its compact sources to the compiled sample of dwarf galaxies in the Census of the Local Universe, an H$\upalpha$ survey with the Palomar Observatory 48 inch Samuel Oschin Telescope.
   We identify 29 over-luminous compact radio sources, evaluate the probability of chance alignment within the sample, investigate the potential nature of these sources, and evaluate their volumetric density and volumetric rate. 
   While optical line-ratio diagnostics on the nebular lines from the host galaxies support a star-formation origin rather than an AGN origin, future high-angular-resolution radio data are necessary to ascertain the origin of the radio sources.
   We discuss planned strategies to differentiate between candidate FRB hosts and intermediate-mass black holes.}


   \maketitle

%-------------------------------------------------------------------

\section{Introduction} \label{sec:intro}

\RI~was the first fast radio burst (FRB) source found to repeat~\citep{Spitler2016Natur.531..202S}. 
Its repetitive nature rules out progenitor models related to cataclysmic explosions for at least a fraction of all FRBs.
\RI~was also the first FRB source to be precisely localized to a host galaxy~\citep{Chatterjee2017Natur.541...58C} from a coordinated observing campaign between the Karl G. Jansky Very Large Array (VLA) and the 305m William E. Gordon Telescope at the Arecibo Observatory. 
The host galaxy is a low-mass, low-metallicity dwarf at a redshift of $z=0.19273\pm0.0008$~\citep{Tendulkar2017ApJ...834L...7T}, and as such has the characteristics of a typical host to long gamma-ray bursts and super-luminous supernovae.
These simple facts suggest that the mechanism driving FRBs may be linked to these latter phenomena.
Moreover, \RI~was the first FRB found to be co-located with a persistent radio source~\citep[PRS;][]{Chatterjee2017Natur.541...58C}, with a luminosity of $L_\mathrm{radio}\sim10^{39}\,\mathrm{erg\,s^{-1}}$, which is more 50 times what would be expected from star formation activity alone.

Very long baseline interferometry (VLBI) observations with the European VLBI Network (EVN) showed that the FRB and the PRS are located within $\lesssim 40\,{\rm pc}$ (transverse distance) of one another~\citep{Marcote2017ApJ...834L...8M}, strongly connecting the two.
Optical and far-infrared observations using the {\it Hubble Space Telescope} revealed that the FRB/PRS location is slightly offset from the centroid of a star formation knot within the host~\citep{Bassa2017ApJ...843L...8B}.
The PRS has a flat spectral index, that is,
$S_\nu\propto\nu^\alpha$, with $\alpha\sim-0.07\pm0.03$ below 10\,GHz~\citep{Resmi2021A&A...655A.102R}, and with a possible turnover at lower frequencies, estimated as $\alpha\sim0.3$ between 433\,MHz and 1.4\,GHz by \citet{Mondal2020MNRAS.498.3863M}. 
The spectral energy distribution (SED) derived from multiwavelength measurements and upper limits matches that of the Crab nebula, though with orders-of-magnitude higher luminosity.
Finally, bursts from \RI~have a high and variable Faraday rotation measure~\citep[RM;][]{Michilli2018Natur.553..182M}, and the dispersion measure (DM) also shows secular changes~\citep{Hessels2019ApJ...876L..23H, Platts2021MNRAS.505.3041P}. 

These measurements combined make a plausible case for an FRB engine that is a young, highly magnetized neutron star embedded in an expanding supernova remnant and powering a pulsar wind nebula~\citep[PWN;][]{Murase2016MNRAS.461.1498M, Margalit2018ApJ...868L...4M} or magnetar wind nebula~\citet[MWN;][]{Margalit2018ApJ...868L...4M, Zhao2021ApJ...923L..17Z}.
Another plausible explanation is that the FRB engine is within the vicinity of a massive black hole, which in turn creates the PRS.
In this context, the FRB source may be a neutron star near a black hole, or may even be a black hole jet.
Other authors have suggested that an accreting compact object engine (e.g., ultra-luminous X-ray sources accreting at highly super-Eddington rates) could explain the PRS \citep{Chen2022arXiv220100999C, Sridhar2022ApJ...937....5S}.
There is only one other  known repeating FRB co-localized to a PRS~\citep[][]{Niu2022Natur.606..873N}.
Like \RI, \RItwin~is hosted in a star-forming dwarf galaxy, with a large $\mathrm{DM}_\mathrm{host}$ contribution, a high repetition rate, and its associated PRS has a shallow spectral index ($\alpha=-0.41\pm0.04$).

Given these two cases of FRB/PRS connection, it seems that PRSs represent an important aspect of some FRBs, even if their nature remains mysterious.
If PRSs are wind nebulae, a limited lifespan during which they can be detected~\citep[$\sim$few centuries;][]{Gaensler2006ARA&A..44...17G} could explain why only a subset of FRBs have a PRS counterpart.
Considering a sample of 15 localized FRBs with radio sensitivity limits that could allow the detection of a PRS (including six repeating FRBs), \citet{Law2022ApJ...927...55L} estimated that PRS occurrence could be as high as 20\% for repeating FRBs given that 2 out of the 6 repeating FRBs in their sample are associated to a PRS. 
Furthermore, given that most FRBs with meaningful PRS limits would also limit PRS emission out to $z\approx0.4$, a distance that includes a similar number of repeating and nonrepeating FRBs ---supporting the argument that PRS emission from repeaters and nonrepeaters is equally well-constrained--- these authors estimate that PRS detectability in repeating FRBs is not strongly biased by distance. 

Separately, \citet{Reines2020ApJ...888...36R} identified a sample of compact radio sources associated with dwarf galaxies and suggested that they may be the long-sought population of intermediate-mass black holes \citep[IMBHs, ${\rm \sim10^2-10^5\,M_\odot}$;][]{Greene2020ARA&A..58..257G} predicted to reside in dwarf galaxies~\citep[e.g.,][]{McConnell2013ApJ...764..184M, Reines2015ApJ...813...82R}. 
Furthermore, the compact radio sample presented by \citet{Reines2020ApJ...888...36R} has been shown to share many similarities with the PRS associated with \RI, with radio luminosities, SEDs, light curves, ratios of radio-to-optical flux, and spatial offsets between the radio source and the host optical center being consistent with arising from the same population~\citep{Eftekhari2020ApJ...895...98E}.

Follow-up observations of the 13 dwarf galaxies with likely accreting IMBHs from \citet{Reines2020ApJ...888...36R} with the Very Long Baseline Array (VLBA) led to four source detections at milliarcsecond (mas) resolution. 
These four sources have the largest offset from their host centroid in the sample, which suggests that they are all likely background AGNs~\citep{Sargent2022ApJ...933..160S}.
Two sources among the VLBA nondetections are associated with spectroscopically confirmed AGNs that are consistent with being located at their galaxy photocenter.
\citeauthor{Sargent2022ApJ...933..160S} also re-examined the radio--star formation relationship and found that a star-formation origin cannot be ruled out for  approximately 5 out of the 9 VLBA nondetections. These authors conclude that these sources are either the posited wandering accreting massive black hole scenario, or simply background AGN.

While the existence of wandering black holes has yet to be definitively proven (IMBHs in dwarf galaxies have a much shallower gravitational potential than in more massive galaxies that could in theory allow for off-nuclear location), the well-documented case of \RI~nevertheless highlights that such off-nuclear compact radio emission can occur, even within a star forming region within the host galaxy.
Although the connection between IMBHs, PRSs, and FRBs remains unclear ---most FRB models prefer a magnetar progenitor~\citep[e.g., see discussion in][]{Eftekhari2020ApJ...895...98E}--- over-luminous compact radio sources (OCRs) in dwarf galaxies are an interesting radio source population in their own right. 

To improve our understanding of PRSs and their potential connection to the FRB and/or IMBH phenomena (e.g., constrain FRB progenitor models), it is imperative to increase the known sample size~\citep{Vohl2023arXiv230311967V}.
Here, we present a targeted search for compact radio sources coincident with dwarf galaxies using the LOFAR Two-Meter Sky Survey (LoTSS) second data release~\citep[DR2;][]{Shimwell2022A&A...659A...1S} --- the most sensitive large-area survey for optically thin synchrotron emission to date--- as our radio reference catalogue, and the Palomar Observatory 48 inch Samuel Oschin Telescope `Census of the Local Universe'~\citep[CLU;][]{Cook2019ApJ...880....7C} as our optical reference catalogue. 

The article is organized as follows: in \S\ref{sec:method} we describe our candidate selection methodology and discuss chance alignment probability. In \S\ref{sec:candidates} we discuss the selected candidates, and complement the information  available for our sample with ancillary survey data at various wavelengths. In \S\ref{sec:discussion} we discuss the potential nature of these sources and our plans for future work, and 
in \S\ref{sec:summary}  we close with a summary.

%-------------------------------------------------------------------

\section{Candidate selection}
\label{sec:method}

\subsection{Sample description} 
\label{subsec:sample}

LoTSS DR2 comprizes 4,396,228 radio sources spanning over 5600\,deg$^2$ of the northern sky.
LoTSS operates at a central frequency of 144\,MHz with 48\,MHz of bandwidth ($120-168$\,MHz).
The survey has a  $\sim6\arcsec\times({\small \frac{144\,\rm{MHz}}{v}})$ angular resolution and a median root mean square (rms) sensitivity of about $80\,\rm{\upmu Jy/beam}$.
Furthermore, with a $0\,\farcs2$ astrometric uncertainty\footnote{Namely, the uncertainty tying the LoTSS radio frame to the PanSTARRS~\citep[PS1;][]{Chambers2016arXiv161205560C} frame, along with the formal error to evaluate the source centroid, given by $\frac{6\arcsec}{\rm{S/N}}$, where S/N is the signal-to-noise ratio. For $\gtrsim$mJy level sources detected at $\gtrsim10\sigma$, the latter is dominant. We refer the reader to \citet{Shimwell2022A&A...659A...1S} for further details about astrometry in LoTSS DR2.} for sources brighter than 20\,mJy ---which is comparable to optical surveys--- and a 90\% point source completeness for sources $\geq0.8$\,mJy/beam, LoTSS DR2 represents an excellent catalogue for our study.

We cross-match LoTSS DR2 to the \clu catalogue, a compilation of all known galaxies out to 200\,Mpc~\citep{Cook2019ApJ...880....7C}.
\clu is an extension to the CLU photometric survey carried out with four narrow-band H$\upalpha$ filters corresponding to redshifts up to $z=0.0471$. 
\clu is a compilation from existing galaxy databases (NASA/IPAC Extragalactic Database (NED, \url{https://ned.ipac.caltech.edu}), Hyperleda, \url{http://leda.univ-lyon1.fr}, Extragalactic Distance Database \url{http://edd.ifa.hawaii.edu}, the Sloan Digital Sky Survey DR12, the 2dF Galaxy Redshift Survey, and The Arecibo Legacy Fast ALFA (ALFALFA)) and is designed to provide the most complete list of galaxies with measured distances in the LIGO sensitivity volume. 
The compiled catalogue comprizes 271,867 sources, of which 95,047 fall within the LoTSS DR2 footprint. 
For brevity, we use the terms CLU and \clu interchangeably throughout the paper. 
Distances based on Tully–Fischer methods were favored over kinematic (i.e., redshift) distances; however, the majority of the distances are based on redshift information. 
For cases where neither Tully-Fischer nor redshift was available, a distance based on the H$\upalpha$ filters is provided. 

In addition to distances, \clu also contains compiled photometric information. 
In particular, sources have been cross-matched to within 4\arcsec~with: 
(i) Sloan Digital Sky Survey (SDSS) data release 12~\citep{Alam2015ApJS..219...12A} for optical fluxes; 
(ii) GALEX all-sky~\citep{Martin2005ApJ...619L...1M, Bianchi2014AdSpR..53..900B} for far- and near-ultraviolet (FUV, NUV) Kron fluxes~\citep{Kron1980ApJS...43..305K}; and
(iii) Wide-field Infrared Survey Explorer~\citep[WISE;][]{Wright2010AJ....140.1868W} for mid-infrared fluxes.
Additionally, \clu aggregates these ancillary data to cull contaminants (i.e., bright stars, high-redshift sources with emission lines shifted to within the CLU H$\upalpha$ bands) and measure several physical properties of galaxies such as their stellar mass and star formation rate (SFR).

We select sources in the mass range corresponding to dwarf galaxies ($10^7 \leq M_*/M_\odot \leq 3\times10^9$), leaving 31,190 sources.
We finally keep only galaxies with a valid SFR measurement, leaving 18,159 galaxies that form our parent sample.
With regards to the LoTSS DR2 catalogue, we select sources with a peak brightness of $\ge0.8$\,mJy/beam, for a total of 2,622,903 sources.
Furthermore, we constrain the matched sample to reject extended sources using the $R_{99.9}$ compactness criterion~\citep[see][Equation~2]{Shimwell2022A&A...659A...1S}, where the ratio of the natural logarithm of the integrated flux density ($S_I$) to peak brightness ($S_P$) is less than or equal to the envelope that encompasses the 99.9 percentile of the $S_I/S_P$ distribution, leaving 2,275,400 sources.
Finally, we limit our source selection to those fitted with a single Gaussian component by the source finder {\texttt pyBDSF}~\citep{Mohan2015ascl.soft02007M}, which was used to produce the LoTSS source catalogue, leaving 2,051,534 sources.

\subsection{Cross-matching and filtering} 
\label{subsec:crossmatch}

We aim to identify OCRs dwelling in dwarf galaxies with luminosity exceeding the contribution expected from star formation alone. 
To find these, we cross-matched our subset of LoTSS and CLU sources (\S~\ref{subsec:sample}) using a radius of $(6+\epsilon)\arcsec$ (the angular resolution of LoTSS, with $\epsilon$ being the astrometric uncertainty for a given source), yielding 708 matches.
We then rejected all sources that lie within three standard deviations of the radio luminosity versus star formation rate (L--SFR) relationship presented by \citet{Gurkan2018MNRAS.475.3010G}\footnote{For consistency with \citet{Gurkan2018MNRAS.475.3010G}, throughout this paper, we use a concordance cosmology with $H_0=70\,{\rm km\,s^{-1}\,Mpc^{-1}}$, $\Omega_m=0.3$, and $\Omega_{\Lambda}=0.7$.}. 
This yields just 32 sources.
It is worth noting that \citet{Smith2021A&A...648A...6S} highlighted a dependence of the L--SFR relation on stellar
mass, with galaxies at $\sim10^{9.5}\,M_*/M_\odot$ following the relation from \citet{Gurkan2018MNRAS.475.3010G}, while those at lower masses tend to have a lower luminosity for a fixed SFR. 
Therefore, given the L-SFR mass dependence and that the stellar mass--metallicity relation indicates a trend where galaxy metallicity increases with increasing stellar mass~\citep[e.g.,][]{Curti2020MNRAS.491..944C}, our sample selected above the $3\sigma$ threshold should be considered as a lower limit, and outliers are not simply due to low metallicity in their host galaxies.

\begin{figure*}
\includegraphics[width=17cm]{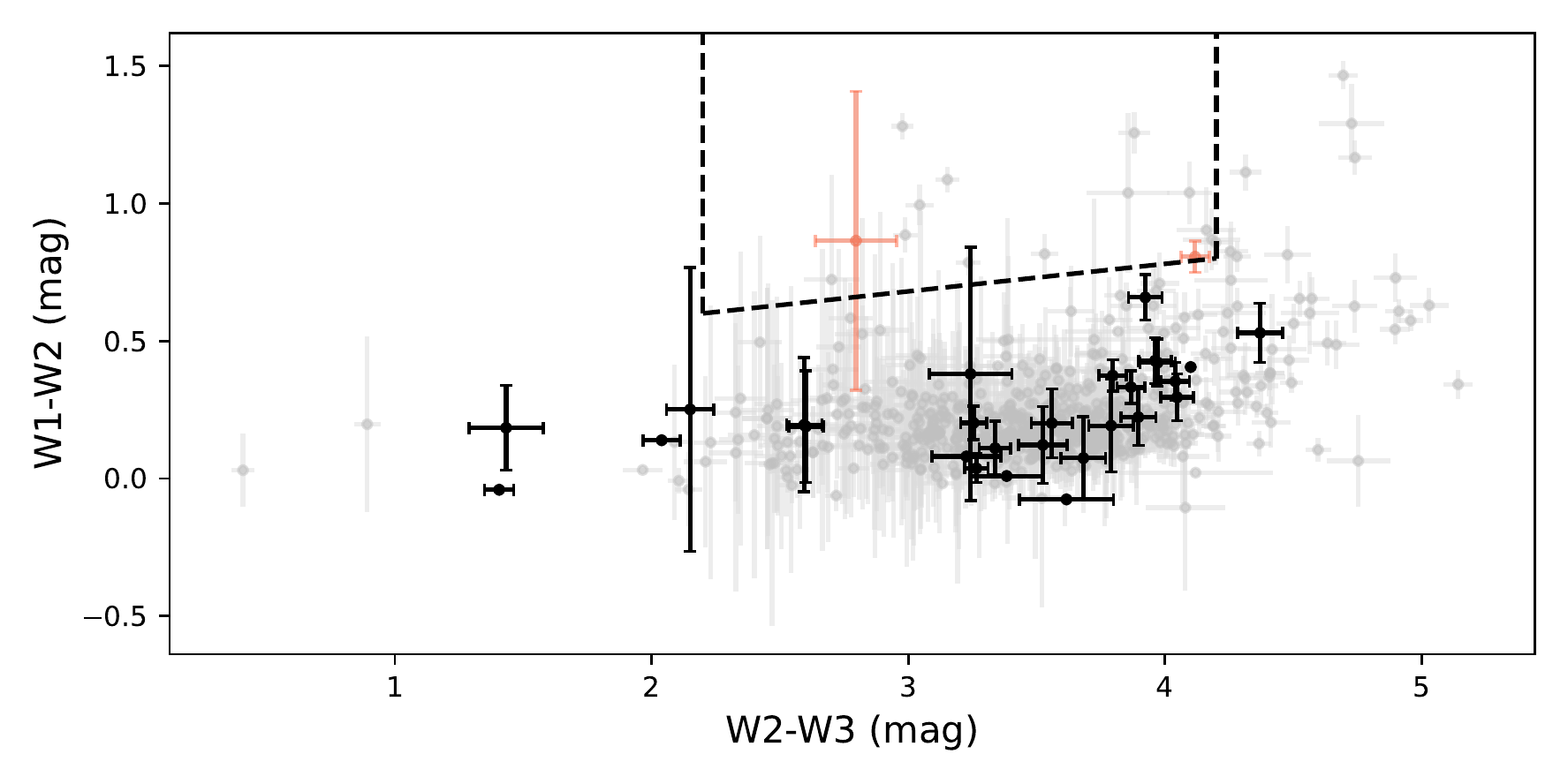}
\caption{WISE color--color plot. 
Photometric ratios in magnitude (mag) between three IR bands (W1${\,\rm_{3.4\upmu\protect\\m}}$, W2${\,\rm_{4.6\upmu\protect\\m}}$, W3${\,\rm_{12\upmu\protect\\m}}$, with W1-W2, W2-W3).
Black markers correspond to dwarf galaxies matched to a compact radio source with luminosity exceeding $3\sigma$ in the L--SFR relation. 
Gray-filled markers indicate dwarf galaxies matched to a radio source. 
We demarcate the two sources falling within the AGN region~\citep{Jarrett2011ApJ...735..112J} marked by dashed lines within our sample as being classified.}

\label{fig:wise}
\end{figure*}

We verify that redshift values listed in CLU match those of other measurements listed in the NED, because redshifted emission lines other than H$\upalpha$ could fall within the CLU filter and masquerade as H$\upalpha$. 
Indeed, objects with incorrectly attributed redshifts will be over-represented in our sample, which is selected based on nonadherence to the L--SFR relationship.
Among the 32 sources, we find that three objects have redshifts that are inconsistent with the median redshift from the NED. 
We therefore remove these from our sample, leaving 29 candidates.
From these, 11 out of the 29 sources have spectra available in the NED.
Next, we check IR magnitudes for any obvious AGNs by evaluating sources whose WISE colors fall within the AGN region prescribed by \citet[][Eq. 1]{Jarrett2011ApJ...735..112J}, represented in Figure~\ref{fig:wise}.
Two candidates fall within this AGN region, leaving 27 sources to be further scrutinized.

\begin{figure*}
\includegraphics[width=17cm]{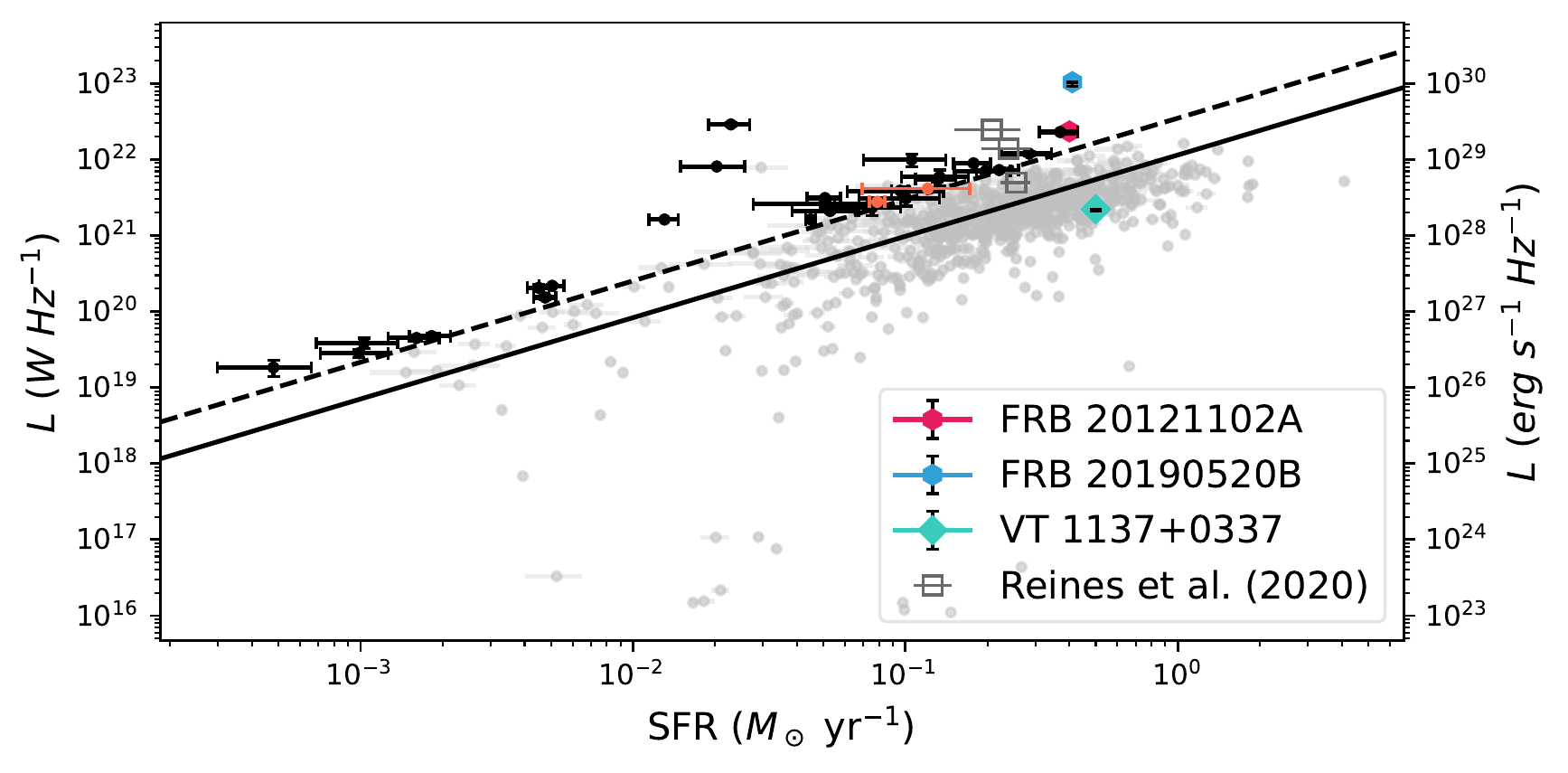}
\caption{Candidate selection via the L--SFR relation. 
Gray-filled markers indicate compact radio sources matched to dwarf galaxies matched. 
Black filled circles correspond to our final selection of OCRs matched to a dwarf galaxy with luminosity exceeding $3\sigma$ (dashed line) on the L--SFR relation by \citet[][and the solid line]{Gurkan2018MNRAS.475.3010G} shows those with validated redshift.
The uncertainty on luminosity is below the marker size.
Orange markers are sources within the AGN region in Figure \ref{fig:wise}.
As a reference, we show PRS luminosity and the SFR of the host for both \RI~\citep{Tendulkar2021ApJ...908L..12T, Law2022ApJ...927...55L} and \RItwin~\citep{Niu2022Natur.606..873N}, with luminosity measurements scaled to 144\,MHz using $\alpha$ from \citet[][evaluated between 2\,MHz and 10\,GHz,]{Resmi2021A&A...655A.102R} and from \citet[][evaluated between 1.5 and 5.5\,GHz]{Niu2022Natur.606..873N}, respectively.
We also show the transient source VT~1137+0337 using values from \citet{Dong2023ApJ...948..119D}.
Finally, we show \citet{Reines2020ApJ...888...36R} galaxies (J0909+5655, J1136+2643, J1220+3020) matched in CLU for which SFR information is available, with luminosity scaled to 144\,MHz using $\alpha$ values fitted between 1.4\,GHz and 9\,GHz by \citet{Eftekhari2020ApJ...895...98E}. 
The large scatter in luminosity is discussed in \S\ref{subsec:chance} and Appendix \ref{app:l_sfr}.
}
\label{fig:selection}
\end{figure*}

We depict the candidate selection on the L--SFR plane in Figure~\ref{fig:selection}.
We discuss the distribution of matched galaxies (including all galaxy masses) along the L--SFR plane in Appendix \ref{app:l_sfr}.
Candidates are summarized in Table~\ref{table:candidates}, and are investigated further in the following sections.
Astrometric uncertainties provided by LoTSS were taken into account during cross-matching. 
Therefore, while a projected offset of $7\arcsec$ is listed for ILT~J125944.53+275800.9, the lower limit on its offset taking into account the radio astrometric uncertainties on right ascension and declination is $5\,\farcs4$. 

\begin{sidewaystable*}
\begin{threeparttable}
\caption{Properties of selected candidates.}
\centering
\begin{tabular}{llrrrrrrrrr}
    \toprule
        Source name & Host name & \multicolumn{2}{r}{Projected offset} & $z$ & $S$ & $\log L$ & $\log {\rm SFR}$ & $\sigma$ & $\alpha$ & $R_g$ \\
        (ILT~J) & ~ & (arcsec) & (pc) & ~  & (mJy) & (${\rm W/Hz}$) & (${\rm M_\odot/yr}$) & ~ & ~ & ~  \\
    \midrule
        003532.36+303008.0 & CLU J003532.3+303008 & $0.7\pm0.1$ & $165$ & $0.0121\pm0.01196$ (n) & $11.89\pm0.35$ & $21.6$ & $-1.02\pm-2.16$ & $3.85$ & $-0.56$ & $1.93$ \\
        125915.34+274604.2 & SSTSL2 J125915.27+274604.1 & $0.8\pm0.4$ & $321$ & $0.0211$ (k) & $2.57\pm0.28$ & $21.4$ & $-1.29\pm-1.63$ & $4.56$ & -- & -- \\
        231715.38+184339.0 & 2MASX J23171540+1843385 & $0.8\pm0.7$ & $355$ & $0.0409\pm0.00034$ (m) & $2.57\pm0.50$ & $22.0$ & $-0.98\pm-1.45$ & $6.09$ & -- & $0.13$ \\
        021835.45+262040.9 & LAMOST J021835.51+262040.7 & $0.9\pm0.3$ & $57$ & $0.0033$ (k) & $6.55\pm0.44$ & $20.2$ & $-2.32\pm-3.35$ & $3.79$ & $-0.53$ & $0.63$ \\
        075257.15+401026.3 & UGC 04068 & $0.9\pm0.2$ & $133$ & {\bf 0.0412$\pm$0.00010} (m) & $7.37\pm0.35$ & $22.5$ & $-1.64\pm-2.40$ & $13.37$ & $-0.54$ & -- \\
        165252.24+391151.7 & CLU J165252.32+391152.0 & $1.0\pm0.8$ & $540$ & $0.0272\pm0.00700$ (k) & $1.83\pm0.38$ & $21.5$ & $-1.00\pm-1.48$ & $3.08$ & -- & $0.53$ \\
        162244.56+321259.3 & 2MASS J16224461+3213007 & $1.0\pm0.1$ & $451$ & {\bf 0.0221$\pm$0.00009} (k) & $8.06\pm0.18$ & $22.0$ & $-0.75\pm-1.56$ & $4.29$ & $-0.63$ & $0.50$ \\
        153943.52+592730.7 & CLU J153943.44+592729.8 & $1.1\pm0.2$ & $883$ & $0.0411\pm0.00695$ (k) & $3.06\pm0.17$ & $22.1$ & $-0.55\pm-1.23$ & $3.72$ & -- & $0.52$ \\
        015915.79+242500.6 & KUG 0156+241 & $1.2\pm0.5$ & $128$ & $0.0130\pm0.00003$ (m) & $4.30\pm0.43$ & $21.2$ & $-1.35\pm-2.73$ & $3.69$ & $-0.01$ & $0.04$ \\
        131858.22+332859.9 & CLU J131858.32+332859.8 & $1.3\pm0.4$ & $74$ & $0.0029\pm0.00285$ (k) & $1.58\pm0.19$ & $19.5$ & $-3.01\pm-3.56$ & $3.77$ & -- & $0.51$ \\
        161439.00+545334.8 & CLU J161439.12+545334.0 & $1.3\pm1.1$ & $76$ & $0.0029\pm0.00285$ (k) & $1.03\pm0.25$ & $19.3$ & $-3.32\pm-3.74$ & $4.69$ & -- & $0.47$ \\
        142859.42+331005.2 & 2MASX J14285953+3310067 & $1.5\pm0.8$ & $863$ & {\bf 0.0291$\pm$0.00009} (k) & $1.97\pm0.29$ & $21.6$ & $-1.00\pm-1.42$ & $3.68$ & -- & $0.08$ \\
        140549.55+365943.9 & CLU J140549.44+365944.5 & $1.5\pm0.3$ & $86$ & $0.0029\pm0.00285$ (k) & $2.66\pm0.21$ & $19.7$ & $-2.74\pm-3.50$ & $3.40$ & -- & $0.73$ \\
        121407.57+423829.2 & KISSR 1246 & $1.8\pm0.1$ & $675$ & $0.0181\pm0.00280$ (k) & $2.86\pm0.11$ & $21.3$ & $-1.28\pm-1.84$ & $3.90$ & -- & $1.06$ \\
        163850.81+352901.0 & CLU J163850.64+352900.9 & $2.0\pm0.7$ & $118$ & {\bf 0.0029$\pm$0.00285} (k) & $2.15\pm0.34$ & $19.6$ & $-2.99\pm-3.47$ & $4.48$ & -- & $0.30$ \\
        110704.14+391812.3 & CLU J110704.32+391811.8 & $2.2\pm0.4$ & $127$ & $0.0029\pm0.00285$ (k) & $2.51\pm0.25$ & $19.7$ & $-2.79\pm-3.46$ & $3.61$ & -- & $0.34$ \\
        \hline
    143050.99+410642.6 & SDSS J143051.12+410640.8 & $2.4\pm0.2$ & $1632$ & {\bf 0.0342$\pm$0.00011} (k) & $2.71\pm0.18$ & $21.9$ & $-0.66\pm-1.41$ & $3.12$ & -- & $0.41$ \\
        220737.01+231516.0 & CLU J220737.20+231515.8 & $2.6\pm0.1$ & $154$ & $0.0029\pm0.00285$ (k) & $12.19\pm0.33$ & $20.3$ & $-2.30\pm-3.27$ & $4.56$ & -- & $1.39$ \\
        090406.54+530314.6 & SDSS J090406.38+530311.8 & $3.1\pm0.0$ & $2410$ & {\bf 0.0386$\pm$0.00007} (k) & $6.69\pm0.13$ & $22.4$ & $-0.43\pm-1.23$ & $4.74$ & $0.18$ & -- \\
        023058.18+232412.6 & CLU J023058.15+232409.3 & $3.3\pm0.6$ & $191$ & $0.0029\pm0.00285$ (k) & $11.49\pm1.31$ & $20.3$ & $-2.35\pm-3.39$ & $4.73$ & $-0.66$ & $0.97$ \\
        140524.35+613358.7 & 2MASS J14052457+6134020 & $3.3\pm0.0$ & $386$ & {\bf 0.0057$\pm$0.00001} (k) & $22.37\pm0.13$ & $21.2$ & $-1.88\pm-2.80$ & $7.23$ & $-0.88$ & $0.55$ \\
        130022.42+281451.7 & 2MASS J13002220+2814499 & $3.7\pm0.8$ & $2002$ & {\bf 0.0266} (k) & $1.47\pm0.33$ & $21.4$ & $-1.12\pm-1.69$ & $3.17$ & -- & $-0.04$ \\
        122250.31+681434.2 & SDSS J122249.71+681431.8 & $3.9\pm0.1$ & $2630$ & {\bf 0.0332$\pm$0.00005} (k) & $2.79\pm0.15$ & $21.8$ & $-0.71\pm-1.34$ & $3.37$ & -- & $0.43$ \\
        091333.83+300056.9 & 2MASX J09133387+3000514 & $5.5\pm0.4$ & $2281$ & {\bf 0.0204$\pm$0.00002} (k) & $5.89\pm0.49$ & $21.7$ & $-0.88\pm-1.66$ & $3.88$ & $-0.97$ & $-0.10$ \\
        113634.77+592533.3 & SBS 1133+597 & $5.6\pm0.0$ & $1182$ & {\bf 0.0104$\pm$0.00011} (k) & $12.82\pm0.12$ & $21.5$ & $-1.30\pm-2.15$ & $5.08$ & $-0.96$ & $0.15$ \\
        125940.18+275123.5 & 2MASX J12594007+2751177 & $5.8\pm0.0$ & $1491$ & {\bf 0.0127$\pm$0.00004} (k) & $22.34\pm0.20$ & $21.9$ & $-1.69\pm-2.27$ & $10.28$ & $-1.00$ & $0.39$ \\
        125944.53+275800.9 & SDSS J125944.76+275807.1 & $7.0\pm1.6$ & $4670$ & $0.0329\pm0.00025$ (k) & $2.38\pm0.56$ & $21.8$ & $-0.87\pm-1.44$ & $4.02$ & -- & -- \\
    \midrule
    \midrule
        161111.24+360401.0 & CLU J161111.2+360400 & $0.3\pm0.1$ & $75$ & $0.0121\pm0.01196$ (n) & $8.45\pm0.17$ & $21.4$ & $-1.10\pm-2.30$ & $3.49$ & $-0.79$ & $1.15$ \\
    143037.30+352053.3 & SDSS J143037.09+352052.8 & $2.6\pm0.6$ & $2036$ & $0.0390\pm0.00000$ (k) & $1.18\pm0.21$ & $21.6$ & $-0.92\pm-1.29$ & $3.35$ & -- & $1.64$ \\
    \bottomrule
    \end{tabular}
    \label{table:candidates}
    
    \begin{tablenotes}
      \small
      \item Columns: 
        Source name in LoTSS DR2; 
        host galaxy name in CLU; 
        projected offset between the radio source and H$\upalpha$ source coordinates in arcsec, and corresponding transverse distance in parsec; 
        redshift ($z$), with distance method indicated in parentheses: (k) kinematic; (m) median (redshift-independent), (n) narrowband (H$\upalpha$); 
        star formation rate ($\log {\rm SFR}$); 
        radio luminosity at 144\,MHz ($\log L$); 
        standard deviation ($\sigma$) above the L-SFR relation; 
        radio spectral index ($\alpha$; \S\ref{subsec:spectral}); radio loudness $R_g$ (\S\ref{sec:discussion}).
        Sources with spectral line measurements in SDSS table galSpecLine~\citep[\S\ref{subsec:bpt};][]{Tremonti2004ApJ...613..898T, Brinchmann2004MNRAS.351.1151B} have their redshifts marked in bold. 
        Candidates below and above an offset value of $(2+\epsilon)\arcsec$ are demarcated with a solid line, with $\epsilon$ being the LoTSS astrometric uncertainty.
        The third (bottom) section includes candidates in the AGN region of Figure \ref{fig:wise}.
    \end{tablenotes}
    \end{threeparttable}
\end{sidewaystable*} 

\subsection{Chance alignment probability}
\label{subsec:chance}

We begin by evaluating the likelihood of matching a galaxy and a background radio source by coincidence due to chance alignment following the two methods described by \citet{Reines2020ApJ...888...36R}.
Firstly, we estimate the cumulative number of compact radio source counts per steradian $N(S_{min})$ with 144\,MHz flux densities greater than $S_{min}=0.8\,\rm{mJy}$, taking into account  only compact sources based on {\tt pyBDSF} single-component cases and the $R_{99.9}$ compactness criterion.
Multiplying $N(S_{min})\approx2,051,534\,{\rm steradian}^{-1}$
by the area confined in a $6\arcsec$ radius circle gives $N_{bk,gal} = 0.0032$, the expected number of background sources for a given galaxy.
Across our entire parent sample of 18,159 
dwarf galaxies, we expect $N_{bk,samp} \approx 58 \pm \sqrt{58} \approx 58 \pm 8$ background sources (where the error is computed for Poisson statistics).
We therefore expect $N_{bk,samp}$ to be present in our original 708 matches, or about 0.8\%.
Performing a similar analysis using a cross-matching radius of $2\arcsec$ leads to $N_{bk,samp} \approx 6 \pm 2$ background sources (with an original number of matches of 573 using a $2\arcsec$ cut), or about $1\%$.
Considering that outliers caused by a false association should be evenly distributed above and below the L--SFR relation (Figure \ref{fig:selection}), it is likely that nearly all candidates below the $(2+\epsilon)\arcsec$ mark and about half of the candidates above this threshold 
are true associations.

\begin{figure}
\resizebox{\hsize}{!}{\includegraphics{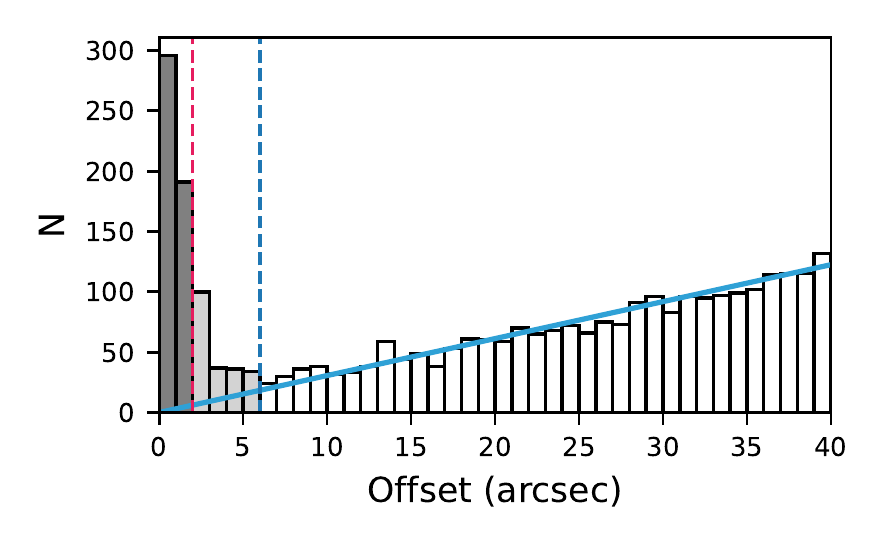}}
\caption{Observed offset distribution after cross-matching CLU and LoTSS surveys, using a match radius of $40\arcsec$.
We select our target galaxies to have offsets of $2\arcsec$ (pink vertical dashed line) and $6\arcsec$ (blue vertical dashed line).
The blue solid line shows the expected number of chance alignments with background sources as a function of offset.}
\label{fig:chance_hist}
\end{figure}

Secondly, we perform an empirical estimate of the expected number of false associations using the data.
We cross-match each source in the CLU subsample to its nearest neighbor in LoTSS, up to a maximal matching radius of $40\arcsec$.
Figure~\ref{fig:chance_hist} shows the observed offset distribution.
The offset probability histogram for chance associations should be equal to zero at an offset of zero and rise linearly for small offsets (blue solid line in Figure~\ref{fig:chance_hist}).
The observed distribution is minimal at a matching radius of $6\arcsec$, beyond which the number of sources per offset, $N(d_{\rm{off}})$, increases linearly as $N(d_{\rm{off}})=3.1d_{\rm{off}}$, with offset $d_{\rm{off}}$ in units of arcsec.
The total estimated number of background sources with offsets of less than $6\arcsec$ is found by integrating $N(d_{\rm{off}})$ from $N(d_{\rm{off}}=0\arcsec$ to $5\arcsec$, which gives $N_{bk,samp} \approx 64 \pm 8$ sources.
This is consistent with our calculation above using known radio source counts.
Here again, putting a cut at $2\arcsec$ (pink vertical line) leads to $N_{bk,samp} \approx 9 \pm 3$, which is also consistent within errors with the analytical method.

From these two analyses, it is unclear whether outliers on the L--SFR relation found in \S\ref{sec:method} are caused by chance alignment, where two unrelated sources lead to an incorrect luminosity calculation (e.g., a radio source is assigned an incorrect redshift).
To resolve this issue, we performed Monte Carlo simulations.
We repeated all cross-matching and outlier selection steps (as described in \S\ref{subsec:sample} and \S\ref{subsec:crossmatch}) after randomly shifting spatial coordinates for all sources within the LoTSS set in each Monte-Carlo run.
For both cases, using $2\arcsec$ and $6\arcsec$ as the cross-matching radius limit, we performed 1000 realizations of the process, selecting spatial shifts from a uniform distribution of $[-10, 10]$ arcmin, keeping track of the number of outliers at each realization.

\begin{figure}
\resizebox{\hsize}{!}{\includegraphics{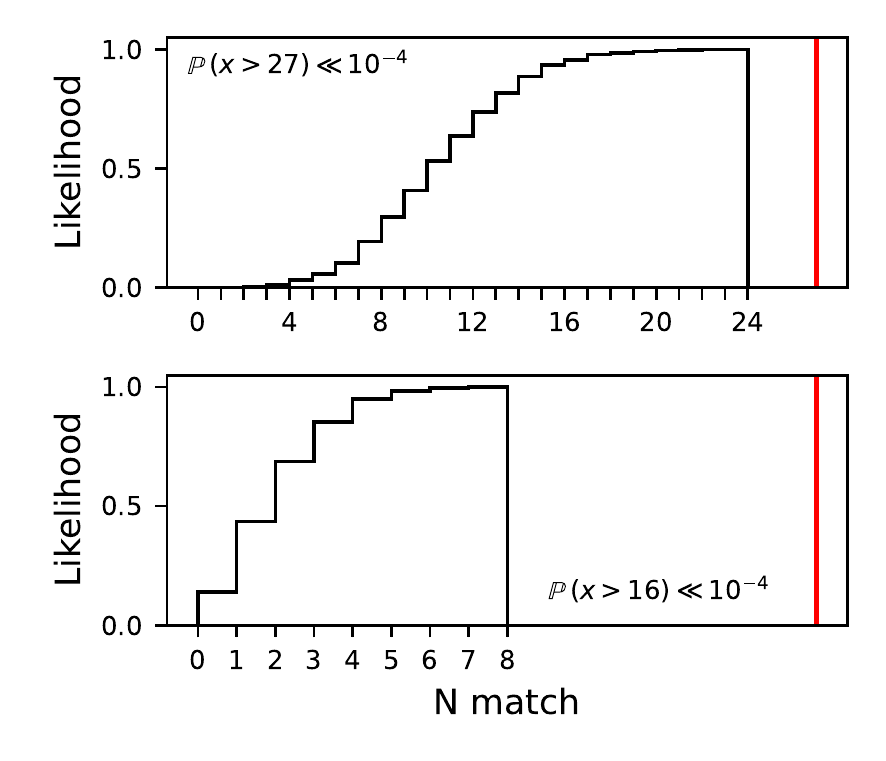}}
\caption{Cumulative distributions from 1000 Monte Carlo realizations of matches exceeding $3\sigma$ on the L--SFR relation from \citet{Gurkan2018MNRAS.475.3010G}. Top panel: $6\arcsec$ matching radius; bottom panel: $2\arcsec$ matching radius.}
\label{fig:montecarlo}
\end{figure}

Cumulative distributions summarising each Monte Carlo simulation are shown in Figure~\ref{fig:montecarlo}.
At $6\arcsec$,  for the entire sample of CLU, on average, it is common to find 10 matches (median) above the $3\sigma$ mark of the L--SFR relation by chance, and is unlikely to find 27 by chance ($p \ll 10^{-4}$). 
At a $2\arcsec$ cutoff, it is common to get two matches by chance, and highly unlikely to find 16.
Given our 16 matches selected below $(2+\epsilon)\arcsec$, we expect a false-positive fraction of 0.125. 
The remaining 11 sources with offsets between $(2+\epsilon)-(6+\epsilon)\arcsec$ may be chance alignments.

While this does not mean that candidates with a projected offset of greater than $2\arcsec$ are necessarily chance alignments ---for example, a small transverse distance may indicate a true association---, assessing whether or not they are real associations remains difficult with the currently available data.
For this reason, we demarcate candidates below and above $(2+\epsilon)\arcsec$ in Table~\ref{table:candidates} and the following sections with a black line.
Nevertheless, we show in the following section that nearly all candidates---both below and above this demarcation---fall within the optical footprint of their respective matched galaxies.

\begin{figure}
\resizebox{\hsize}{!}{\includegraphics{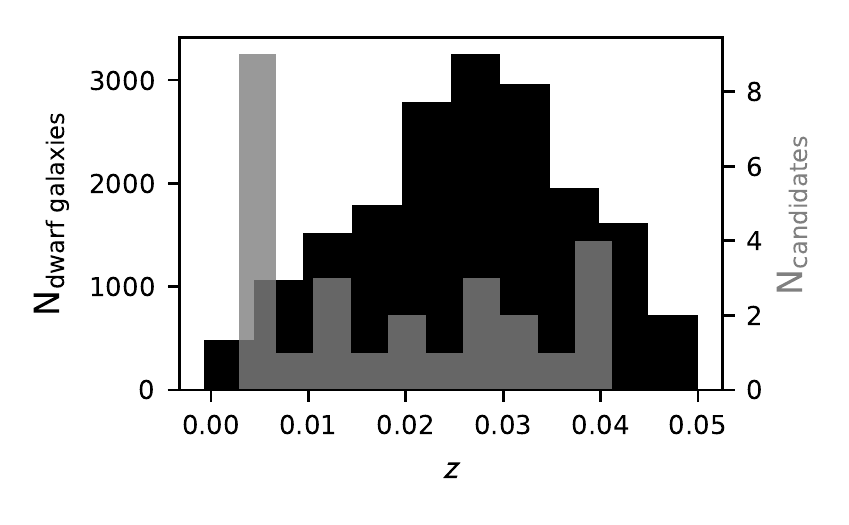}}
\caption{Redshift distributions of selected candidates listed in Table \ref{table:candidates} (gray) and for all dwarf galaxies from CLU in the LoTSS field (black). 
A two-sample KS test between these two distributions gives a $p$-value of $\ll 0.005$.}
\label{fig:ks-test}
\end{figure}

Finally, the redshift distribution of the selected candidates listed in Table \ref{table:candidates}  is vastly different from that of all dwarf galaxies from CLU in the LoTSS field (Figure \ref{fig:ks-test}).
A two-sample Kolmogorov-Smirnov (KS) test between these two distributions lends a $p$-value of $1.08 \times 10^{-5}$, indicating the selected candidates do not simply track the redshift distribution of the whole CLU dwarf galaxy sample. 

%-------------------------------------------------------------------

\section{Candidates}
\label{sec:candidates}

\begin{figure*}
\centering
\begin{tabular}{cccc}
\includegraphics[width=41mm]{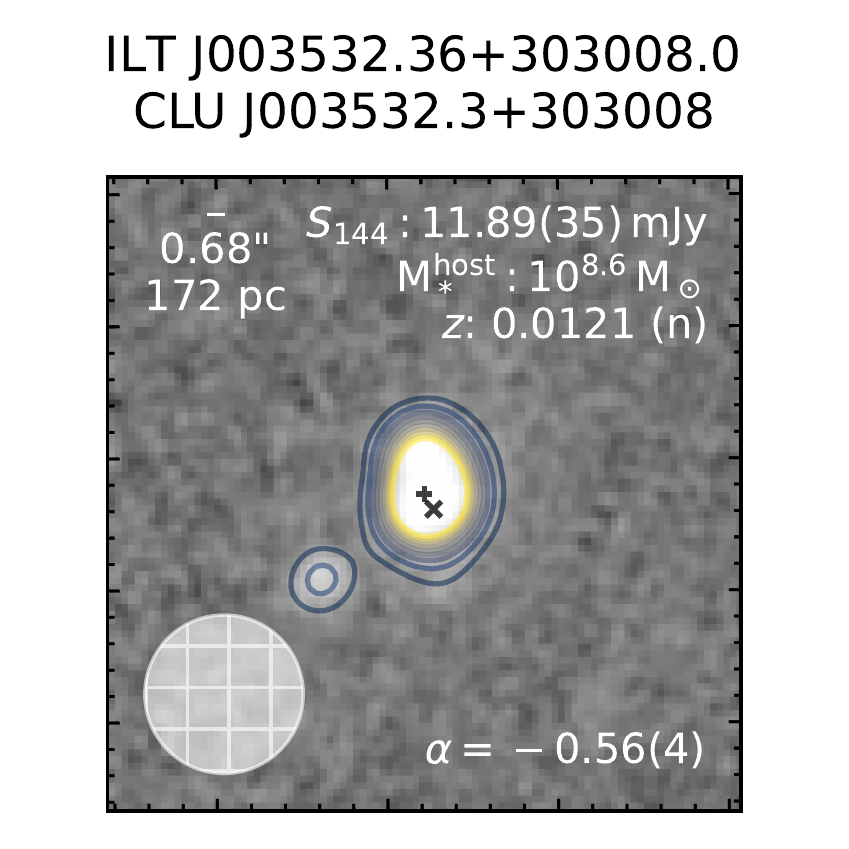} &
\includegraphics[width=41mm]{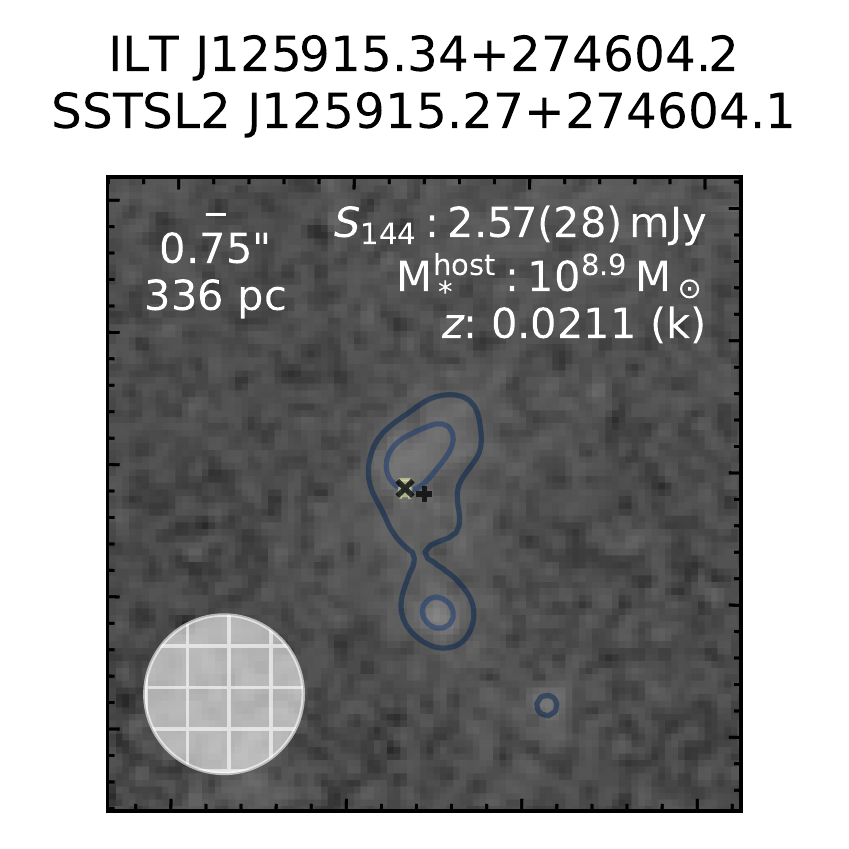} &
\includegraphics[width=41mm]{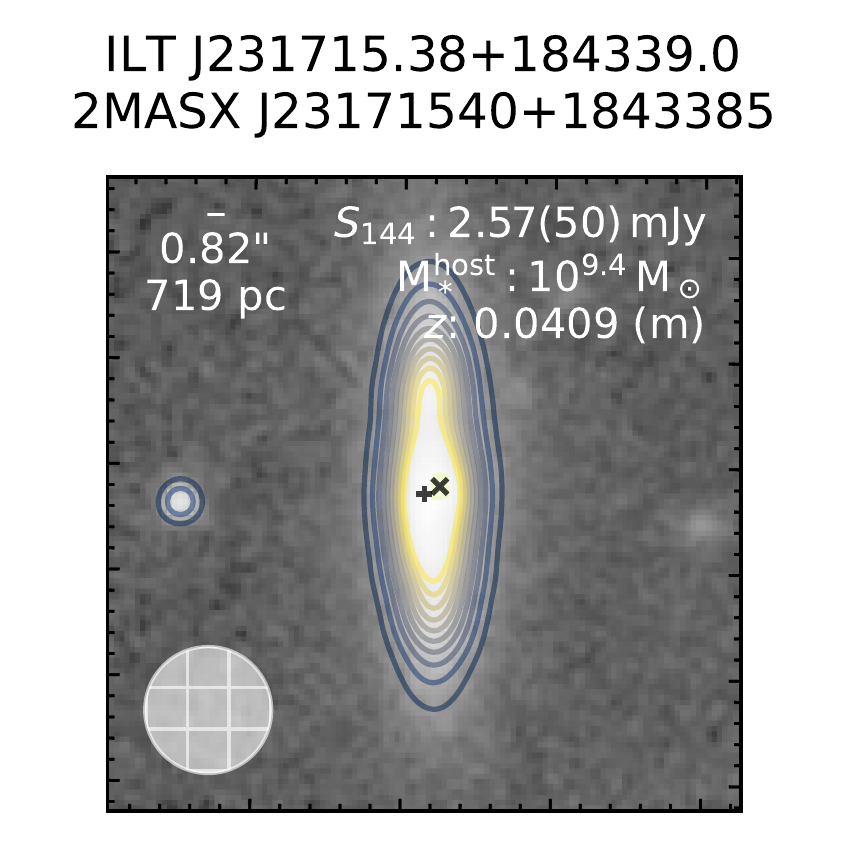} &
\includegraphics[width=41mm]{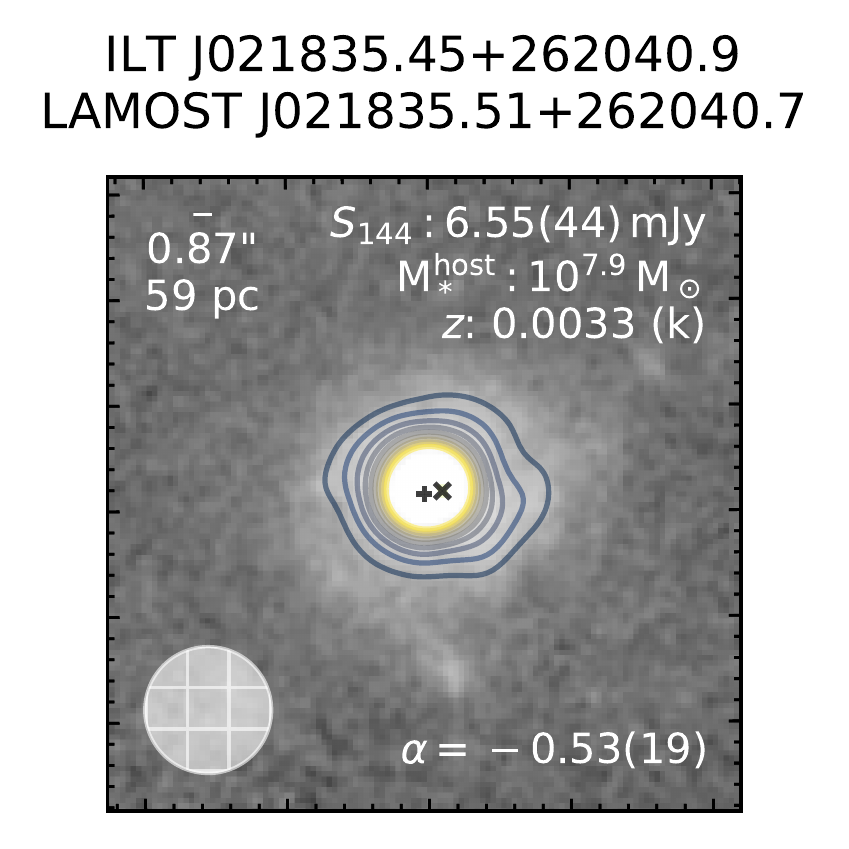} \\[0.2cm]
\includegraphics[width=41mm]{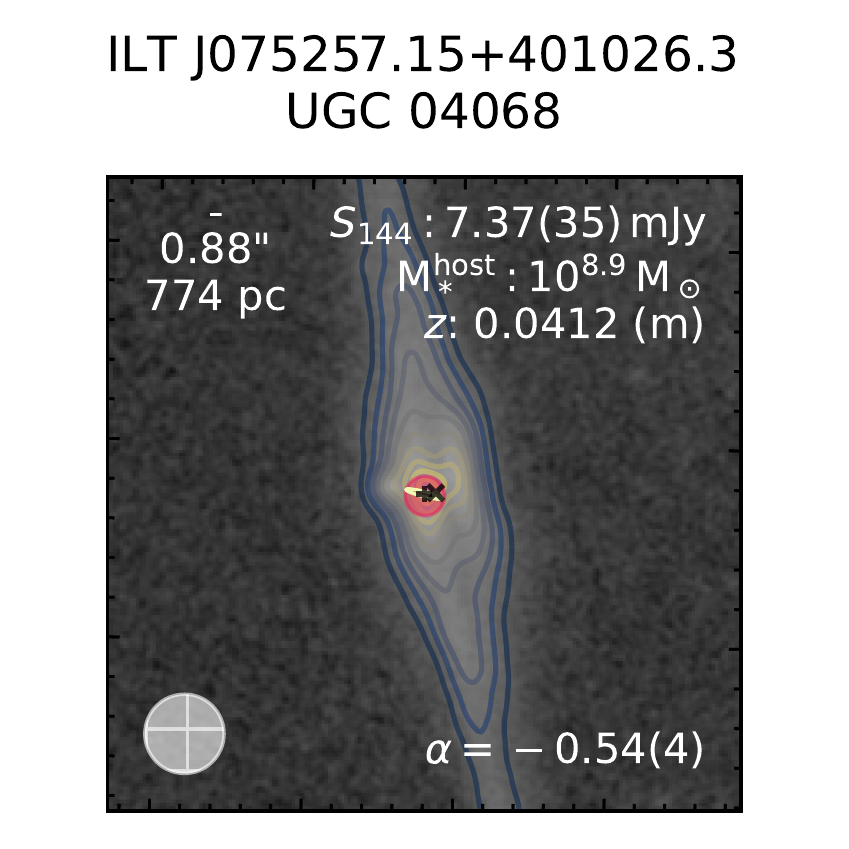} &
\includegraphics[width=41mm]{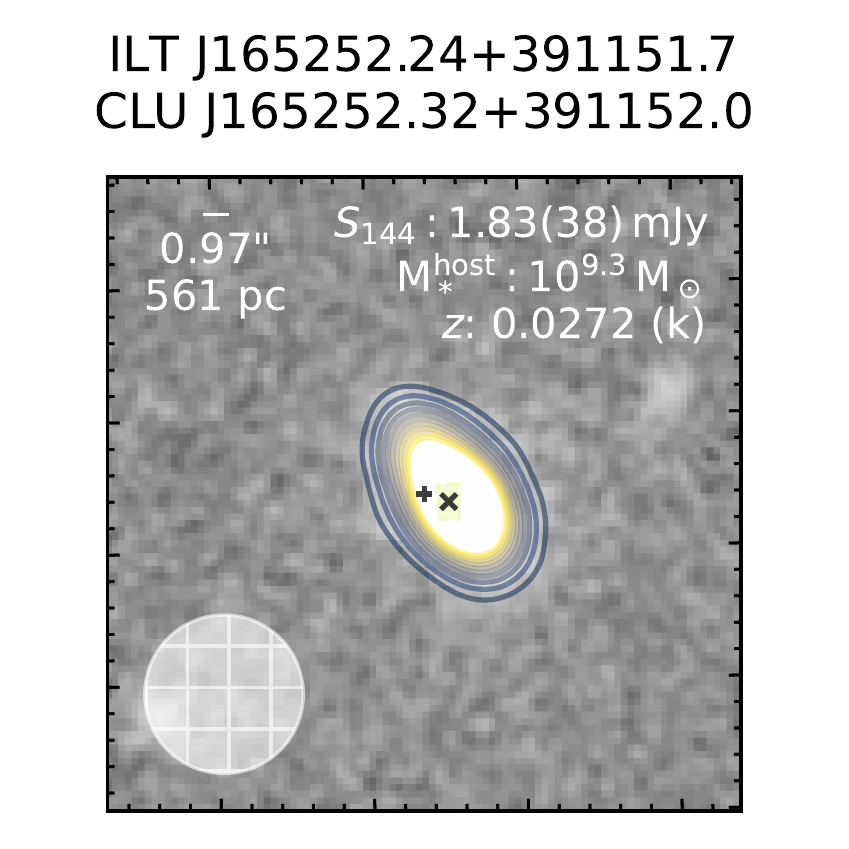} &
\includegraphics[width=41mm]{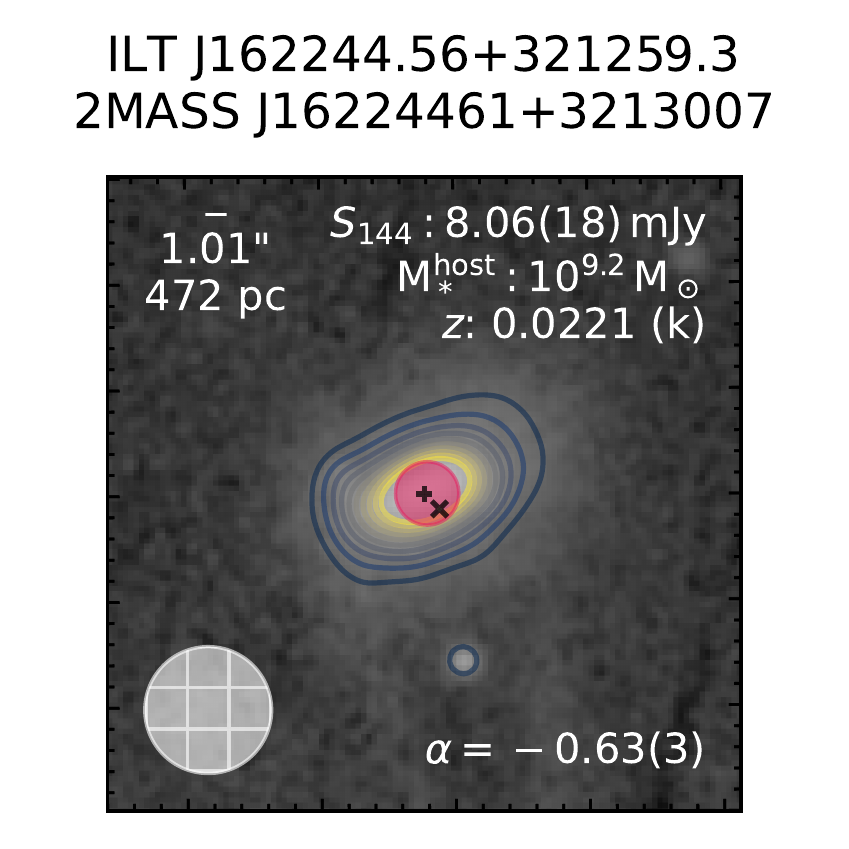} &
\includegraphics[width=41mm]{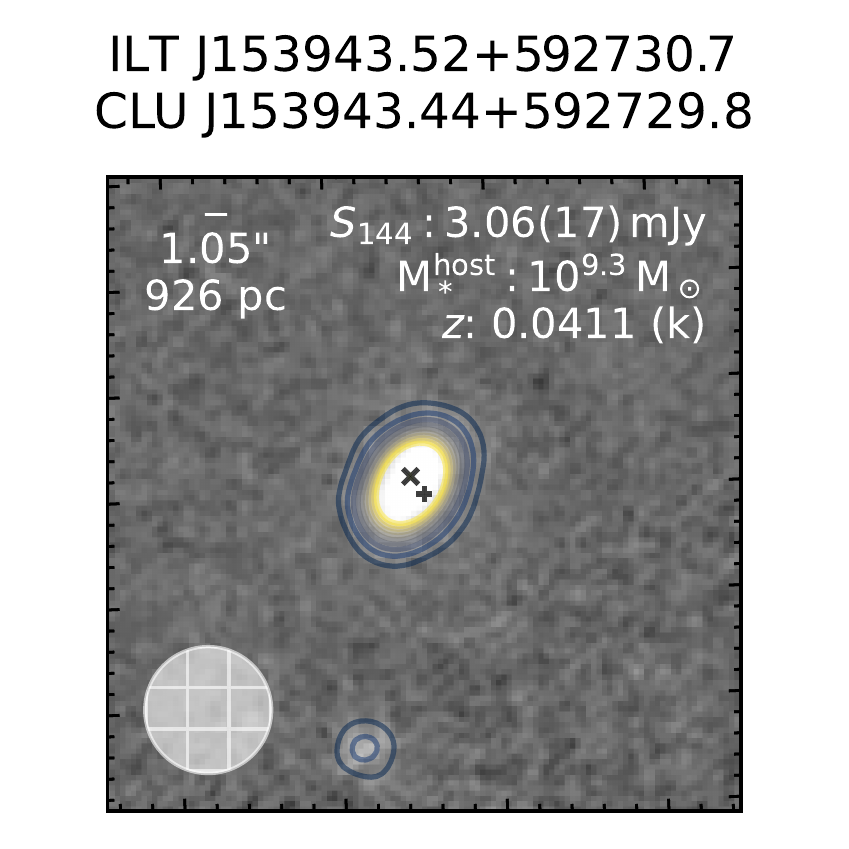} \\[0.2cm]
\includegraphics[width=41mm]{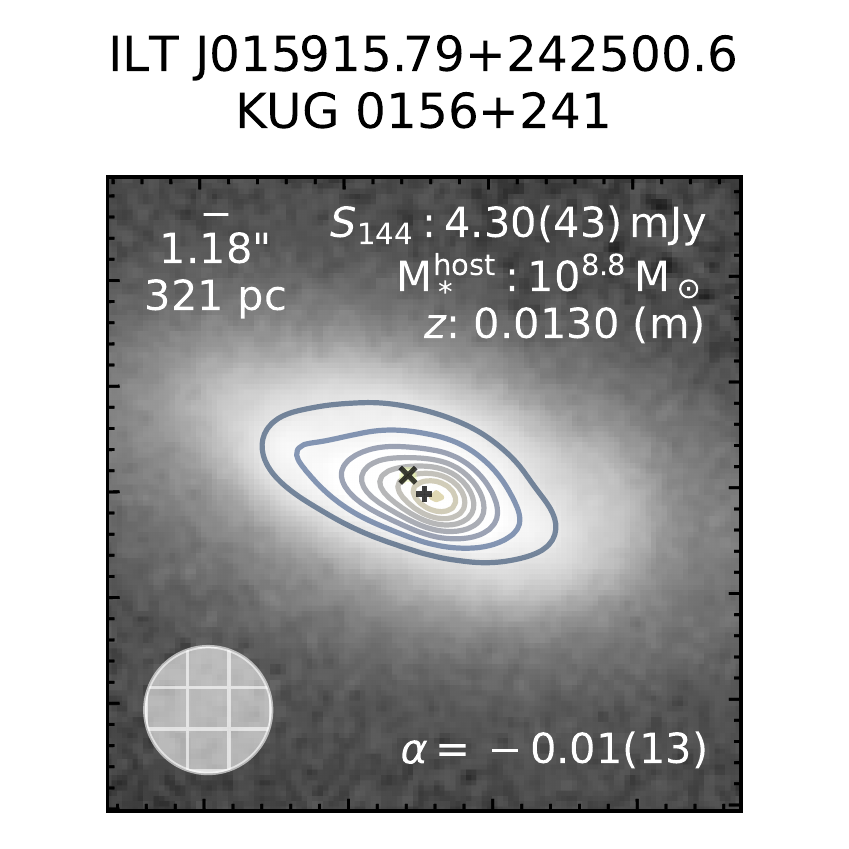} &
\includegraphics[width=41mm]{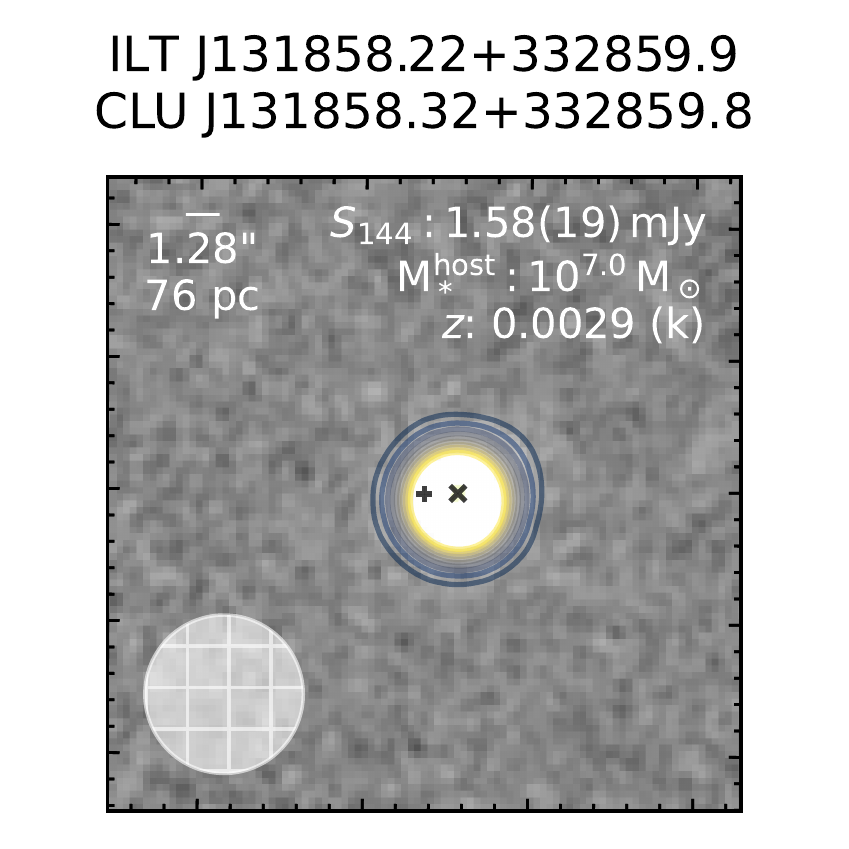} &
\includegraphics[width=41mm]{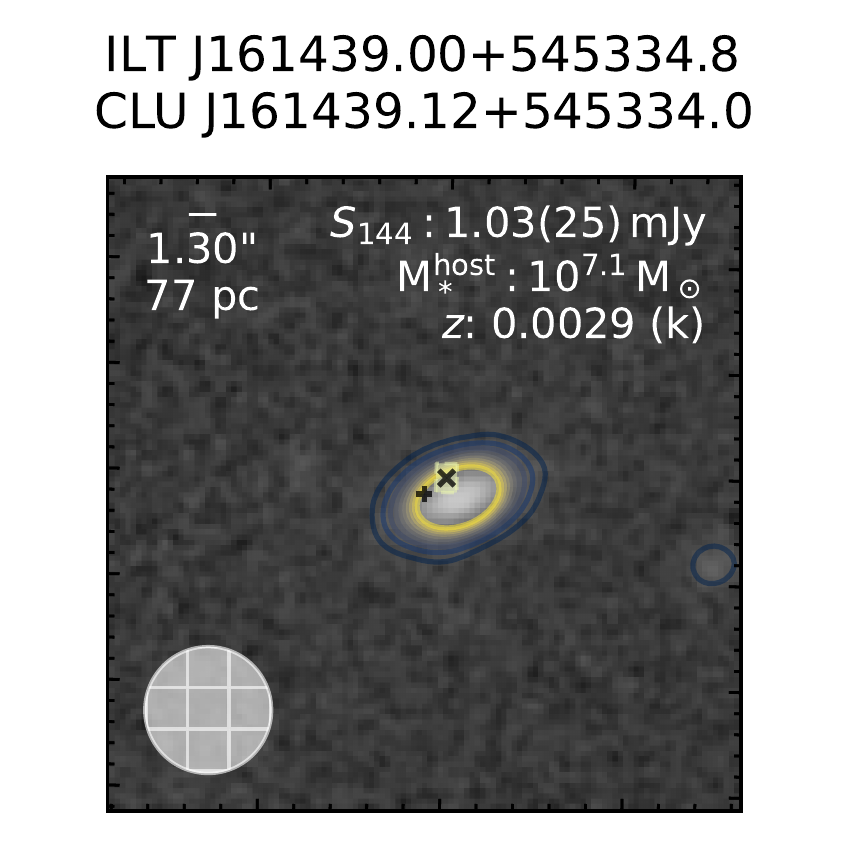} &
\includegraphics[width=41mm]{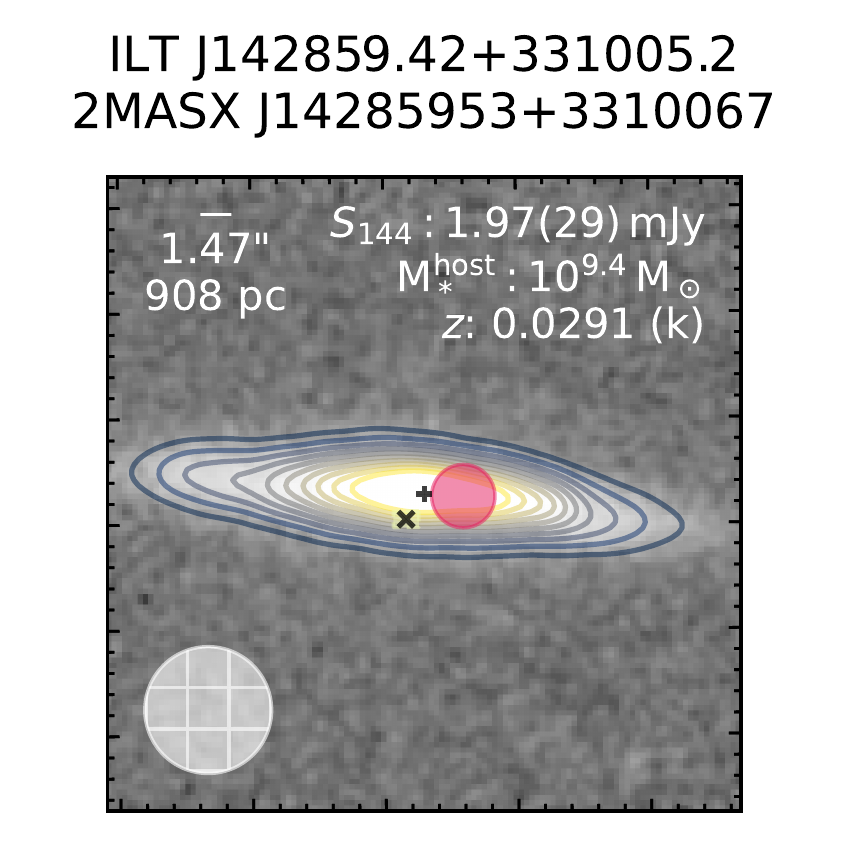} \\[0.2cm]
\includegraphics[width=41mm]{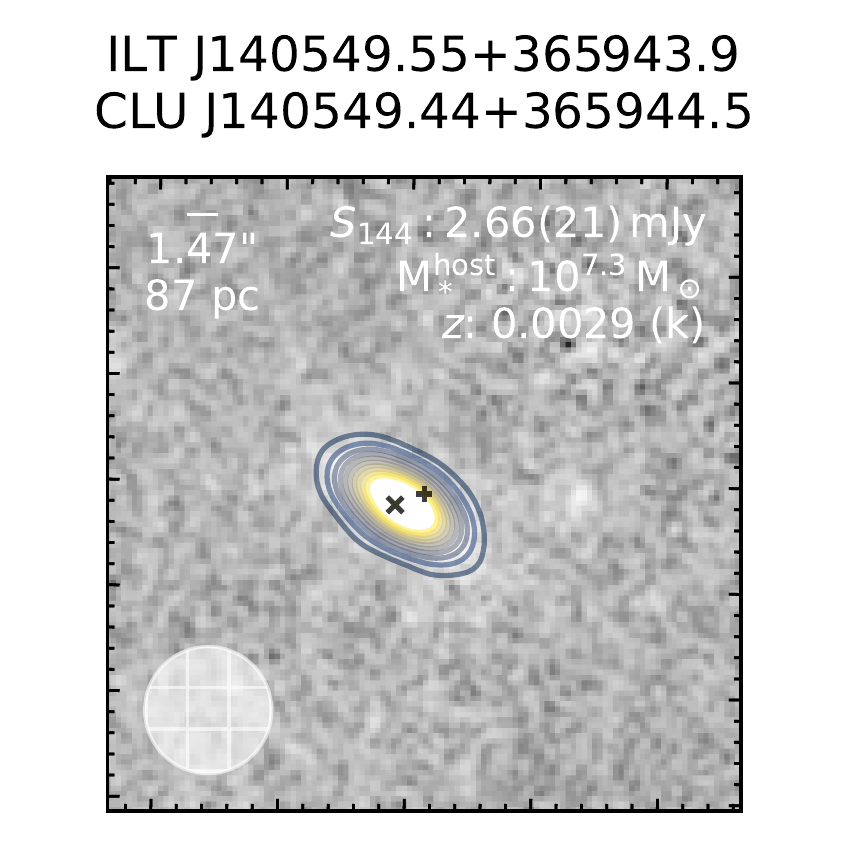} &
\includegraphics[width=41mm]{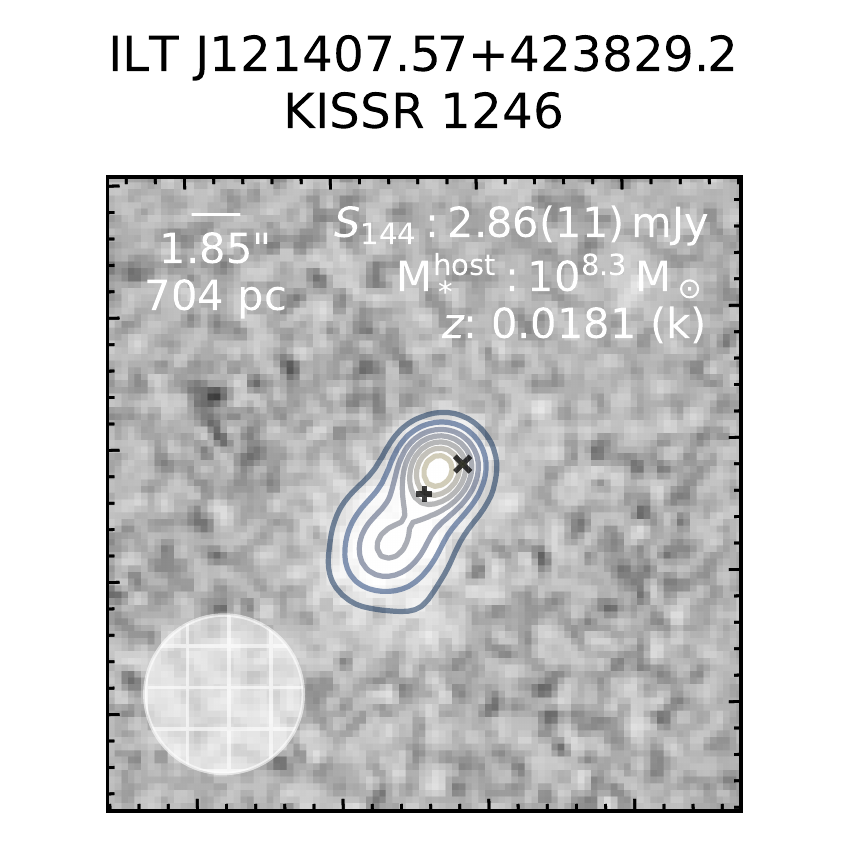} &
\includegraphics[width=41mm]{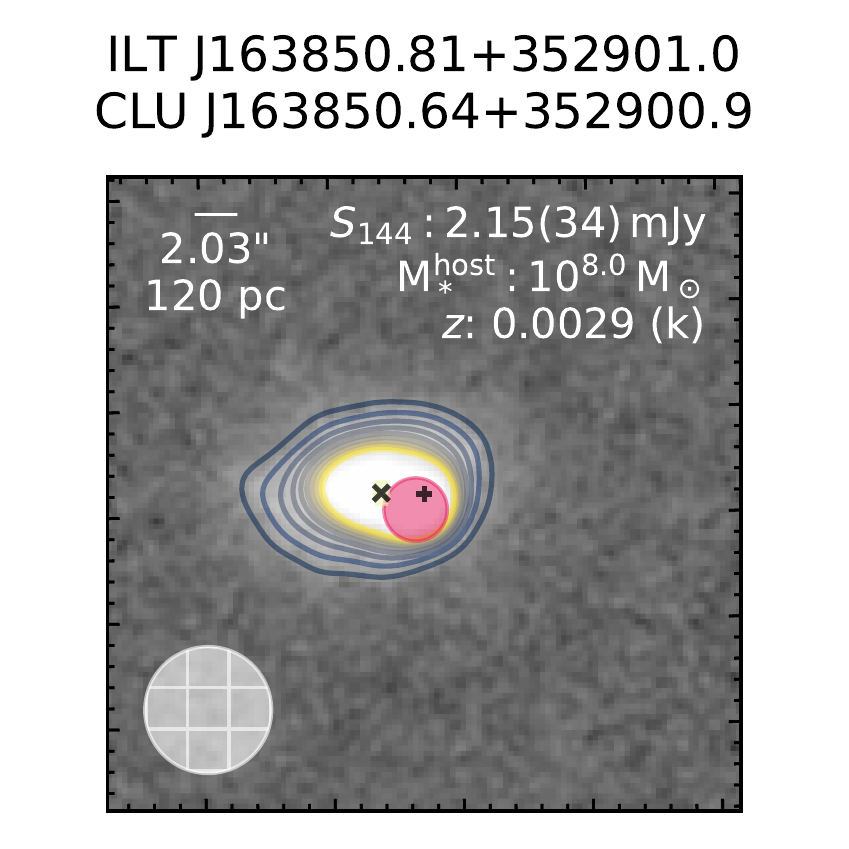} &
\includegraphics[width=41mm]{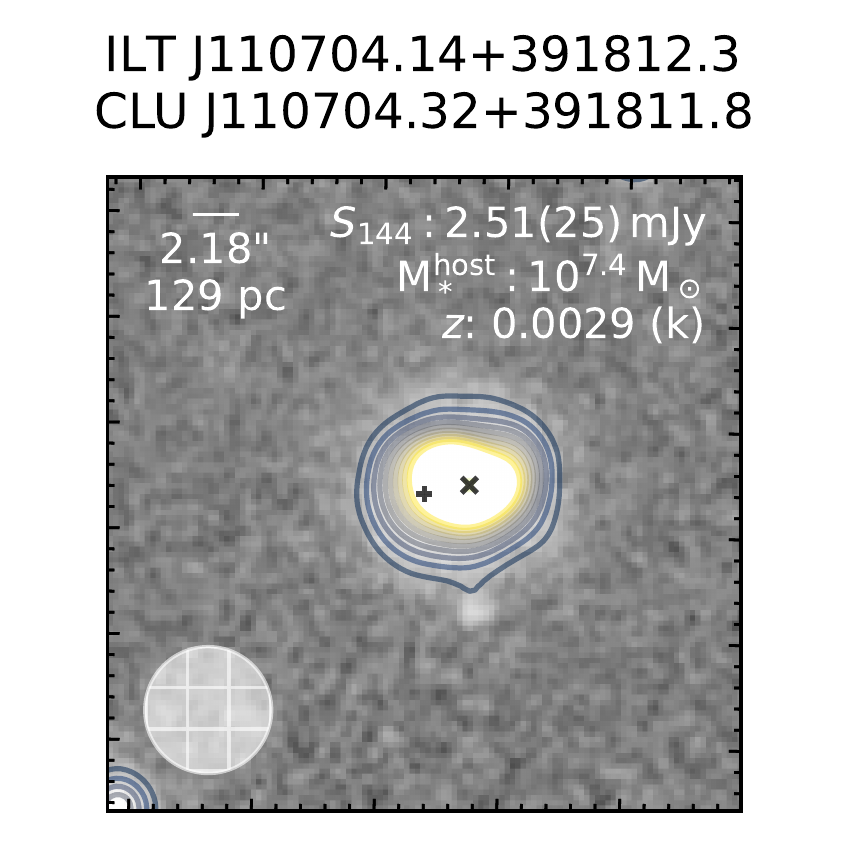} 
\end{tabular}
\caption{
    Over-luminous compact radio sources with a projected offset of less than $(2+\epsilon)\arcsec$. Each panel indicates the source name from LoTSS and the host galaxy name above a PS1 R filter image in logarithmic gray scale. 
    Contours indicate optical data from PS1~\citep[][]{Chambers2016arXiv161205560C, Flewelling2020ApJS..251....7F} at $\mu+[1, 2, ..., 10]\sigma$ levels, with $\mu$ and $\sigma$ being the median and standard deviation, respectively. 
    The black crosses and plus symbols ($\times$, +) indicate the central coordinates of matched objects from LoTSS and CLU, respectively. 
    LoTSS astrometric uncertainty is marked as a yellow box.
    Where available in the CLU catalog, we indicate a yellow ellipse corresponding to the H$\upalpha$ detection isophote (D25).
    For each panel, in the top left, we indicate the spatial offset ($\arcsec$) between the $\times$ and + markers as a white bar, along with corresponding transverse distance (pc). 
    Top right: LoTSS radio flux at 144\,MHz in mJy, with uncertainty on the last digit in parenthesis, host galaxy stellar mass (${\rm M^{\rm host}_*}$) in ${\rm M_\odot}$, and redshift ($z$), with distance method indicated in parentheses. Here, (k) means kinematic, (m) median (redshift-independent), and (n) narrowband (H$\upalpha$). 
    Where available, we indicate the spectral index ($\alpha$; Figure~\ref{fig:spectral_indices}; \S\ref{subsec:spectral}) and the location of the SDSS spectroscopic fiber as a filled yellow circle (Figure \ref{fig:bpt}; \S\ref{subsec:bpt}, \S\ref{fig:spectral_indices}).
    Bottom left: White circles indicate the LoTSS $6\arcsec$ beam, noting that the restoring beam used in DDFacet~\citep{Tasse2018A&A...611A..87T} for each image product type is kept constant over the data release region and that all image products are made with a $uv$-minimum of 100\,m with the $uv$-maximum varied to provide images at different resolutions; the highest resolution $6\arcsec$ images use baselines up to 120\,km (i.e., all LOFAR stations within the Netherlands).
}
\label{fig:family_plot}
\end{figure*}

\begin{figure*}
\centering
\begin{tabular}{cccc}
\includegraphics[width=41mm]{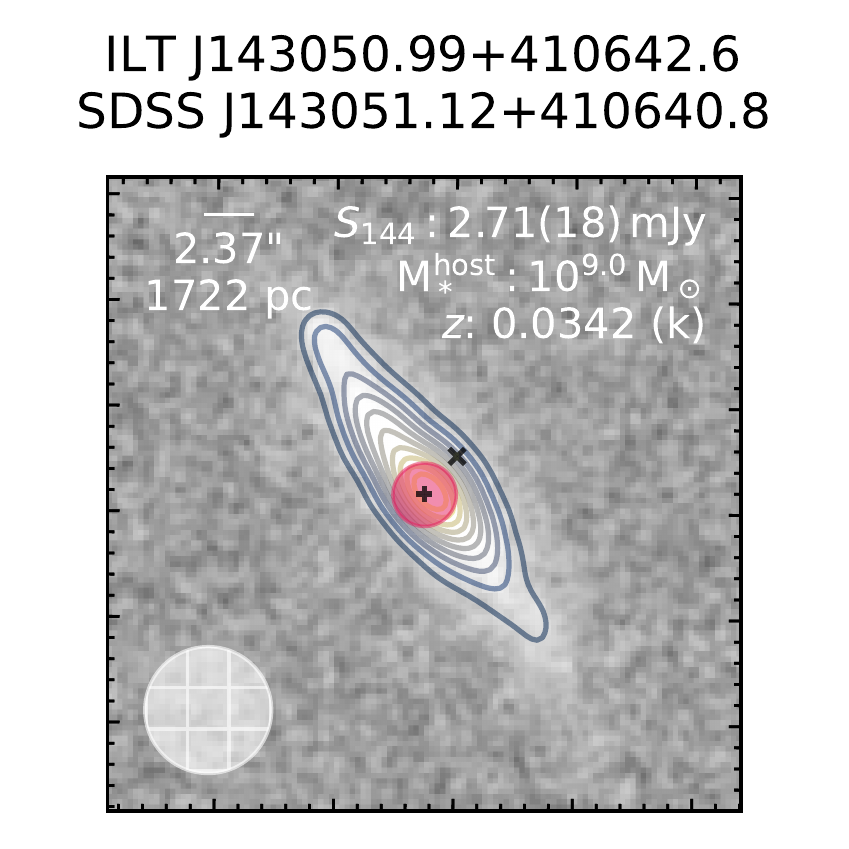} &
\includegraphics[width=41mm]{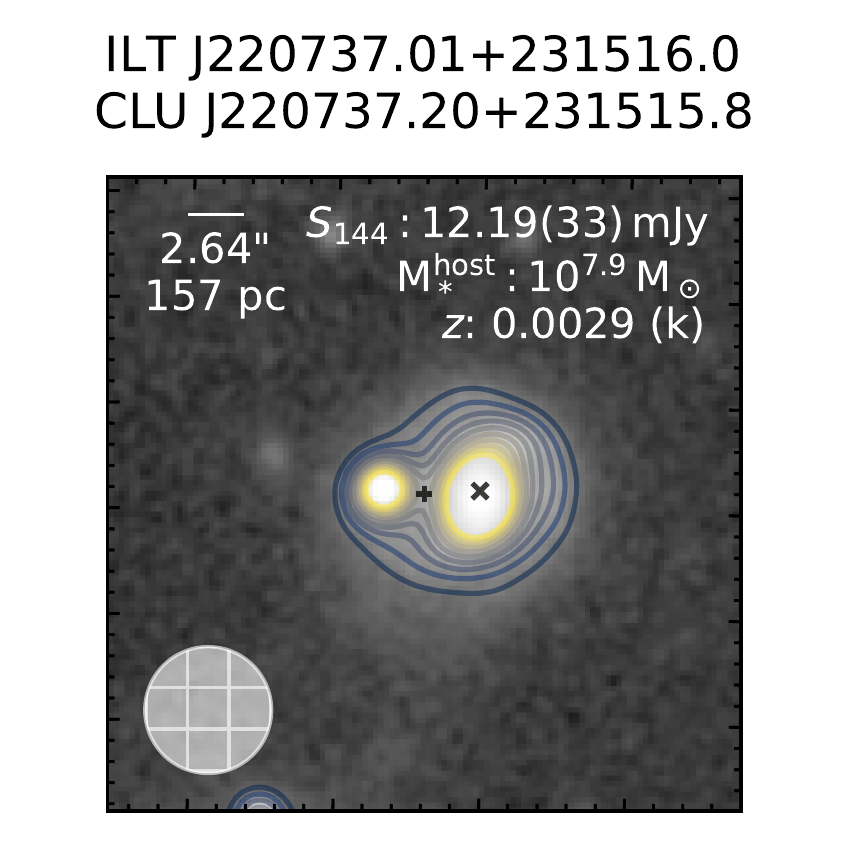} &
\includegraphics[width=41mm]{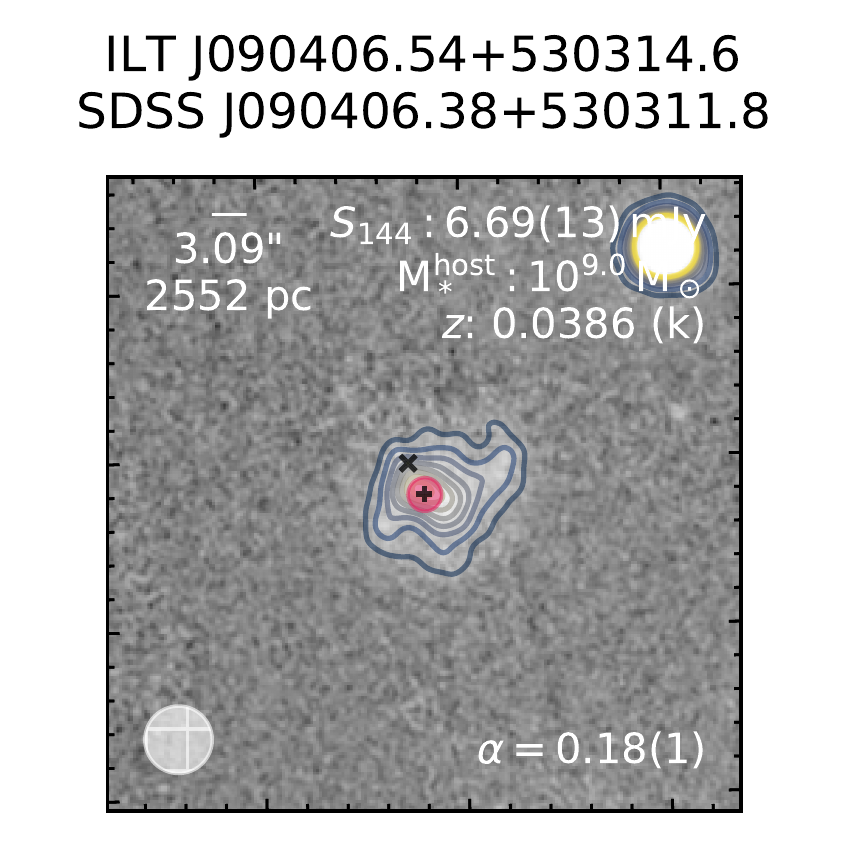} &
\includegraphics[width=41mm]{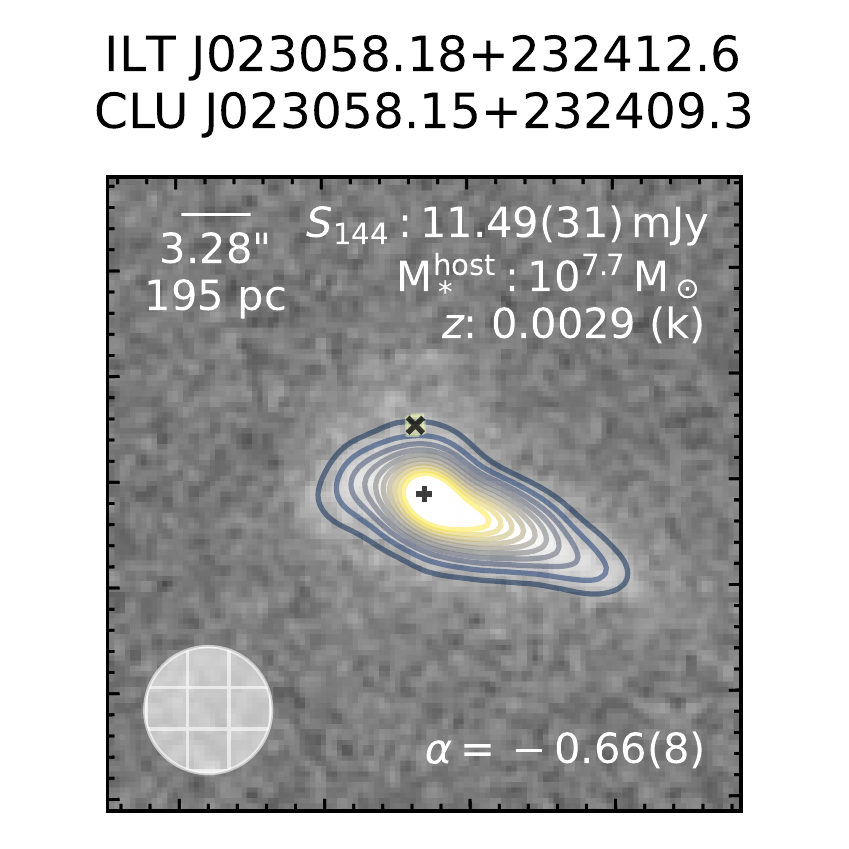} \\[0.2cm]
\includegraphics[width=41mm]{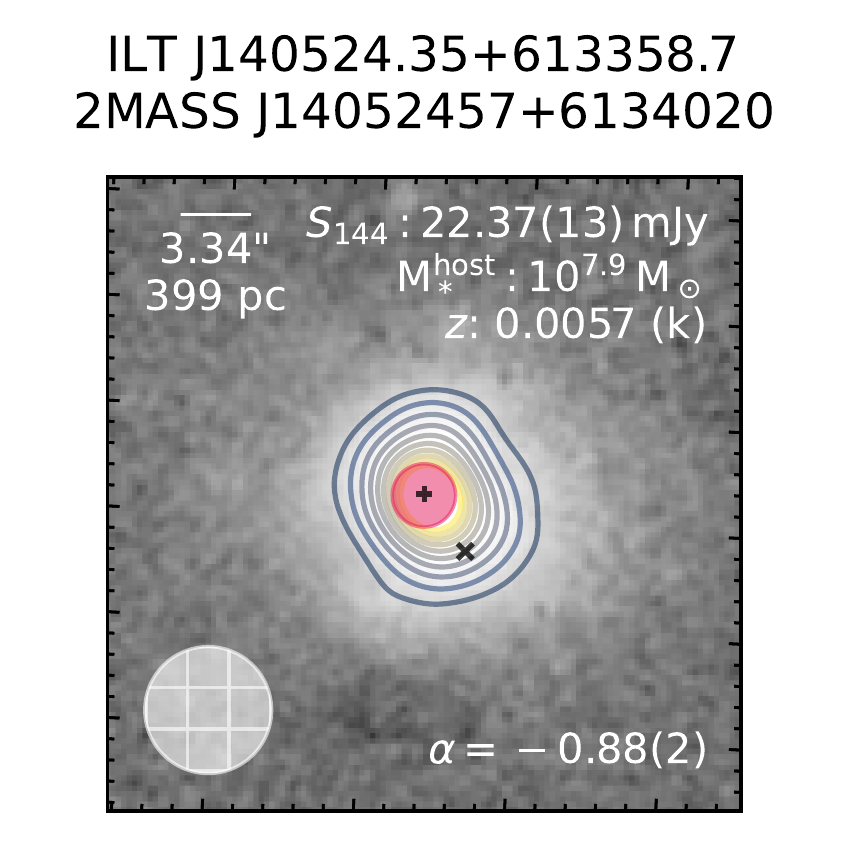} &
\includegraphics[width=41mm]{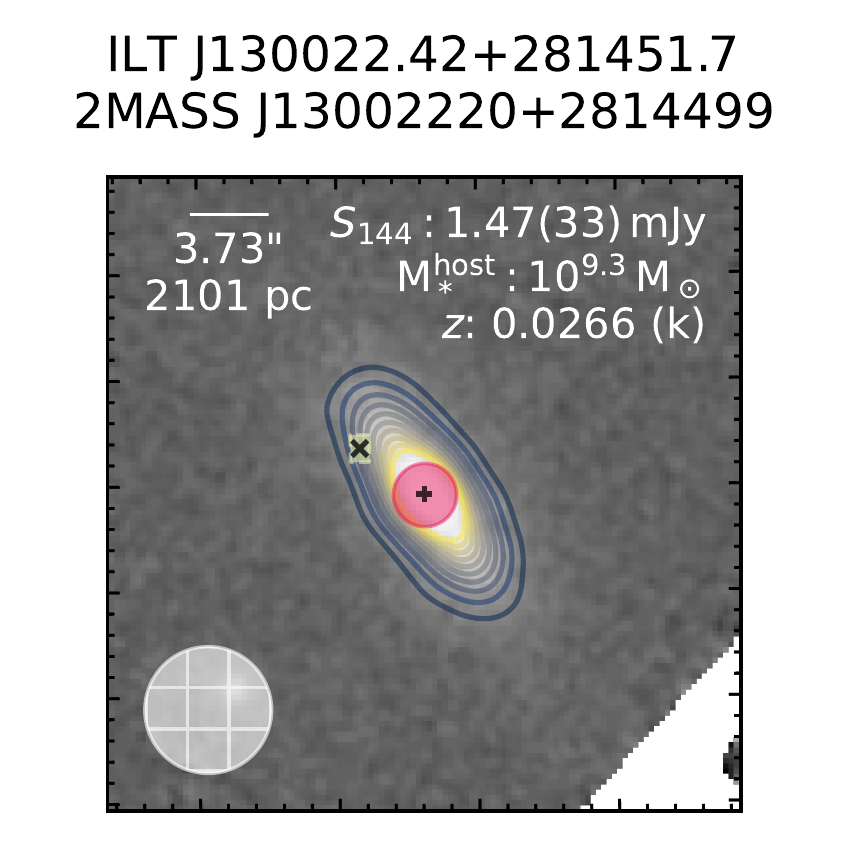} &
\includegraphics[width=41mm]{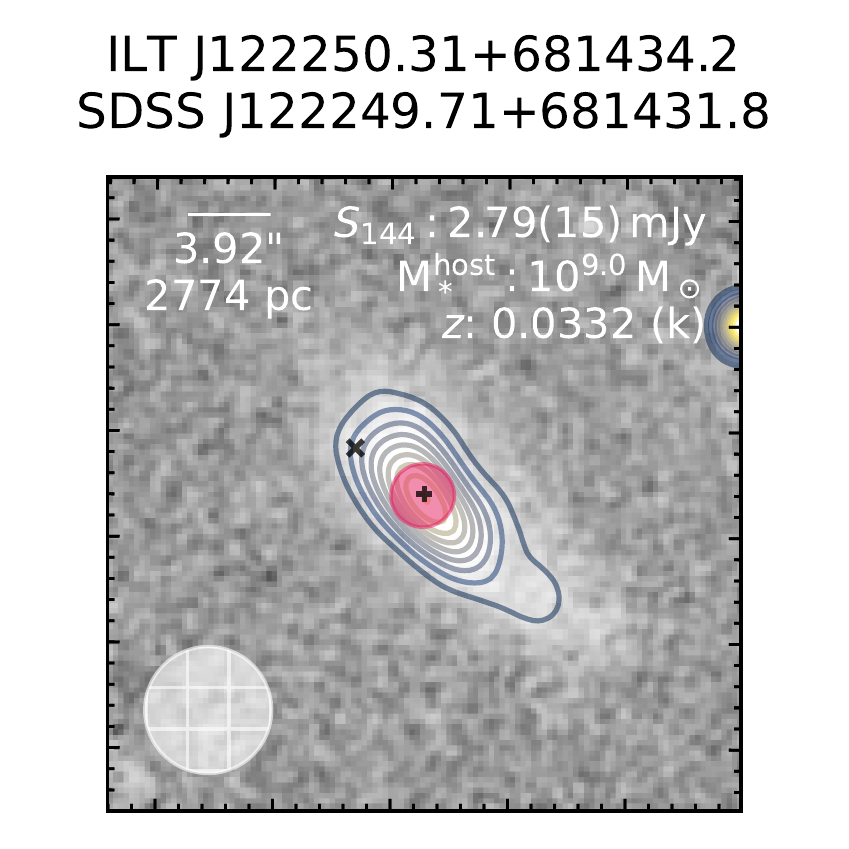} &
\includegraphics[width=41mm]{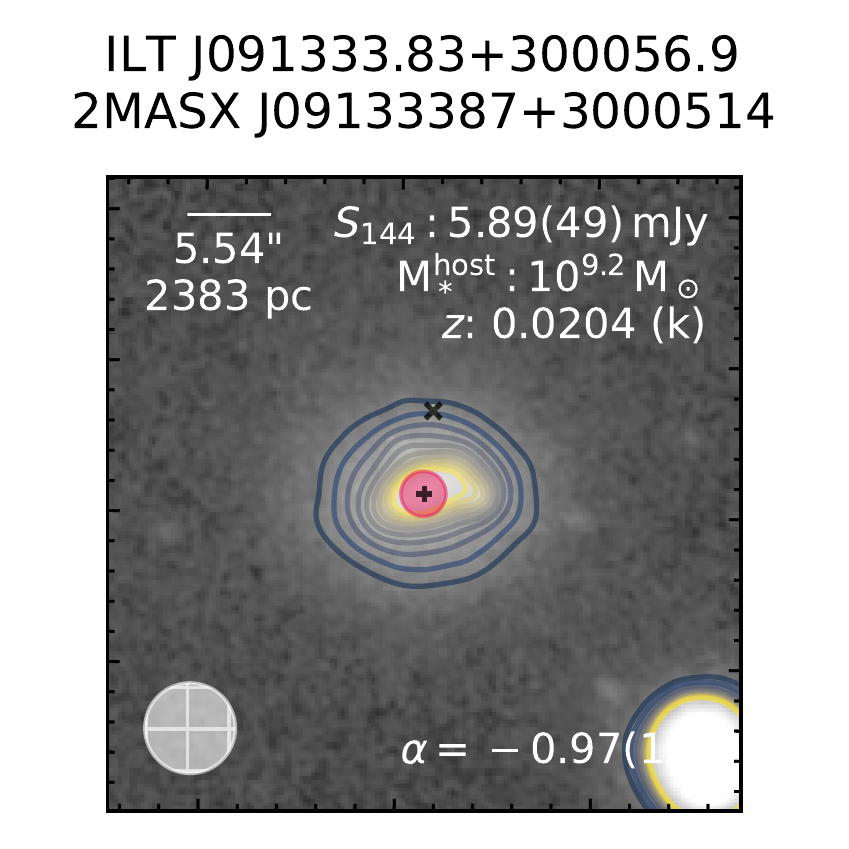} \\[0.2cm]
\end{tabular}
\begin{tabular}{ccc}
\includegraphics[width=41mm]{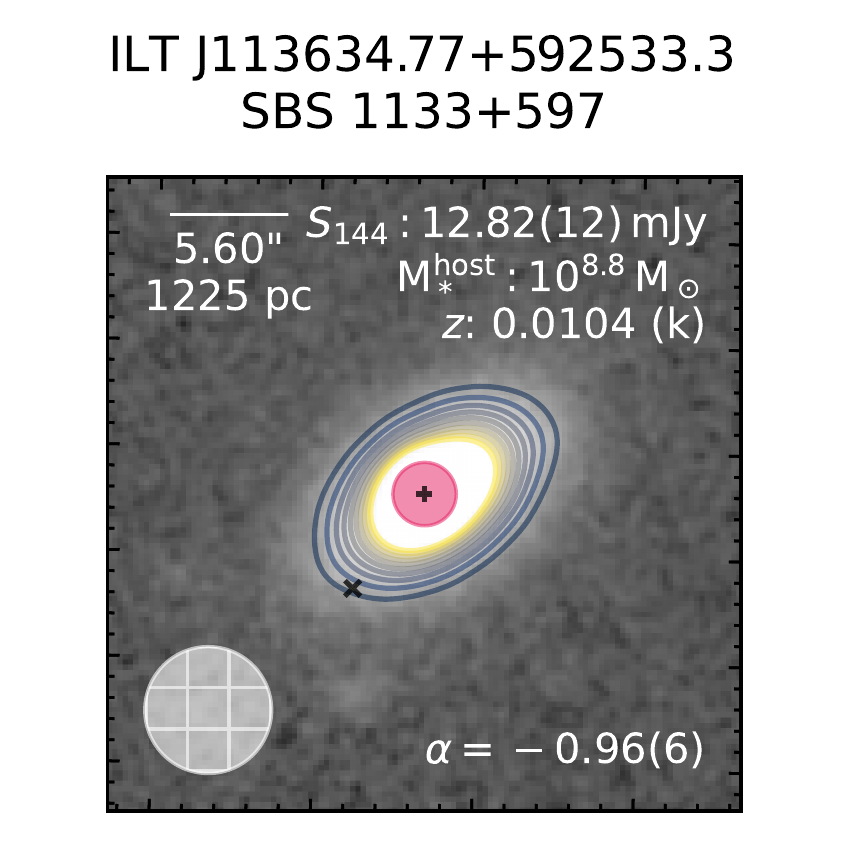} &
\includegraphics[width=41mm]{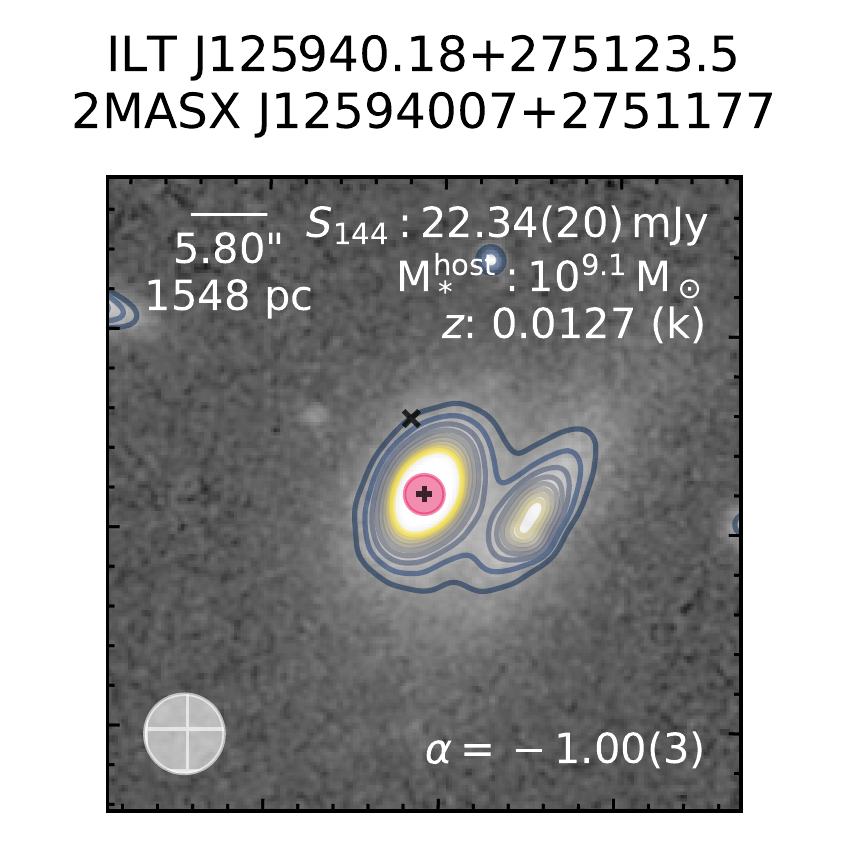} &
\includegraphics[width=41mm]{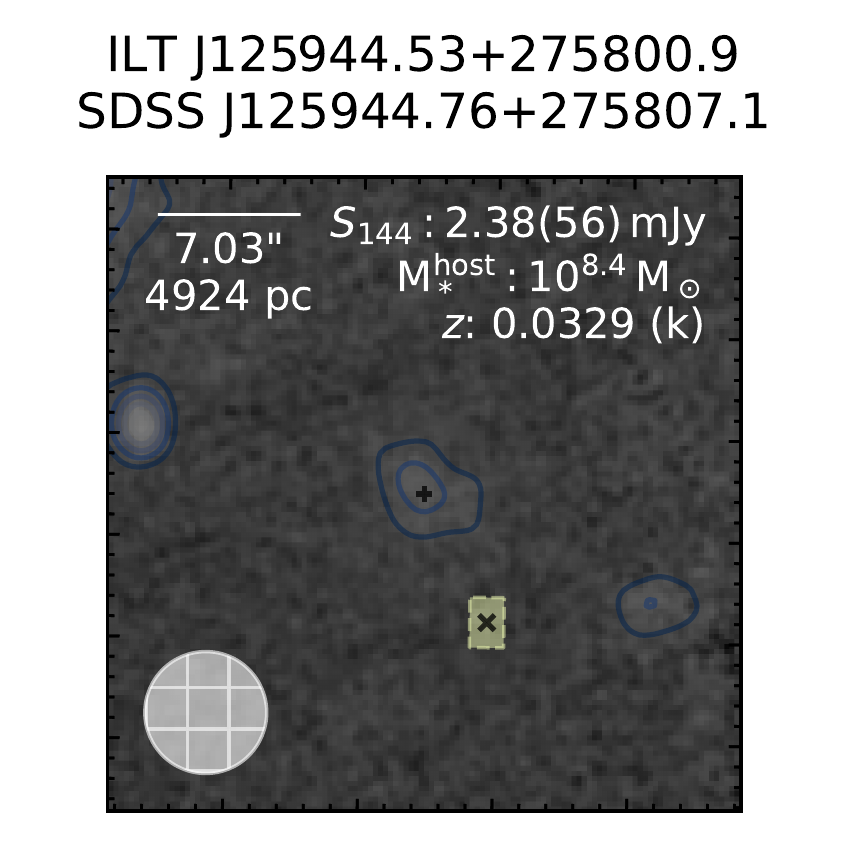} \\[0.2cm]
\end{tabular}
\begin{tabular}{cc}
\hline
\hline
\includegraphics[width=41mm]{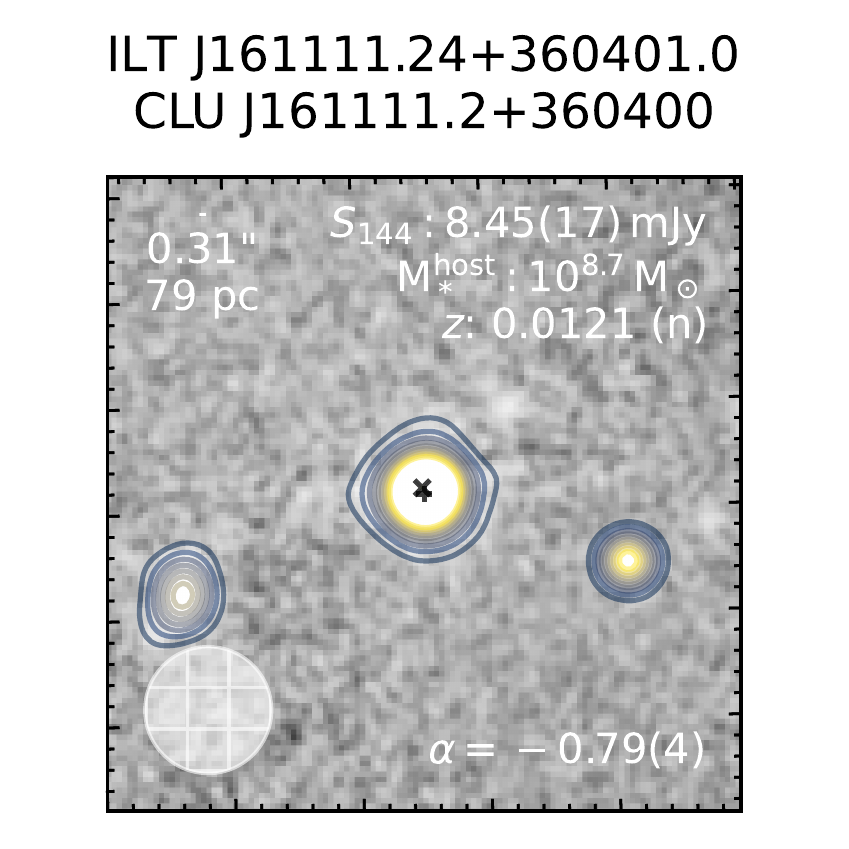} &
\includegraphics[width=41mm]{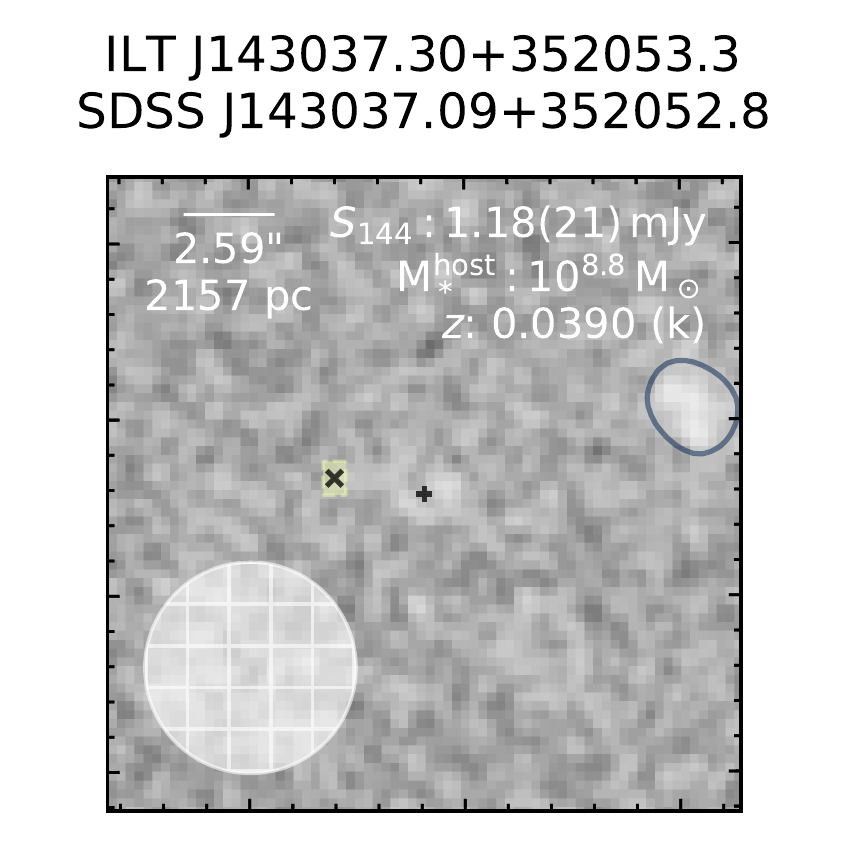} 
\end{tabular}
\caption{
    Over-luminous compact radio sources with projected offset greater than $(2+\epsilon)\arcsec$. 
    OCRs within the AGN region in Figure \ref{fig:wise} are shown below demarcation lines.
    See Figure~\ref{fig:family_plot} for component descriptions.
}
\label{fig:family_plot_continued}
\end{figure*}

The final candidate list includes 27 sources, with 16 having a projected offset of lower than $(2+\epsilon)\arcsec$. 
Composite images for each candidate in ascending order of offset are presented in Figure~\ref{fig:family_plot} (offset $\lesssim2\arcsec$) and Figure~\ref{fig:family_plot_continued} ($\gtrsim2\arcsec$ and OCRs within the AGN region of Figure \ref{fig:wise}), and candidates are detailed in Table~\ref{table:candidates}. 
Given that all H$\upalpha$ filters used by CLU fall within the frequency range covered by the PanSTARRS~\citep[PS1;][]{Chambers2016arXiv161205560C, Flewelling2020ApJS..251....7F} r filter, we use PS1 to generate image cutouts around the coordinates of  our candidates.
The central coordinates of the matched objects from LoTSS and CLU are indicated by black crosses and plus symbols,  respectively. 

All candidates shown in Figure~\ref{fig:family_plot} and most candidates in Figure~\ref{fig:family_plot_continued} fall well within the galaxy contours. 
ILT~J125940.18+275123.5 falls within the low-surface-brightness regions of its host galaxy, while the associated host galaxy of ILT~J125940.18+275123.5 (candidate with largest offset) is of extremely low surface density, and likely a chance alignment (further discussed in Appendix \ref{sec:appendix1}).
SDSS~J143037.09+352052.8 (matched to the AGN candidate ILT~143037.30+352053.3) is also extremely faint in the PS1 r filter image and may also be a chance alignment. 
However, data from the JVO Subaru/Suprime-Cam~\citep{Aihara2019PASJ...71..114A} show the radio source to be within the galaxy (Appendix \ref{sec:appendix1}).

We note that galaxies
2MASS J13002220+2814499 (ILT~J130022.42+281451.7), 
SDSS J125944.76+275807.1 (ILT~J125944.53+275800.9), 
2MASX J12594007+2751177 (ILT~J125940.18+275123.5), and 
SSTSL2 J125915.27+274604.1 (ILT~J125915.34+274604.2) are part of the cluster of galaxies ACO~1656 (Appendix \ref{sec:appendix2}).
Similarly, 2MASX~J23171540+1843385 (ILT~J231715.38+184339.0) is part of the compact group of galaxies HCG~94. 
Finally, the \href{http://simbad.cds.unistra.fr/simbad/sim-id?Ident=HCG++37&NbIdent=query_hlinks&Coord=09+13+35.6%2B30+00+51&children=5&submit=children&hlinksdisplay=h_all}{Simbad} service\footnote{\url{http://simbad.cds.unistra.fr/simbad}} indicates 2MASX~J09133387+3000514 (ILT~J091333.83+300056.9) as having a 75\% probability of being a hierarchical member of the galaxy cluster HCG~37.

In the following subsections, we investigate the possible nature of the candidates by considering complementary information in ancillary surveys; for example, optical spectra and flux density measurements at other radio wavelengths. 

\subsection{Main source of ionization in host galaxies}
\label{subsec:bpt}

\begin{figure*}
\includegraphics[width=17cm]{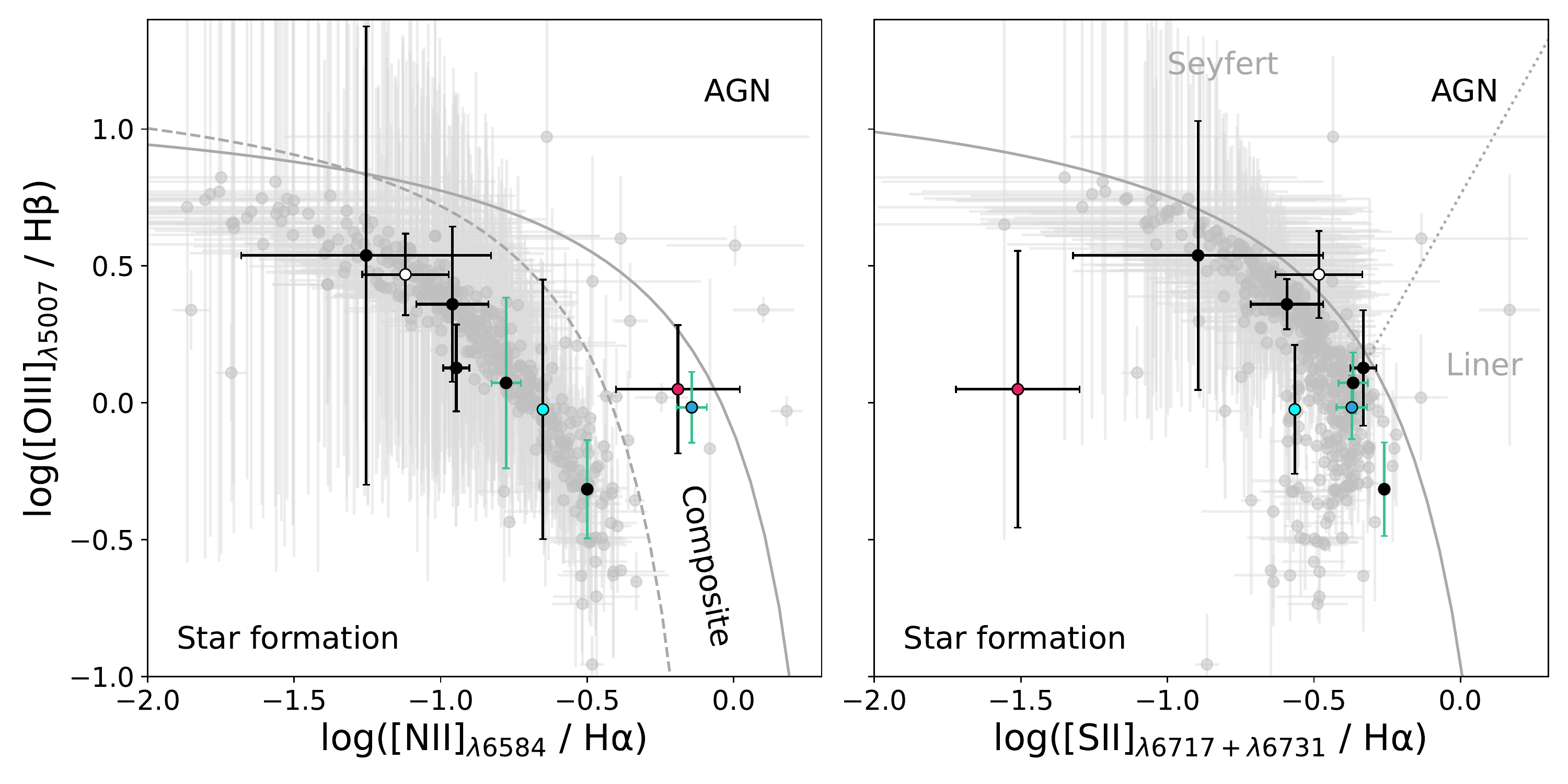}
\caption{BPT diagrams for sources with SDSS spectra showing the main source of ionization. 
Green and black markers correspond to sources with projected separation of less than or greater than $(2+\epsilon)\arcsec$, respectively. 
Gray circles are all ``compact radio source-dwarf galaxy'' matches with spectral line information in SDSS.
Markers with a colored inner circle (colors other than black) occupy interesting parts of parameter space and are discussed in \S\ref{sec:discussion}.
The galaxy marked with a white inner circle lies above the gray line in the ${\rm S{\sc II}/H\upalpha}$ panel, while being well within the star formation range in the ${\rm [N{\sc II}]/H\upalpha}$ panel. 
Blue and pink inner circles fall within the composite region in the ${\rm [N{\sc II}]/H\upalpha}$ panel, where we can expect a contribution from both star formation ($H{\sc II}$ regions) and AGN activity. 
Finally, the galaxy marked by a cyan inner circle was classified as an AGN candidate by \citet{Truebenbach2017MNRAS.468..196T} by selecting sources detected in the AllWISE and FIRST catalogs but not detected in 2MASS or SDSS DR7 and DR9. 
}
\label{fig:bpt}
\end{figure*}

A useful tool for distinguishing between galaxies with different prevailing photo-ionization sources is the family of emission line ratio diagnostic diagrams introduced by \citet*[hereafter, BPT]{Baldwin1981PASP...93....5B} in which the source location is determined by a pair of low-ionization, emission-line intensity ratios.
We searched archival data\footnote{Data from: SDSS~\citep{Abdurrouf2022ApJS..259...35A} DR 17, CALIFA~\citep[Calar Alto Legacy Integral Field Area;][]{Gonzalez2015A&A...581A.103G} survey, and LAMOST~\citep[Large Sky Area Multi-Object Fibre Spectroscopic Telescope;][]{Guo2022yCat..36670044G} survey.} for optical spectra to within $6\arcsec$ of the CLU coordinates. 
In SDSS, we found 11 matches within our candidates, and 483 out of the total 708 matched dwarf galaxies. 
However, it is worth noting that $\sim40$\% of nearby radio-loud AGNs are too gas poor and optically inactive to be detected via optical line ratio~\citep{Gereb2015A&A...580A..43G}.
Furthermore, low-luminosity dwarf galaxies with low metallicities ---particularly those with high star formation rates--- are susceptible to contamination from stellar processes that can potentially hide AGN indicators~\citep{Reines2013ApJ...775..116R, Molina2021ApJ...910....5M}.
Using the normalized emission line measurements from the MPA-JHU\footnote{A collaboration of researchers from the Max Planck Institute for Astrophysics (MPA) and the Johns Hopkins University (JHU).} spectroscopic reanalysis\footnote{Values are taken from the table galSpecLine. A table description is available at \href{http://skyserver.sdss.org/dr17/MoreTools/browser/}{this url} (last visited 7 September 2022).}~\citep{Tremonti2004ApJ...613..898T, Brinchmann2004MNRAS.351.1151B}, we evaluated the ratios of measured line fluxes (Table~\ref{table:BPT}) for $\rm {[O{\sc III}]/H\upbeta}$ against ${\rm [N{\sc II}]/H\upalpha}$ and ${\rm [S{\sc II}]/H\upalpha}$. 
We show the results in Figure~\ref{fig:bpt}. 

\begin{table*}
\begin{threeparttable}
\centering
\caption{Emission line ratio measurements for host galaxies observed by SDSS.}
\begin{tabular}{llrrrrr}
\toprule
           Source name & Host name & $\log~[{\rm N{\sc II}]/H}\upalpha$ & $\log~[{\rm O{\sc II}I]/H}\upbeta$  & $\log~[{\rm S{\sc II}]/H}\upalpha$ & $\alpha$ & $R_g$ \\
\midrule
        ILT~J075257.15+401026.3 & UGC~04068 & $-0.143(129)$ & $-0.017(052)$ & $-0.371(115)$ & $-0.54$ & -- \\
        ILT~J162244.56+321259.3 & 2MASS~J16224461+3213007 & $-0.777(311)$ & $0.073(050)$ & $-0.366(111)$ & $-0.63$ & $0.50$ \\
        ILT~J142859.42+331005.2 & 2MASX~J14285953+3310067 & $-0.500(179)$ & $-0.316(005)$ & $-0.260(170)$ & -- & $0.08$ \\
    \hline
        ILT~J143050.99+410642.6 & SDSS~J143051.12+410640.8 & $-0.946(158)$ & $0.127(044)$ & $-0.331(210)$ & -- & $0.41$ \\
        ILT~J090406.54+530314.6 & SDSS~J090406.38+530311.8 & $-1.120(148)$ & $0.469(147)$ & $-0.483(159)$ & $0.18$ & -- \\
        ILT~J140524.35+613358.7 & 2MASS~J14052457+6134020 & $-1.254(837)$ & $0.538(426)$ & $-0.895(491)$ & $-0.88$ & $0.55$ \\
        ILT~J122250.31+681434.2 & SDSS~J122249.71+681431.8 & $-0.960(283)$ & $0.360(123)$ & $-0.592(091)$ & -- & $0.43$ \\
        ILT~J113634.77+592533.3 & SBS~1133+597 & $-0.651(474)$ & $-0.024(005)$ & $-0.565(235)$ & $-0.96$ & $0.15$ \\
        ILT~J125940.18+275123.5 & 2MASX~J12594007+2751177 & $-0.190(234)$ & $0.049(211)$ & $-1.510(505)$ & $-1.00$ & $0.39$ \\
\bottomrule
\end{tabular}
\label{table:BPT}
\end{threeparttable}
\end{table*}

\subsection{Spectral indices in the radio band}
\label{subsec:spectral}

A radio continuum spectrum dominated by nonthermal synchrotron emission has a characteristic power-law slope, $S_v \propto v^\alpha$. 
To evaluate the spectral index $\alpha$ of our candidates, we cross-matched other radio surveys to within a $6\arcsec$ radius from our source center.
In particular, we searched the Rapid ASKAP Continuum Survey~\citep[RACS;][1.25\,GHz]{McConnell2020PASA...37...48M}, the Faint Images of the Radio Sky at Twenty centimeters survey~\citep[FIRST;][1.4\,GHz]{Becker1995ApJ...450..559B}, the NRAO VLA Sky Survey~\citep[NVSS;][1.4\,GHz]{Condon1998AJ....115.1693C}, and the Very Large Array Sky Survey~\citep[VLASS;][3\,GHz]{Lacy2020PASP..132c5001L}. 
The coverage of each survey can be found in Appendix \ref{app:coverage}.

\begin{figure*}
    \centering
    \includegraphics[width=17cm]{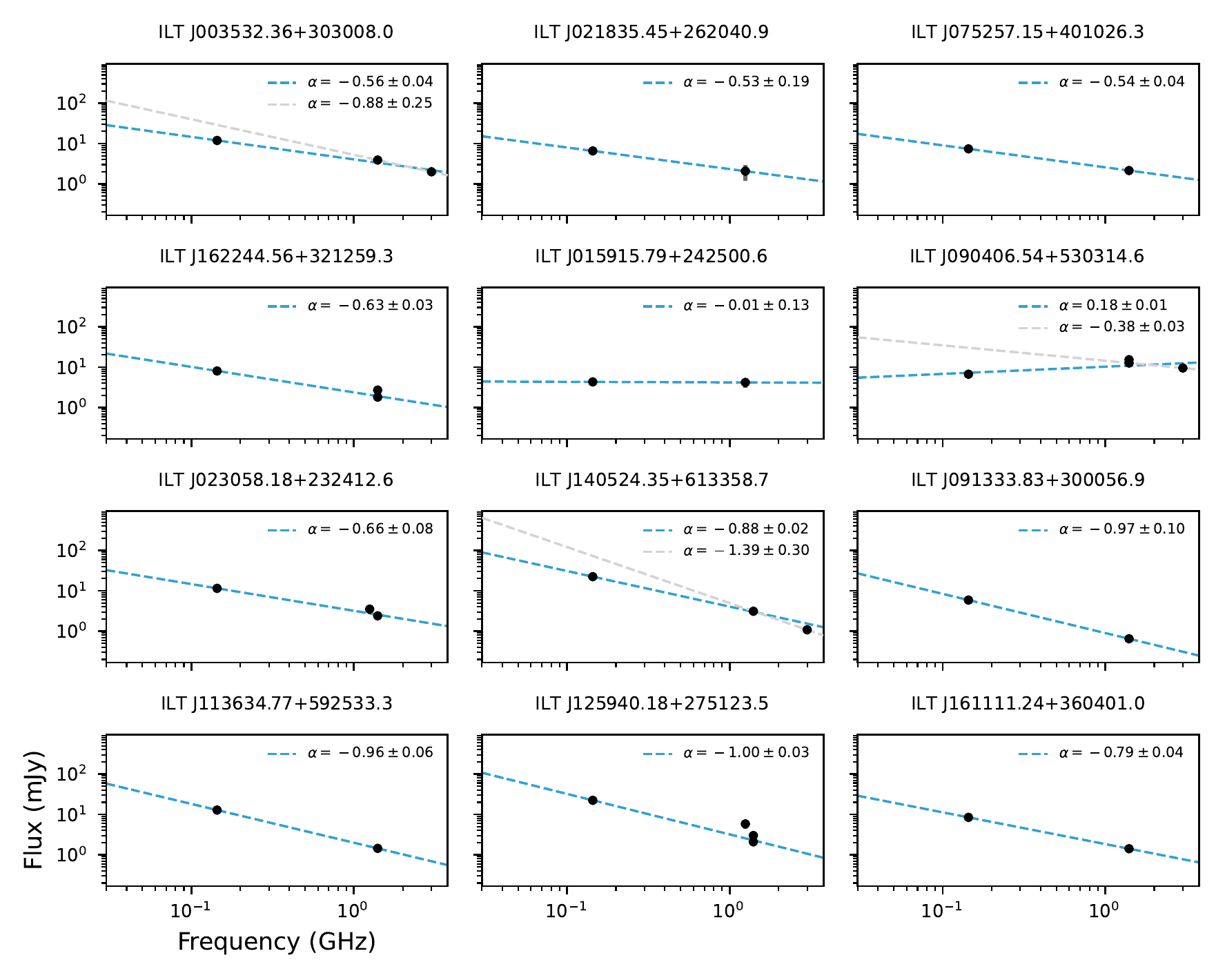} 
    \caption{Radio spectra of candidates matched in one or more of the RACS, FIRST, NVSS, and VLASS surveys, with central frequencies of 1.25\,GHz, 1.4\,GHz, 1.4\,GHz, and 3\,GHz, respectively. LoTSS observes at a central frequency of 144\,MHz. The source name in LoTSS is indicated for each candidate. Black markers show flux measurements with uncertainties (small enough not to be visible), and the dashed blue lines show the best power-law fit over all frequencies, with power-law index $\alpha$ also indicated. Where applicable, a gray dashed line shows the best power-law fit at higher frequencies (${\rm \sim1-3}$\,GHz. }
    \label{fig:spectral_indices}
\end{figure*}

We find a total of 12 matches among the 29 source candidates.
Spatial coverage by the various surveys is discussed in Appendix \ref{app:coverage}. 
Table~\ref{table:fluxes}  lists the multifrequency flux measurements used to evaluate spectral indices ($\alpha$), following the order of sources in Table~\ref{table:candidates}. 
Figure~\ref{fig:spectral_indices} presents for each matched source a radio spectrum and best model from fitting a single power law to flux measurements including uncertainties (dashed line), with corresponding $\alpha$. 
The resulting spectral indices range between $0.2\pm0.01$ and $-1.0\pm0.10$. 

\begin{figure*}
    \centering
    \includegraphics[width=17cm]{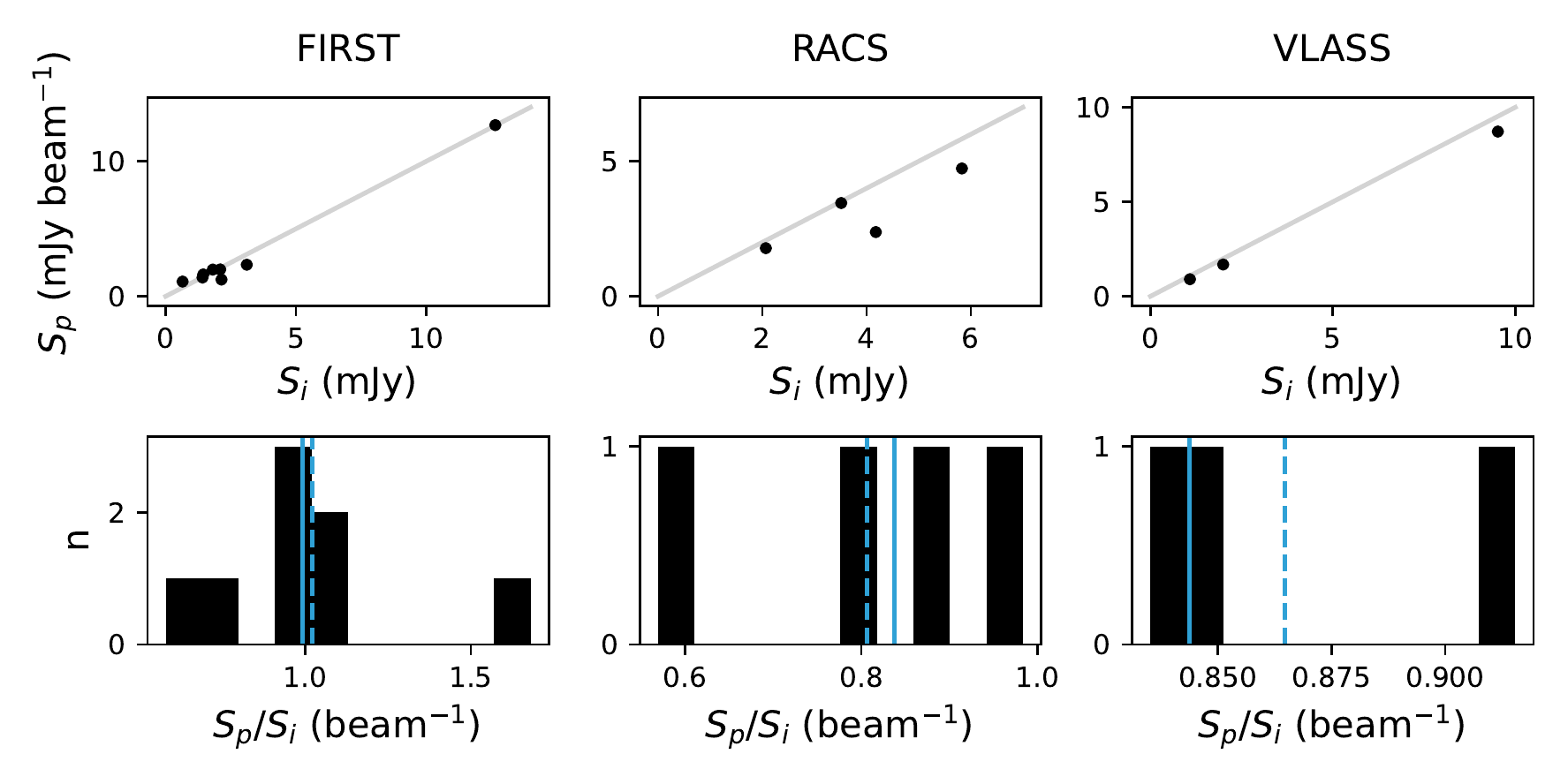} 
    \caption{Comparing peak flux ($S_p$) and integrated flux ($S_i$, Tables \ref{table:candidates} and \ref{table:fluxes}) measurements in the various surveys. Gray lines in the upper panel indicate a 1:1 ratio. In the lower panel, the solid and dashed blue lines indicate the median and mean of each ratio distribution, respectively. NVSS does not provide peak flux information.}
    \label{fig:s_peak_int}
\end{figure*}

It is worth noting that if our sources are partially resolved by some surveys then our spectral index estimates will be systematically biased. The beam sizes of LoTSS (6\arcsec) and FIRST (5\arcsec) are comparable, while the beam of VLASS is slightly smaller ($\sim$3\arcsec), and is larger in both NVSS and and RACS ($\sim$15\arcsec).  We checked whether or not the sources were point-like by comparing the peak flux density to the integrated flux density in RACS, VLASS and LoTSS (NVSS does not provide peak flux information) in Fig. \ref{fig:s_peak_int}. In most cases, flux ratios in FIRST, RACS and VLASS point toward compactness, with median ratios of 1.0, 0.8, and 0.8, respectively (with a minimal ratio of 0.6, 0.6, and 0.8; and a maximal ratio of 1.7, 1.0, and 0.9). Despite the point-source-like nature of our candidates, we caution the reader that the possible presence of diffuse synchrotron emission from the host galaxy could have biased our flux estimates. This effect is especially apparent at very large beam size ratios \citep[e.g.,][]{Venkattu2023arXiv230702365V}. Unfortunately, we do not yet have VLBI detection of our sources to test this hypothesis:  we cross-matched the candidates with the Radio Fundamental Catalog\footnote{\url{http://astrogeo.org/rfc/}, last visited 27 June 2023.} and mJIVE-20~\citep{DellerMiddelberg2014AJ....147...14D}---two VLBI datasets---, but found no matches.

\begin{table*}
\begin{threeparttable}
    \centering
    \caption{Fluxes for candidates matched in at least one ancillary radio survey.}
    \begin{tabular}{lrrrrrrr}
    \toprule
        Source name & $S_{\rm LoTSS}$ & $S_{\rm RACS}$ & $S_{\rm FIRST}$ & $S_{\rm NVSS}$ & $S_{\rm VLASS}$ & $\alpha$ & $\alpha_{\rm HF}$   \\
        (ILT~J) & (mJy) & (mJy) & (mJy) & (mJy) & (mJy) & ~ & ~ \\
    \midrule
        003532.36+303008.0 & $11.89\pm0.35$ & -- & -- & $3.90\pm0.50$ & $1.99\pm0.28$ & $-0.6\pm0.04$ & $-0.9\pm0.25$ \\
        021835.45+262040.9 & $6.55\pm0.44$ & $2.07\pm0.86$ & -- & -- & -- & $-0.5\pm0.19$ & -- \\
        075257.15+401026.3 & $7.37\pm0.35$ & -- & $2.14\pm0.15$ & -- & [$0.50\pm0.05$] & $-0.5\pm0.04$ & [$-1.9\pm0.13$] \\
        162244.56+321259.3 & $8.06\pm0.18$ & -- & $1.81\pm0.15$ & $2.70\pm0.40$ & [$1.20\pm0.05$] & $-0.6\pm0.03$ & [$-0.8\pm0.05$] \\
    015915.79+242500.6 & $4.30\pm0.43$ & $4.18\pm1.13$ & -- & -- & -- & $-0.0\pm0.13$ & -- \\
    \hline
        090406.54+530314.6 & $6.69\pm0.13$ & -- & $12.65\pm0.15$ & $15.30\pm0.90$ & $9.52\pm0.23$ & $0.2\pm0.01$ & $-0.4\pm0.03$ \\
        023058.18+232412.6 & $11.49\pm1.31$ & $3.51\pm0.89$ & -- & $2.40\pm0.40$ & -- & $-0.7\pm0.08$ & -- \\
        140524.35+613358.7 & $22.37\pm0.13$ & -- & $3.11\pm0.14$ & -- & $1.08\pm0.24$ & $-0.9\pm0.02$ & $-1.4\pm0.30$ \\
        091333.83+300056.9 & $5.89\pm0.49$ & -- & $0.65\pm0.13$ & -- & -- & $-1.0\pm0.10$ & -- \\
        113634.77+592533.3 & $12.82\pm0.12$ & -- & $1.44\pm0.20$ & -- & -- & $-1.0\pm0.06$ & -- \\
        125940.18+275123.5 & $22.34\pm0.20$ & $5.83\pm1.38$ & $2.09\pm0.18$ & $3.00\pm0.40$ & -- & $-1.0\pm0.03$ & -- \\
    \midrule
    \midrule
    161111.24+360401.0 & $8.45\pm0.17$ & -- & $1.41\pm0.14$ & -- & [$0.50\pm0.05$] & $-0.8\pm0.04$ & [$-1.4\pm0.18$] \\
    \bottomrule
    \end{tabular}
    \label{table:fluxes}
    
    \begin{tablenotes}
      \small
      \item Measurements from LoTSS, RACS, VLA FIRST, NVSS, and VLASS at 144\,MHz, 1.25\,GHz, 1.4\,GHz, 1.4\,GHz, and 3\,GHz, respectively. Values in square parentheses are estimated based on cutout image from \href{http://cutouts.cirada.ca}{CIRADA Image Cutout Web Service}. Only spectral indices in brackets include estimates from CIRADA Cutouts.
\end{tablenotes}
\end{threeparttable}
\end{table*}

 \subsection{ X-ray and gamma ray bands} 
 \label{subsec:xray}
 We searched the ancillary data from the Chandra and XMM-Newton (namely Chandra \href{https://cxc.cfa.harvard.edu/csc/}{Source Catalog Release 2.0}, XMM-EPIC, and XMM-EPIC-STACK) which we query via the \href{https://astroquery.readthedocs.io/en/latest/esasky/esasky.html}{astroquery (v0.4.6) ESA Sky module}, and the Burst Alert Spectroscopic Survey \citep[BASS;][]{Koss2017ApJ...850...74K} to within 15\arcsec. 
 Chandra SC2 observed 7 out of the 29 candidates, and XMM-Newton only 2.  
 The only match (found in Chandra SC2) corresponds to the source 2CXO~J125943.4+275802. 
 The X-ray source covers the cluster of galaxies ACO~1656 mentioned in \S\ref{sec:candidates}, and can be attributed to free-free emission (bremsstrahlung) at the cluster scale, and therefore not directly attributable to our compact source. 
 It is worth noting that X-ray observations toward \RI~have not yet been constraining~\citep{Chatterjee2017Natur.541...58C, Marcote2017ApJ...834L...8M}.  
 Similarly, we search the Fermi 4FGL second data release to within 6 arcmin for gamma-ray emission (also with {\tt astroquery.esasky}). 
 All candidate fields have been observed; we did not find any match.

 We find it difficult to interpret the X-ray nondetections using the fundamental plane of BH activity, that is, the empirical correlation between the continuum X-ray, radio emission, and mass of an accreting black hole~\citep{Gultekin2019ApJ...871...80G}. 
 Firstly, the 6\arcsec spatial resolution data from LoTSS can only provide upper limits as it is not possible to isolate the nuclear core flux with good point-source sensitivity (which would require subarcsecond spatial resolution). 
 Furthermore, LoTSS and any X-ray data mentioned in this section were not obtained simultaneously, and the timescale for variations in flux would be short for IMBHs, that is, scaling with the size of the ``event horizon'' and therefore with the black hole mass.
 
 We obtained flux upper limits in the hard state for both Chandra (accessed with {\tt CSCview}\footnote{\url{https://cxc.harvard.edu/csc/gui/intro.html}, last accessed 18 July 2023.}, using values in the 2--7\,keV range)\ and XMM-Newton (accessed with {\tt flix}\footnote{\url{https://www.ledas.ac.uk/flix/flix.html}, last accessed 18 July 2023.} for XMM-Newton in the 2--10\,keV range).
 We evaluate the upper limits using the prescription from \citet{Gultekin2019ApJ...871...80G}, which typically uses radio flux at 5\,GHz, and X-ray flux at 2--10\,keV. 
 We scale radio flux to 5\,GHz using either $\alpha=-0.7$ or using spectral indices from Table \ref{table:fluxes} where available. 
 Results are listed in Table \ref{table:bh_upperlimit}, with estimated BH mass limits ranging from below $\sim10^7\,{\rm M_\odot}$ to $\sim10^8\,{\rm M_\odot}$.
 Given the reasons highlighted above, we advise the reader to use these upper limits with caution. 
 
 \begin{table*}
 \begin{threeparttable}
    \centering
    \caption{BH mass upper limits estimated from radio flux and X-ray flux limits.}
    \begin{tabular}{lrrrrrr}
    \toprule
        Source name & M$_{*, {\rm host}}$ & L$_{\rm r}$ & L$_{\rm x, XMM}$ & L$_{\rm x, Chandra}$ & M$_{\sbullet, {\rm XMM}}$ & M$_{\sbullet, {\rm Chandra}}$  \\
        (ILT~J) & (${\rm M_\odot}$) & (${\rm erg\,s^{-1}}$) & (${\rm erg\,s^{-1}}$) & (${\rm erg\,s^{-1}}$) & (${\rm M_\odot}$) & (${\rm M_\odot}$) \\
    \midrule
153943.52+592730.7 & $2.0 \times 10^9$ & $5.0 \times 10^{37}$ & $<3.0 \times 10^{41}$ & --                     & $<2.2 \times 10^7$  & --                  \\
125940.18+275123.5 & $1.4 \times 10^9$ & $3.4 \times 10^{37}$ & $<3.5 \times 10^{40}$ & $<1.1 \times 10^{39}$  & $<5.2 \times 10^7$  & $<4.0 \times 10^8$ \\
125915.34+274604.2 & $8.2 \times 10^8$ & $1.1 \times 10^{37}$ & --                    & $<1.3 \times 10^{40}$  & --                  & $<2.6 \times 10^7$ \\
125944.53+275800.9 & $2.7 \times 10^8$ & $2.5 \times 10^{37}$ & --                    & $<7.2 \times 10^{39}$  & --                  & $<9.4 \times 10^7$ \\
142859.42+331005.2 & $2.7 \times 10^9$ & $1.6 \times 10^{37}$ & --                    & $<2.2 \times 10^{40}$  & --                  & $<3.0 \times 10^7$ \\
091333.83+300056.9 & $1.7 \times 10^9$ & $2.3 \times 10^{37}$ & --                    & $<2.9 \times 10^{39}$  & --                  & $<1.5 \times 10^8$ \\
231715.38+184339.0 & $2.3 \times 10^9$ & $4.1 \times 10^{37}$ & --                    & $<6.9 \times 10^{39}$  & --                  & $<1.7 \times 10^8$ \\
143037.30+352053.3 & $6.3 \times 10^8$ & $1.7 \times 10^{37}$ & --                    & $<2.6 \times 10^{40}$  & --                  & $<3.0 \times 10^7$ \\
     \bottomrule
    \end{tabular}
    \label{table:bh_upperlimit}
    
\begin{tablenotes}
   \small
   \item Radio luminosity from LoTSS are scaled to 5\,GHz,  X-ray data from XMM-Newton (2--10\,keV) and Chandra (2--7\,keV).
\end{tablenotes}
\end{threeparttable}
\end{table*}
 
%-------------------------------------------------------------------

\section{Discussion}
\label{sec:discussion}

\subsection{Cause of radio emission}
We here investigate what can be ascertained about the potential progenitors of our selected candidates given emission line ratios and spectral index measurements presented in \S\ref{sec:candidates}. 
Markers below and to the left of the solid and dashed gray lines in Figure \ref{fig:bpt} indicate that the emission lines are due to star formation and not to AGN activity~\citep{Kewley2001ApJ...556..121K, Kauffmann2003MNRAS.346.1055K}. 
Measurement uncertainties cannot definitively rule out an AGN contribution in three cases. 
For galaxies falling within the star formation region of this parameter space, ionising flux is primarily provided by hot, massive, young stars and associated supernovae surrounded by H{\sc II} regions~\citep{Zajacek2019A&A...630A..83Z}.

A few cases occupy interesting regions of parameter space. 
ILT~J090406.54+530314.6 (white inner circles) sits above the gray line in the ${\rm S{\sc II}/H\upalpha}$ panel, while sitting well within the star formation range in the ${\rm [N{\sc II}]/H\upalpha}$ panel. 
ILT~J075257.15+401026.3 (blue inner circles) and ILT~J125940.18+275123.5 (pink inner circles) fall within the composite region between models from \citet{Kewley2001ApJ...556..121K} and \citet{Kauffmann2003MNRAS.346.1055K} in the ${\rm [N{\sc II}]/H\upalpha}$ panel. 
For these, we can expect a contribution from both star formation (H{\sc II} regions) and AGN activity. 

Similarly, the galaxy UGC~04068, which hosts ILT~J075257.15+401026.3 has been classified as an AGN by \citet{Veron2010A&A...518A..10V}, while the classification based on the SDSS spectrum is simply ``galaxy'' (rather than other considered classes in SDSS nomenclature, such as QSO). 
There is a bright star located near the centroid of the galaxy (slightly leftward in Figure \ref{fig:family_plot}) that may impact the overall flux observed in the spectrum, especially given the location of the SDSS spectrograph fiber (pink circle) almost exactly between the galaxy centroid and that of the bright star. 
It is also the only case within the candidates where the CLU catalog contains a fitted H$\upalpha$ D25 measurement. 
CLU~J163850.64+352900.9 and 2MASX~J09133387+3000514 are unclassified in SDSS, while all other matches are classified as galaxies.

Finally, we note that ILT~J113634.77+592533.3 (cyan inner circles) was classified as an AGN candidate by \citet{Truebenbach2017MNRAS.468..196T} by selecting sources detected in the AllWISE and FIRST catalogs, but not detected in 2MASS or SDSS DR7 and DR9. 
However, we note that the source may be matched to the galaxy SBS~1133+597, which has been observed by SDSS and for which the BPT diagram rather indicates that the driving source of ionization in the galaxy can be attributed to star formation. 

The value of $\alpha$ helps distinguish between optically thin and optically thick emission mechanisms.
We note that flux measurements from the archival surveys we used span several decades of observations, with FIRST and NVSS being the oldest, and RACS, VLASS and LoTSS being contemporaneous but not simultaneous. 
Using archival data spanning many decades comes with the caveat that measurements may be affected by time-dependent phenomena like scintillation or source evolution. 
In cases where radio emission is powered by an AGN, variability at various timescales (from days to years) can be expected with potential flux variations of the order of 100 mJy at GHz frequencies, and in rare cases varying by factors of several in flux density over timescales measured in decades ~\citep{Nyland2020AAS...23512901N}.
Almost no variability is expected from star formation on the timescale covered by our data.

The range of spectral indices covered by various source types is known to differ.
Pulsars have spectral indices of $\lesssim-1.2$~\citep{Bates2013MNRAS.431.1352B}. 
Given the range of values of our candidates, they are unlikely to be  pulsars. 
Moreover, we can assume that if an OCR (detected by LoTSS/FIRST/NVSS) is a radio pulsar, it would have to be Galactic, and would therefore be an unrelated foreground object in a chance alignment with the background galaxy as LoTSS should not be sensitive to extragalactic pulsars. 
Supernova remnants tend to have spectral indices ranging between $-0.1$ and $-0.8$~\citep{Kothes2006A&A...457.1081K, Alvarez2001A&A...372..636A}. 
Six of our candidates fall within this range, including five cases with offset below $(2+\epsilon)\arcsec$. 
\citet{Planck2011A&A...536A..15P} showed that the spectral indices of AGNs at low frequencies ($1.1-\leq70$\,GHz) are fairly flat, with an average of $-0.06$. 
Their distribution is narrow, with 91\% of the indices being in the range $\alpha\in[$-0.5$, 0.5]$. 
However, a few sources have remarkably steep spectra $\leq-0.8$,  while others have inverted spectra ($\alpha=0.86$).
Although our spectral indices are calculated at lower frequencies than these (where possible, we also computed the spectral index at higher frequency between $1.4-3$\,GHz), they all fall within this broad range. 

Comparing spectral indices between 4.85 and 10.45\,GHz from a distribution of radio sources with optical counterparts, \citet{Zajacek2019A&A...630A..83Z} showed that the ionization potential of sources with an inverted radio spectrum ($\alpha>-0.4$) is weaker than that of sources with a steep radio spectrum ($\alpha<-0.7$). 
In particular, simultaneous two-point $\alpha$ measurement at 4.85 and 10.45\,GHz at Effelsberg highlighted that decreasing spectral indices from steep to flat ($-0.7<\alpha<-0.4$) to inverted leads to a decrease in typical line ratios (BPT diagram), particularly ${\rm [O{\sc III}]/H\upbeta}$.

\citet{Zajacek2019A&A...630A..83Z} considered radio loudness $R_g$ in addition to $\alpha$ and ionization ratio in order to highlight three distinct classes of radio emitters resulting from recurrent nuclear jet activity distributed along the transition from Seyfert to LINER sources in the optical diagnostic, namely sources with: 
    (class 1) steep $\alpha$, high ionization ratio, and high radio loudness; 
    (class 2) flat $\alpha$, lower ionization ratio, and intermediate radio loudness; and 
    (class 3) inverted $\alpha$, low ionization ratio, and low radio loudness. 

To compare our results to those of \citet{Zajacek2019A&A...630A..83Z}, we computed $R_g$ using the flux density from LoTSS, $F_{144}$. 
We converted $F_{144}$ into the $AB_\nu$ radio magnitude system of \citet{Oke1983ApJ...266..713O}, according to \citet{Ivezic2002AJ....124.2364I}: $m_{1.4} = -2.5\log{F_{1.4} / 3631\,{\rm Jy}}$, in which the zero point 3631\,Jy does not depend on wavelength, and scaled the fluxes from 144\,MHz to 1.4\,GHz using either our fitted spectral indices or $\alpha=-0.7$ otherwise. 
Subsequently, the radio loudness can be calculated as the ratio of the radio flux density to the optical flux density, $R_g \equiv \log{F_{\rm radio}/F_{\rm optical}} = 0.4 (g-m_{144})$, with $g$ corresponding to the magnitude in the optical $g$-band.
We use SDSS magnitudes in the g band where available, and g-band Kron magnitudes from PS1 where available otherwise.

\begin{figure}
\resizebox{\hsize}{!}{\includegraphics{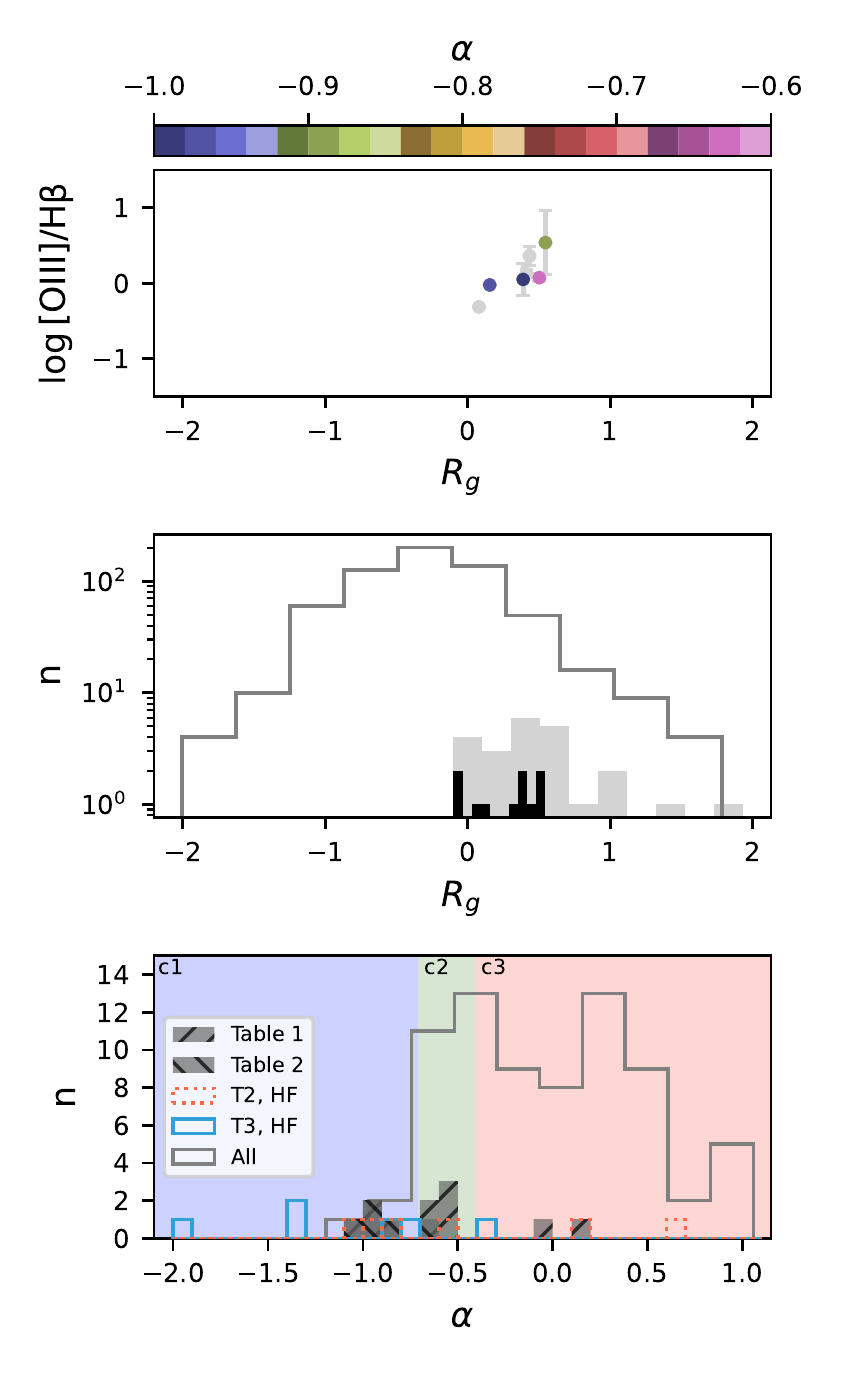}}
\caption{Comparing radio loudness $R_g$, emission line ratio ${\rm [O{\sc III}]/H\upbeta,}$ and spectral index, where available. 
From the sample of 26 $R_g$ values (Table \ref{table:candidates}), 7 also have ${\rm [O{\sc III}]/H\upbeta}$ values available (Table \ref{table:BPT}, alias T2; black bars in central panel), and 9 have fitted $\alpha$ values (Table \ref{table:fluxes}, alias T3). 
HF indicates spectral indices fitted between 1.4 and 3\,GHz.
Class 1, 2, and 3 (c1, c2, c3) as defined by \citet{Eckart1986A&A...168...17E}.
Unfilled histograms in middle and lower panels represent all available ``compact radio source-dwarf galaxy'' matches below 3$\sigma$ on the L--SFR relation. 
}
\label{fig:loudness}
\end{figure}

We list values of $R_g$ in Table~\ref{table:candidates}. 
For our whole set of candidates, $R_g$ ranges between 0.05 and 2.48, with mean, median, and standard deviation of 1.21, 1.16, and 0.48, respectively.
Sources for which we can evaluate $R_g$ and emission line ratio between ${\rm [O{\sc III}]}$ and ${\rm H\upbeta}$ are shown in the upper panel of Figure~\ref{fig:loudness}, displaying $\alpha$ where possible using the color map\footnote{We note that the point with $\alpha \sim -0.6$ shown in pink is calculated using detections at all available frequencies, and that its equivalent at higher frequencies corresponds to $-0.88$, as shown in Table~\ref{table:fluxes}.} and in gray otherwise. 
In addition, we show distributions of $R_g$ (central panel) and $\alpha$ (lower panel) for candidates with and without available optical spectra in gray and black, respectively.
 In the lower panel, we mark the regions defined by \citet{Eckart1986A&A...168...17E} separating steep ($\alpha<-0.7$, class 1), flat ($-0.7 \leq \alpha \leq -0.4$, class 2), and inverted ($\alpha>-0.4$, class 3) spectrum sources in blue, green and pink, respectively, in order to reflect the distributions of the spectral index for samples of radio-loud galaxies in the range of ${\rm 1.6-5\,GHz}$ ---regions of parameter space also used to classify sources by \citet{Zajacek2019A&A...630A..83Z}.

The region occupied in the emission line ratio--loudness plane by our matched sources corresponds to that of sources classified as class 3 by \citet[][e.g., their Figure~12]{Zajacek2019A&A...630A..83Z}, though with much lower $R_g$ (the lower bound on $R_g$ used by  \citeauthor{Zajacek2019A&A...630A..83Z}  being $\sim0.7$). 
The sample used by these latter authors covered supermassive black holes ($\rm >10^5\,M_\odot$), explaining the loudness discrepancy where dwarf galaxies could be hosting IMBHs instead.
Their spectral index distribution (black, lower panel) rather points toward class 1 or 2, even when considering $\alpha$ evaluated at higher frequencies ($1.4-3$\,GHz; orange dotted border showing sources with associated spectrum; blue border showing all cases matched at high frequencies).
Unfortunately, given that only four sources shown in the top panel of Figure~\ref{fig:loudness} were matched in other surveys to evaluate a spectral index, these results provide only small number statistics.
Nevertheless, the initial information they carry points toward a mismatch between the primarily AGN-related source studied by \citet{Zajacek2019A&A...630A..83Z}, which mainly fall along the demarcation line between Seyfert and LINER, and our candidates primarily located well within the star formation region of the BPT diagram, which strengthens our hypothesis that the OCRs in our sample are not AGNs. 
Follow-up observations at higher frequencies similar to those observed by \citeauthor{Zajacek2019A&A...630A..83Z} would allow this tension to be further scrutinized.

\subsection{Volumetric density and rate of over-luminous compact radio sources}
\label{subsec:volumetric}

The CLU catalog consists of galaxies selected to be within 200\,Mpc and is complete to a flux limit of $10^{-14}\,$erg\,s$^{-1}$\,cm$^{-2}$. 
This limit corresponds to a dust-unobscured star formation rate of $\approx 1$\,M$_\odot$\,yr$^{-1}$ \citep{Cook2019ApJ...880....7C}.
The 200\,Mpc distance is small enough for us to assume a Euclidean geometry in our volume density calculations. 
Based on the 29 sources in Table 1, we can summarily compute a lower limit of $856\pm150$ sources\,Gpc$^{-3}$ ($1\sigma$ Poisson bounds in parentheses) for compact radio sources (on arcsecond-scales) above 0.8\,mJy at 144\, MHz that deviate by more than $3\sigma$ from the radio--SFR relationship. 
The limit is preliminary because VLBI observations are necessary to conclusively rule out a star-formation origin for our sources.

We can compare this rate to that suggested by \citet{Law2022ApJ...927...55L} for persistent radio sources associated with FRB progenitors. 
These latter authors compute a volume density of $50-10,000$\,Gpc$^{-3}$ for sources with a 1.4\,GHz radio luminosity of greater than $10^{29}\,$\,ergs\,s$^{-1}$\,Hz$^{-1}$. 
A separate volume density computed by \citet{Ofek2017ApJ...846...44O} is close to the upper end of the density computed by \citet{Law2022ApJ...927...55L}. 
A source of $10^{29}$\,erg\,s$^{-1}$\,Hz$^{-1}$  luminosity placed at our survey horizon of 200\,Mpc would have a flux density of 2.2\,mJy. 
Given our survey completeness of 0.8\,mJy, such a source would be detected in our survey if it was optically thin or if it had an inverted spectrum with a spectral index shallower than $\approx 0.4$ (i.e., a relatively flat spectrum). 
However, our survey is also sensitive to closer sources with much lower flux densities. 

To account for this, we assumed that the FRB-related PRS sources follow a Schechter luminosity function \citep{1976ApJ...203..297S} with an exponent of $-1$ and a cut-off luminosity that is ten times the \citet{Law2022ApJ...927...55L} normalization point of $10^{29}$\,erg\,s$^{-1}$\,Hz$^{-1}$. 
We also assumed that the sources have a flat spectrum and computed the number of sources that would likely exceed the 0.9\,mJy completeness limit of our survey.
We then numerically computed the expected number of detectable sources within a 200\,Mpc horizon.
For the $50-10,000$\,Gpc$^{-3}$ range specified in \citet{Law2022ApJ...927...55L}, we expect to detect $0.3-58$ sources, which is consistent with our yield of 28 candidates. 
Values of cut-off luminosity in excess of $10^{30.5}\,$erg\,s$^{-1}$\,Hz$^{-1}$ are necessary to create a tension between our yield and the \citet{Law2022ApJ...927...55L} rates. 
While the consistency is encouraging, we caution against drawing firm conclusions because of the unknown FRB beaming fraction on which the \citet{Law2022ApJ...927...55L} estimate is based, along with the disparate selection filters applied in the analysis of these latter authors and ours.

\subsection{Star formation versus active (intermediate-mass) black holes}

\citet{Condon2019ApJ...872..148C} evaluated the luminosity functions for sources whose radio emission is dominated by star formation and AGN, respectively. 
After scaling the luminosity of our candidates to 1.4\,GHz using either the spectral indices evaluated in \S\ref{subsec:spectral} or $\alpha\approx-0.7$ for typical synchrotron spectra of optically thin radio sources for the remaining sources, we find luminosity values ranging between $10^{17.9}$ and $10^{22.7}$\,$\mathrm{W\,Hz^{-1}}$, with a median of $10^{19.9}$\,$\mathrm{W\,Hz^{-1}}$. 
Compared to the luminosity functions for star formation and AGN of \citet{Condon2019ApJ...872..148C}, the radio emission from our candidates is more likely attributable to star formation. 

A critical step towards establishing the candidates presented in previous sections as potential FRB hosts is to conclusively determine the compactness of these sources. 
To this end, we obtained time on the European VLBI network (EVN) and e-MERLIN to observe the most likely candidates.
Compactness in LoTSS images only ensures a brightness temperature of $\gtrsim 10^4\,{\rm K}$, which is insufficient to rule out unusually radio-bright star formation as the cause of the radio emission. 
Moreover, our sources may have a significant component of their radio flux attributed to star formation with the rest in a compact source component. 
VLBI at $\lesssim 10\,$mas resolution is therefore the best observational technique to totally eliminate (i.e., resolve out) the star-formation component and establish the presence of a compact source. 
Furthermore, including the e-MERLIN array should allow us to distinguish between compact and star-formation components, if present.
In addition, we are in the process of reimaging the archival LoTSS radio data on these sources while including the international stations from LOFAR. 
The resulting images should have a resolution of about $0\farcs25$ \citep[LOFAR-VLBI;][]{Morabito2022A&A...658A...1M}.
These higher-angular-resolution images should inform us about the following possible outcomes.

If a target were not detected at very high resolution then it would confirm the star-formation hypothesis. 
This would be a rather unusual conclusion as the selected targets all violate the radio--SFR relationship, and so a nondetection at very high resolution would cast doubt on the canonical radio-AGN selection technique that is widely used~\citep{Davis2022MNRAS.511.4109D}.
It is possible that AGN-related flux is present on intermediate scales of a few hundred\,milliarcseconds, which are inaccessible to the EVN, but should be accessible by the intermediate scale of e-MERLIN. 
Moreover, in such a case, AGN-related flux should also be detectable with LOFAR long-baseline data.

The detection of a core--jet structure would confirm the AGN-like IMBH hypothesis. 
Although the radio detection of black-hole jet candidates in dwarf galaxies based on the procedure mentioned in \S\ref{sec:method} is now becoming feasible \citep{Davis2022MNRAS.511.4109D}, confirmatory VLBI detection of the jet (or structure thereof) is rare \citep{Paragi2014ApJ...791....2P, Yang2020MNRAS.495L..71Y, Eftekhari2020ApJ...895...98E}. 
As such, confirmation of the AGN hypothesis will have interesting scientific consequences for studies of feedback in dwarf galaxies.

If an unresolved point is detected, although it would rule out the star-formation hypothesis, both the PWN and unresolved AGN would remain plausible---even if the source proves to be slightly ($\ll 1\arcsec$) offset from the optical stellar light centroid. 
Based on the known properties of starburst galaxies, a detection on EVN long baselines should exclude star formation as the cause of the bulk of the radio emission~\citep{Condon1991ApJ...378...65C}.
Here, a path forward would be to follow-up such sources to model their broad-band SED (e.g., with optical spectroscopy directly on-source to search for canonical AGN signatures, and with radio observations at C, X, and K bands) in order to decipher between the PWN and AGN hypotheses.

\subsection{Future search for FRBs}

Finally, we also plan to search these targets for millisecond-duration bursts with the 25 meter Westerbork Synthesis Radio Telescope. 
Starting with the hypothesis that some are similar in nature to currently known PRSs, we can expect these to be repeating FRB sources.
Furthermore, given the periodic activity of some FRBs, such as \RI~\citep{Cruces2021MNRAS.500..448C, Rajwade2020MNRAS.495.3551R} and FRB\,20180916B~\citep{Chime2020Natur.582..351C}, it is plausible that a subset of our candidates  could also display on/off phases of FRB emission.

%-------------------------------------------------------------------

\section{Summary}
\label{sec:summary}

In this paper, we present a targeted search for OCRs coincident with dwarf galaxies up to $z\lesssim 0.05$. 
\begin{enumerate}
\item We identified candidate compact sources with luminosity exceeding $3\sigma$ relative to the L--SFR relation~\citep{Gurkan2018MNRAS.475.3010G}.
\item Through ancillary surveys, we investigated the possible nature of the candidates.
\item Emission line ratios from SDSS spectra show the main source of ionization in the host galaxies where the candidates are located is likely star formation, and not AGN activity. 
\item Spectral indices suggest that our candidates could be SNRs or AGNs---although, the combination of $\alpha$ with emission line ratios and radio loudness provides evidence that we may be observing sources other than typical AGNs. 
\item A comparison to the luminosity functions for star formation and AGN of \citet{Condon2019ApJ...872..148C} indicates that radio emission from our candidates is more likely attributable to star formation.
\item We derive a preliminary lower limit of $856\pm150$ sources\,Gpc$^{-3}$ ($1\sigma$ Poisson bounds in parentheses) for compact radio sources (on arcsecond-scales) above 0.8\,mJy at 144\, MHz that deviate by more than $3\sigma$ from the radio--SFR relationship.
\item For the $50-10,000$\,Gpc$^{-3}$ range specified in \citet{Law2022ApJ...927...55L}, we expect to detect $0.3-58$ sources within the 200\,Mpc horizon of our optical parent sample, which is consistent with our yield of 28 candidates. 

\end{enumerate}

Follow-up high-angular-resolution imaging should allow us to further describe these outlying radio sources.
Furthermore, searching these sources for high-time-resolution bursts may inform us about the FRB progenitor.
If only a subsample of our candidates turn out to be active sources of persistent radio emission associated to FRBs, as is the case for those presented by \citet[][]{Chatterjee2017Natur.541...58C} and \citet[][]{Niu2022Natur.606..873N}, it would be possible to evaluate whether or not these are indeed calorimeters allowing us to estimate the energy output of the central FRB engine. 

We end by noting that due to the unprecedented sensitivity of  LoTSS to optically thin synchrotron sources in a wide-angle survey, we have been able to select interesting radio sources in dwarf galaxies. 
The proposed VLBI observations are a crucial step toward the discovery of a new population of either wind nebulae or black holes in nearby dwarf galaxies---both outcomes being scientifically interesting.  

%-------------------------------------------------------------------

\begin{acknowledgements}
This work was carried out in part through funding from the European Open Science Cloud (EOSC) Future, an EU-funded H2020 project. 
DV thanks Benito Marcote, Shivani Bhandari, Kenzie Nimmo, and Betsey Adams for valuable discussions. We thank our referee for constructive comments. 
%LOFAR
LOFAR is the Low Frequency Array designed and constructed by ASTRON. It has observing, data processing, and data storage facilities in several countries, which are owned by various parties (each with their own funding sources), and which are collectively operated by the ILT foundation under a joint scientific policy. The ILT resources have benefited from the following recent major funding sources: CNRS-INSU, Observatoire de Paris and Université d'Orléans, France; BMBF, MIWF-NRW, MPG, Germany; Science Foundation Ireland (SFI), Department of Business, Enterprise and Innovation (DBEI), Ireland; NWO, The Netherlands; The Science and Technology Facilities Council, UK; Ministry of Science and Higher Education, Poland; The Istituto Nazionale di Astrofisica (INAF), Italy. 
%PS1
The Pan-STARRS1 Surveys (PS1) and the PS1 public science archive have been made possible through contributions by the Institute for Astronomy, the University of Hawaii, the Pan-STARRS Project Office, the Max-Planck Society and its participating institutes, the Max Planck Institute for Astronomy, Heidelberg and the Max Planck Institute for Extraterrestrial Physics, Garching, The Johns Hopkins University, Durham University, the University of Edinburgh, the Queen's University Belfast, the Harvard-Smithsonian Center for Astrophysics, the Las Cumbres Observatory Global Telescope Network Incorporated, the National Central University of Taiwan, the Space Telescope Science Institute, the National Aeronautics and Space Administration under Grant No. NNX08AR22G issued through the Planetary Science Division of the NASA Science Mission Directorate, the National Science Foundation Grant No. AST-1238877, the University of Maryland, Eotvos Lorand University (ELTE), the Los Alamos National Laboratory, and the Gordon and Betty Moore Foundation.
%VLASS
This research has made use of the VLASS QLimage cutout server at URL cutouts.cirada.ca, operated by the Canadian Initiative for Radio Astronomy Data Analysis (CIRADA). CIRADA is funded by a grant from the Canada Foundation for Innovation 2017 Innovation Fund (Project 35999), as well as by the Provinces of Ontario, British Columbia, Alberta, Manitoba and Quebec, in collaboration with the National Research Council of Canada, the US National Radio Astronomy Observatory and Australia’s Commonwealth Scientific and Industrial Research Organisation.
%Subaru
This research is based in part on data collected at Subaru Telescope, which is operated by the National Astronomical Observatory of Japan. We are honored and grateful for the opportunity of observing the Universe from Maunakea, which has the cultural, historical and natural significance in Hawaii.
Part of the data are retrieved from the JVO portal (http://jvo.nao.ac.jp/portal) operated by the NAOJ.
Software: Astropy~\citep{astropy:2013, astropy:2018, astropy:2022}, Astroquery~\citep{Ginsburg2019AJ....157...98G}, ~\citep{Robitaille2012ascl.soft08017R}, Aladin~\citep{Boch2011ASPC..442..683B}, Topcat~\citep{Taylor2011ascl.soft01010T}, askap\_surveys\footnote{\url{https://github.com/askap-vast/askap_surveys}, last visited 10 July 2023.} (David Kaplan and Andrew O'Brien (UW-Milwaukee)), matchmaker~\citep{vohl_d_2023_7986693}. The custom code used to generate these results is publicly available at \url{https://doi.org/10.5281/zenodo.10018524}.

\end{acknowledgements}

%-------------------------------------------------------------------

\bibliography{references}

\begin{thebibliography}{93}
\expandafter\ifx\csname natexlab\endcsname\relax\def\natexlab#1{#1}\fi

\bibitem[{{Abdurro'uf} {et~al.}(2022){Abdurro'uf}, {Accetta}, {Aerts}, {Silva
  Aguirre}, {Ahumada}, {Ajgaonkar}, {Filiz Ak}, {Alam}, {Allende Prieto},
  {Almeida}, {Anders}, {Anderson}, {Andrews}, {Anguiano}, {Aquino-Ort{\'\i}z},
  {Arag{\'o}n-Salamanca}, {Argudo-Fern{\'a}ndez}, {Ata}, {Aubert},
  {Avila-Reese}, {Badenes}, {Barb{\'a}}, {Barger}, {Barrera-Ballesteros},
  {Beaton}, {Beers}, {Belfiore}, {Bender}, {Bernardi}, {Bershady}, {Beutler},
  {Bidin}, {Bird}, {Bizyaev}, {Blanc}, {Blanton}, {Boardman}, {Bolton},
  {Boquien}, {Borissova}, {Bovy}, {Brandt}, {Brown}, {Brownstein}, {Brusa},
  {Buchner}, {Bundy}, {Burchett}, {Bureau}, {Burgasser}, {Cabang}, {Campbell},
  {Cappellari}, {Carlberg}, {Wanderley}, {Carrera}, {Cash}, {Chen}, {Chen},
  {Cherinka}, {Chiappini}, {Choi}, {Chojnowski}, {Chung}, {Clerc}, {Cohen},
  {Comerford}, {Comparat}, {da Costa}, {Covey}, {Crane}, {Cruz-Gonzalez},
  {Culhane}, {Cunha}, {Dai}, {Damke}, {Darling}, {Davidson}, {Davies},
  {Dawson}, {De Lee}, {Diamond-Stanic}, {Cano-D{\'\i}az}, {S{\'a}nchez},
  {Donor}, {Duckworth}, {Dwelly}, {Eisenstein}, {Elsworth}, {Emsellem},
  {Eracleous}, {Escoffier}, {Fan}, {Farr}, {Feng}, {Fern{\'a}ndez-Trincado},
  {Feuillet}, {Filipp}, {Fillingham}, {Frinchaboy}, {Fromenteau}, {Galbany},
  {Garc{\'\i}a}, {Garc{\'\i}a-Hern{\'a}ndez}, {Ge}, {Geisler}, {Gelfand},
  {G{\'e}ron}, {Gibson}, {Goddy}, {Godoy-Rivera}, {Grabowski}, {Green},
  {Greener}, {Grier}, {Griffith}, {Guo}, {Guy}, {Hadjara}, {Harding},
  {Hasselquist}, {Hayes}, {Hearty}, {Hern{\'a}ndez}, {Hill}, {Hogg},
  {Holtzman}, {Horta}, {Hsieh}, {Hsu}, {Hsu}, {Huber}, {Huertas-Company},
  {Hutchinson}, {Hwang}, {Ibarra-Medel}, {Chitham}, {Ilha}, {Imig}, {Jaekle},
  {Jayasinghe}, {Ji}, {Johnson}, {Jones}, {J{\"o}nsson}, {Katkov}, {Khalatyan},
  {Kinemuchi}, {Kisku}, {Knapen}, {Kneib}, {Kollmeier}, {Kong}, {Kounkel},
  {Kreckel}, {Krishnarao}, {Lacerna}, {Lane}, {Langgin}, {Lavender}, {Law},
  {Lazarz}, {Leung}, {Leung}, {Lewis}, {Li}, {Li}, {Lian}, {Liang}, {Lin},
  {Lin}, {Lin}, {Lintott}, {Long}, {Longa-Pe{\~n}a}, {L{\'o}pez-Cob{\'a}},
  {Lu}, {Lundgren}, {Luo}, {Mackereth}, {de la Macorra}, {Mahadevan},
  {Majewski}, {Manchado}, {Mandeville}, {Maraston}, {Margalef-Bentabol},
  {Masseron}, {Masters}, {Mathur}, {McDermid}, {Mckay}, {Merloni},
  {Merrifield}, {Meszaros}, {Miglio}, {Di Mille}, {Minniti}, {Minsley},
  {Monachesi}, {Moon}, {Mosser}, {Mulchaey}, {Muna}, {Mu{\~n}oz}, {Myers},
  {Myers}, {Nadathur}, {Nair}, {Nandra}, {Neumann}, {Newman}, {Nidever},
  {Nikakhtar}, {Nitschelm}, {O'Connell}, {Garma-Oehmichen}, {Luan Souza de
  Oliveira}, {Olney}, {Oravetz}, {Ortigoza-Urdaneta}, {Osorio}, {Otter},
  {Pace}, {Padilla}, {Pan}, {Pan}, {Parikh}, {Parker}, {Peirani}, {Pe{\~n}a
  Ram{\'\i}rez}, {Penny}, {Percival}, {Perez-Fournon}, {Pinsonneault},
  {Poidevin}, {Poovelil}, {Price-Whelan}, {B{\'a}rbara de Andrade Queiroz},
  {Raddick}, {Ray}, {Rembold}, {Riddle}, {Riffel}, {Riffel}, {Rix}, {Robin},
  {Rodr{\'\i}guez-Puebla}, {Roman-Lopes}, {Rom{\'a}n-Z{\'u}{\~n}iga}, {Rose},
  {Ross}, {Rossi}, {Rubin}, {Salvato}, {S{\'a}nchez}, {S{\'a}nchez-Gallego},
  {Sanderson}, {Santana Rojas}, {Sarceno}, {Sarmiento}, {Sayres}, {Sazonova},
  {Schaefer}, {Schiavon}, {Schlegel}, {Schneider}, {Schultheis}, {Schwope},
  {Serenelli}, {Serna}, {Shao}, {Shapiro}, {Sharma}, {Shen}, {Shetrone}, {Shu},
  {Simon}, {Skrutskie}, {Smethurst}, {Smith}, {Sobeck}, {Spoo}, {Sprague},
  {Stark}, {Stassun}, {Steinmetz}, {Stello}, {Stone-Martinez},
  {Storchi-Bergmann}, {Stringfellow}, {Stutz}, {Su}, {Taghizadeh-Popp},
  {Talbot}, {Tayar}, {Telles}, {Teske}, {Thakar}, {Theissen}, {Tkachenko},
  {Thomas}, {Tojeiro}, {Hernandez Toledo}, {Troup}, {Trump}, {Trussler},
  {Turner}, {Tuttle}, {Unda-Sanzana}, {V{\'a}zquez-Mata}, {Valentini},
  {Valenzuela}, {Vargas-Gonz{\'a}lez}, {Vargas-Maga{\~n}a}, {Alfaro},
  {Villanova}, {Vincenzo}, {Wake}, {Warfield}, {Washington}, {Weaver},
  {Weijmans}, {Weinberg}, {Weiss}, {Westfall}, {Wild}, {Wilde}, {Wilson},
  {Wilson}, {Wilson}, {Wolf}, {Wood-Vasey}, {Yan}, {Zamora}, {Zasowski},
  {Zhang}, {Zhao}, {Zheng}, {Zheng}, \& {Zhu}}]{Abdurrouf2022ApJS..259...35A}
{Abdurro'uf}, {Accetta}, K., {Aerts}, C., {et~al.} 2022, \apjs, 259, 35

\bibitem[{{Aihara} {et~al.}(2019){Aihara}, {AlSayyad}, {Ando}, {Armstrong},
  {Bosch}, {Egami}, {Furusawa}, {Furusawa}, {Goulding}, {Harikane}, {Hikage},
  {Ho}, {Hsieh}, {Huang}, {Ikeda}, {Imanishi}, {Ito}, {Iwata}, {Jaelani},
  {Kakuma}, {Kawana}, {Kikuta}, {Kobayashi}, {Koike}, {Komiyama}, {Li},
  {Liang}, {Lin}, {Luo}, {Lupton}, {Lust}, {MacArthur}, {Matsuoka}, {Mineo},
  {Miyatake}, {Miyazaki}, {More}, {Murata}, {Namiki}, {Nishizawa}, {Oguri},
  {Okabe}, {Okamoto}, {Okura}, {Ono}, {Onodera}, {Onoue}, {Osato}, {Ouchi},
  {Shibuya}, {Strauss}, {Sugiyama}, {Suto}, {Takada}, {Takagi}, {Takata},
  {Takita}, {Tanaka}, {Terai}, {Toba}, {Uchiyama}, {Utsumi}, {Wang}, {Wang}, \&
  {Yamada}}]{Aihara2019PASJ...71..114A}
{Aihara}, H., {AlSayyad}, Y., {Ando}, M., {et~al.} 2019, \pasj, 71, 114

\bibitem[{{Alam} {et~al.}(2015){Alam}, {Albareti}, {Allende Prieto}, {Anders},
  {Anderson}, {Anderton}, {Andrews}, {Armengaud}, {Aubourg}, {Bailey}, {Basu},
  {Bautista}, {Beaton}, {Beers}, {Bender}, {Berlind}, {Beutler}, {Bhardwaj},
  {Bird}, {Bizyaev}, {Blake}, {Blanton}, {Blomqvist}, {Bochanski}, {Bolton},
  {Bovy}, {Shelden Bradley}, {Brandt}, {Brauer}, {Brinkmann}, {Brown},
  {Brownstein}, {Burden}, {Burtin}, {Busca}, {Cai}, {Capozzi}, {Carnero
  Rosell}, {Carr}, {Carrera}, {Chambers}, {Chaplin}, {Chen}, {Chiappini},
  {Chojnowski}, {Chuang}, {Clerc}, {Comparat}, {Covey}, {Croft}, {Cuesta},
  {Cunha}, {da Costa}, {Da Rio}, {Davenport}, {Dawson}, {De Lee}, {Delubac},
  {Deshpande}, {Dhital}, {Dutra-Ferreira}, {Dwelly}, {Ealet}, {Ebelke},
  {Edmondson}, {Eisenstein}, {Ellsworth}, {Elsworth}, {Epstein}, {Eracleous},
  {Escoffier}, {Esposito}, {Evans}, {Fan}, {Fern{\'a}ndez-Alvar}, {Feuillet},
  {Filiz Ak}, {Finley}, {Finoguenov}, {Flaherty}, {Fleming}, {Font-Ribera},
  {Foster}, {Frinchaboy}, {Galbraith-Frew}, {Garc{\'\i}a},
  {Garc{\'\i}a-Hern{\'a}ndez}, {Garc{\'\i}a P{\'e}rez}, {Gaulme}, {Ge},
  {G{\'e}nova-Santos}, {Georgakakis}, {Ghezzi}, {Gillespie}, {Girardi},
  {Goddard}, {Gontcho}, {Gonz{\'a}lez Hern{\'a}ndez}, {Grebel}, {Green},
  {Grieb}, {Grieves}, {Gunn}, {Guo}, {Harding}, {Hasselquist}, {Hawley},
  {Hayden}, {Hearty}, {Hekker}, {Ho}, {Hogg}, {Holley-Bockelmann}, {Holtzman},
  {Honscheid}, {Huber}, {Huehnerhoff}, {Ivans}, {Jiang}, {Johnson},
  {Kinemuchi}, {Kirkby}, {Kitaura}, {Klaene}, {Knapp}, {Kneib}, {Koenig},
  {Lam}, {Lan}, {Lang}, {Laurent}, {Le Goff}, {Leauthaud}, {Lee}, {Lee},
  {Licquia}, {Liu}, {Long}, {L{\'o}pez-Corredoira}, {Lorenzo-Oliveira},
  {Lucatello}, {Lundgren}, {Lupton}, {Mack}, {Mahadevan}, {Maia}, {Majewski},
  {Malanushenko}, {Malanushenko}, {Manchado}, {Manera}, {Mao}, {Maraston},
  {Marchwinski}, {Margala}, {Martell}, {Martig}, {Masters}, {Mathur},
  {McBride}, {McGehee}, {McGreer}, {McMahon}, {M{\'e}nard}, {Menzel},
  {Merloni}, {M{\'e}sz{\'a}ros}, {Miller}, {Miralda-Escud{\'e}}, {Miyatake},
  {Montero-Dorta}, {More}, {Morganson}, {Morice-Atkinson}, {Morrison},
  {Mosser}, {Muna}, {Myers}, {Nandra}, {Newman}, {Neyrinck}, {Nguyen},
  {Nichol}, {Nidever}, {Noterdaeme}, {Nuza}, {O'Connell}, {O'Connell},
  {O'Connell}, {Ogando}, {Olmstead}, {Oravetz}, {Oravetz}, {Osumi}, {Owen},
  {Padgett}, {Padmanabhan}, {Paegert}, {Palanque-Delabrouille}, {Pan},
  {Parejko}, {P{\^a}ris}, {Park}, {Pattarakijwanich}, {Pellejero-Ibanez},
  {Pepper}, {Percival}, {P{\'e}rez-Fournon}, {P{\'e}rez-R{\`a}fols},
  {Petitjean}, {Pieri}, {Pinsonneault}, {Porto de Mello}, {Prada}, {Prakash},
  {Price-Whelan}, {Protopapas}, {Raddick}, {Rahman}, {Reid}, {Rich}, {Rix},
  {Robin}, {Rockosi}, {Rodrigues}, {Rodr{\'\i}guez-Torres}, {Roe}, {Ross},
  {Ross}, {Rossi}, {Ruan}, {Rubi{\~n}o-Mart{\'\i}n}, {Rykoff},
  {Salazar-Albornoz}, {Salvato}, {Samushia}, {S{\'a}nchez}, {Santiago},
  {Sayres}, {Schiavon}, {Schlegel}, {Schmidt}, {Schneider}, {Schultheis},
  {Schwope}, {Sc{\'o}ccola}, {Scott}, {Sellgren}, {Seo}, {Serenelli}, {Shane},
  {Shen}, {Shetrone}, {Shu}, {Silva Aguirre}, {Sivarani}, {Skrutskie},
  {Slosar}, {Smith}, {Sobreira}, {Souto}, {Stassun}, {Steinmetz}, {Stello},
  {Strauss}, {Streblyanska}, {Suzuki}, {Swanson}, {Tan}, {Tayar}, {Terrien},
  {Thakar}, {Thomas}, {Thomas}, {Thompson}, {Tinker}, {Tojeiro}, {Troup},
  {Vargas-Maga{\~n}a}, {Vazquez}, {Verde}, {Viel}, {Vogt}, {Wake}, {Wang},
  {Weaver}, {Weinberg}, {Weiner}, {White}, {Wilson}, {Wisniewski},
  {Wood-Vasey}, {Ye`che}, {York}, {Zakamska}, {Zamora}, {Zasowski}, {Zehavi},
  {Zhao}, {Zheng}, {Zhou}, {Zhou}, {Zou}, \& {Zhu}}]{Alam2015ApJS..219...12A}
{Alam}, S., {Albareti}, F.~D., {Allende Prieto}, C., {et~al.} 2015, \apjs, 219,
  12

\bibitem[{{Alvarez} {et~al.}(2001){Alvarez}, {Aparici}, {May}, \&
  {Reich}}]{Alvarez2001A&A...372..636A}
{Alvarez}, H., {Aparici}, J., {May}, J., \& {Reich}, P. 2001, \aap, 372, 636

\bibitem[{{Astropy Collaboration} {et~al.}(2022){Astropy Collaboration},
  {Price-Whelan}, {Lim}, {Earl}, {Starkman}, {Bradley}, {Shupe}, {Patil},
  {Corrales}, {Brasseur}, {N{"o}the}, {Donath}, {Tollerud}, {Morris},
  {Ginsburg}, {Vaher}, {Weaver}, {Tocknell}, {Jamieson}, {van Kerkwijk},
  {Robitaille}, {Merry}, {Bachetti}, {G{"u}nther}, {Aldcroft},
  {Alvarado-Montes}, {Archibald}, {B{'o}di}, {Bapat}, {Barentsen}, {Baz{'a}n},
  {Biswas}, {Boquien}, {Burke}, {Cara}, {Cara}, {Conroy}, {Conseil}, {Craig},
  {Cross}, {Cruz}, {D'Eugenio}, {Dencheva}, {Devillepoix}, {Dietrich},
  {Eigenbrot}, {Erben}, {Ferreira}, {Foreman-Mackey}, {Fox}, {Freij}, {Garg},
  {Geda}, {Glattly}, {Gondhalekar}, {Gordon}, {Grant}, {Greenfield}, {Groener},
  {Guest}, {Gurovich}, {Handberg}, {Hart}, {Hatfield-Dodds}, {Homeier},
  {Hosseinzadeh}, {Jenness}, {Jones}, {Joseph}, {Kalmbach}, {Karamehmetoglu},
  {Ka{l}uszy{'n}ski}, {Kelley}, {Kern}, {Kerzendorf}, {Koch}, {Kulumani},
  {Lee}, {Ly}, {Ma}, {MacBride}, {Maljaars}, {Muna}, {Murphy}, {Norman},
  {O'Steen}, {Oman}, {Pacifici}, {Pascual}, {Pascual-Granado}, {Patil},
  {Perren}, {Pickering}, {Rastogi}, {Roulston}, {Ryan}, {Rykoff}, {Sabater},
  {Sakurikar}, {Salgado}, {Sanghi}, {Saunders}, {Savchenko}, {Schwardt},
  {Seifert-Eckert}, {Shih}, {Jain}, {Shukla}, {Sick}, {Simpson},
  {Singanamalla}, {Singer}, {Singhal}, {Sinha}, {Sip{H{o}}cz}, {Spitler},
  {Stansby}, {Streicher}, {{{S}}umak}, {Swinbank}, {Taranu}, {Tewary},
  {Tremblay}, {Val-Borro}, {Van Kooten}, {Vasovi{'c}}, {Verma}, {de Miranda
  Cardoso}, {Williams}, {Wilson}, {Winkel}, {Wood-Vasey}, {Xue}, {Yoachim},
  {Zhang}, {Zonca}, \& {Astropy Project Contributors}}]{astropy:2022}
{Astropy Collaboration}, {Price-Whelan}, A.~M., {Lim}, P.~L., {et~al.} 2022,
  \apj, 935, 167

\bibitem[{{Astropy Collaboration} {et~al.}(2018){Astropy Collaboration},
  {Price-Whelan}, {Sip{\H{o}}cz}, {G{\"u}nther}, {Lim}, {Crawford}, {Conseil},
  {Shupe}, {Craig}, {Dencheva}, {Ginsburg}, {Vand erPlas}, {Bradley},
  {P{\'e}rez-Su{\'a}rez}, {de Val-Borro}, {Aldcroft}, {Cruz}, {Robitaille},
  {Tollerud}, {Ardelean}, {Babej}, {Bach}, {Bachetti}, {Bakanov}, {Bamford},
  {Barentsen}, {Barmby}, {Baumbach}, {Berry}, {Biscani}, {Boquien}, {Bostroem},
  {Bouma}, {Brammer}, {Bray}, {Breytenbach}, {Buddelmeijer}, {Burke},
  {Calderone}, {Cano Rodr{\'\i}guez}, {Cara}, {Cardoso}, {Cheedella}, {Copin},
  {Corrales}, {Crichton}, {D'Avella}, {Deil}, {Depagne}, {Dietrich}, {Donath},
  {Droettboom}, {Earl}, {Erben}, {Fabbro}, {Ferreira}, {Finethy}, {Fox},
  {Garrison}, {Gibbons}, {Goldstein}, {Gommers}, {Greco}, {Greenfield},
  {Groener}, {Grollier}, {Hagen}, {Hirst}, {Homeier}, {Horton}, {Hosseinzadeh},
  {Hu}, {Hunkeler}, {Ivezi{\'c}}, {Jain}, {Jenness}, {Kanarek}, {Kendrew},
  {Kern}, {Kerzendorf}, {Khvalko}, {King}, {Kirkby}, {Kulkarni}, {Kumar},
  {Lee}, {Lenz}, {Littlefair}, {Ma}, {Macleod}, {Mastropietro}, {McCully},
  {Montagnac}, {Morris}, {Mueller}, {Mumford}, {Muna}, {Murphy}, {Nelson},
  {Nguyen}, {Ninan}, {N{\"o}the}, {Ogaz}, {Oh}, {Parejko}, {Parley}, {Pascual},
  {Patil}, {Patil}, {Plunkett}, {Prochaska}, {Rastogi}, {Reddy Janga},
  {Sabater}, {Sakurikar}, {Seifert}, {Sherbert}, {Sherwood-Taylor}, {Shih},
  {Sick}, {Silbiger}, {Singanamalla}, {Singer}, {Sladen}, {Sooley},
  {Sornarajah}, {Streicher}, {Teuben}, {Thomas}, {Tremblay}, {Turner},
  {Terr{\'o}n}, {van Kerkwijk}, {de la Vega}, {Watkins}, {Weaver}, {Whitmore},
  {Woillez}, {Zabalza}, \& {Astropy Contributors}}]{astropy:2018}
{Astropy Collaboration}, {Price-Whelan}, A.~M., {Sip{\H{o}}cz}, B.~M., {et~al.}
  2018, \aj, 156, 123

\bibitem[{{Astropy Collaboration} {et~al.}(2013){Astropy Collaboration},
  {Robitaille}, {Tollerud}, {Greenfield}, {Droettboom}, {Bray}, {Aldcroft},
  {Davis}, {Ginsburg}, {Price-Whelan}, {Kerzendorf}, {Conley}, {Crighton},
  {Barbary}, {Muna}, {Ferguson}, {Grollier}, {Parikh}, {Nair}, {Unther},
  {Deil}, {Woillez}, {Conseil}, {Kramer}, {Turner}, {Singer}, {Fox}, {Weaver},
  {Zabalza}, {Edwards}, {Azalee Bostroem}, {Burke}, {Casey}, {Crawford},
  {Dencheva}, {Ely}, {Jenness}, {Labrie}, {Lim}, {Pierfederici}, {Pontzen},
  {Ptak}, {Refsdal}, {Servillat}, \& {Streicher}}]{astropy:2013}
{Astropy Collaboration}, {Robitaille}, T.~P., {Tollerud}, E.~J., {et~al.} 2013,
  \aap, 558, A33

\bibitem[{{Baldwin} {et~al.}(1981){Baldwin}, {Phillips}, \&
  {Terlevich}}]{Baldwin1981PASP...93....5B}
{Baldwin}, J.~A., {Phillips}, M.~M., \& {Terlevich}, R. 1981, \pasp, 93, 5

\bibitem[{{Bassa} {et~al.}(2017){Bassa}, {Tendulkar}, {Adams}, {Maddox},
  {Bogdanov}, {Bower}, {Burke-Spolaor}, {Butler}, {Chatterjee}, {Cordes},
  {Hessels}, {Kaspi}, {Law}, {Marcote}, {Paragi}, {Ransom}, {Scholz},
  {Spitler}, \& {van Langevelde}}]{Bassa2017ApJ...843L...8B}
{Bassa}, C.~G., {Tendulkar}, S.~P., {Adams}, E.~A.~K., {et~al.} 2017, \apjl,
  843, L8

\bibitem[{{Bates} {et~al.}(2013){Bates}, {Lorimer}, \&
  {Verbiest}}]{Bates2013MNRAS.431.1352B}
{Bates}, S.~D., {Lorimer}, D.~R., \& {Verbiest}, J.~P.~W. 2013, \mnras, 431,
  1352

\bibitem[{{Becker} {et~al.}(1995){Becker}, {White}, \&
  {Helfand}}]{Becker1995ApJ...450..559B}
{Becker}, R.~H., {White}, R.~L., \& {Helfand}, D.~J. 1995, \apj, 450, 559

\bibitem[{{Bianchi} {et~al.}(2014){Bianchi}, {Conti}, \&
  {Shiao}}]{Bianchi2014AdSpR..53..900B}
{Bianchi}, L., {Conti}, A., \& {Shiao}, B. 2014, Advances in Space Research,
  53, 900

\bibitem[{{Boch} {et~al.}(2011){Boch}, {Oberto}, {Fernique}, \&
  {Bonnarel}}]{Boch2011ASPC..442..683B}
{Boch}, T., {Oberto}, A., {Fernique}, P., \& {Bonnarel}, F. 2011, in
  Astronomical Society of the Pacific Conference Series, Vol. 442, Astronomical
  Data Analysis Software and Systems XX, ed. I.~N. {Evans}, A.~{Accomazzi},
  D.~J. {Mink}, \& A.~H. {Rots}, 683

\bibitem[{{Bonnarel} {et~al.}(2000){Bonnarel}, {Fernique}, {Bienaym{\'e}},
  {Egret}, {Genova}, {Louys}, {Ochsenbein}, {Wenger}, \&
  {Bartlett}}]{Bonnarel2000A&AS..143...33B}
{Bonnarel}, F., {Fernique}, P., {Bienaym{\'e}}, O., {et~al.} 2000, \aaps, 143,
  33

\bibitem[{{Brinchmann} {et~al.}(2004){Brinchmann}, {Charlot}, {White},
  {Tremonti}, {Kauffmann}, {Heckman}, \&
  {Brinkmann}}]{Brinchmann2004MNRAS.351.1151B}
{Brinchmann}, J., {Charlot}, S., {White}, S.~D.~M., {et~al.} 2004, \mnras, 351,
  1151

\bibitem[{{Chambers} {et~al.}(2016){Chambers}, {Magnier}, {Metcalfe},
  {Flewelling}, {Huber}, {Waters}, {Denneau}, {Draper}, {Farrow}, {Finkbeiner},
  {Holmberg}, {Koppenhoefer}, {Price}, {Rest}, {Saglia}, {Schlafly}, {Smartt},
  {Sweeney}, {Wainscoat}, {Burgett}, {Chastel}, {Grav}, {Heasley}, {Hodapp},
  {Jedicke}, {Kaiser}, {Kudritzki}, {Luppino}, {Lupton}, {Monet}, {Morgan},
  {Onaka}, {Shiao}, {Stubbs}, {Tonry}, {White}, {Ba{\~n}ados}, {Bell},
  {Bender}, {Bernard}, {Boegner}, {Boffi}, {Botticella}, {Calamida},
  {Casertano}, {Chen}, {Chen}, {Cole}, {Deacon}, {Frenk}, {Fitzsimmons},
  {Gezari}, {Gibbs}, {Goessl}, {Goggia}, {Gourgue}, {Goldman}, {Grant},
  {Grebel}, {Hambly}, {Hasinger}, {Heavens}, {Heckman}, {Henderson}, {Henning},
  {Holman}, {Hopp}, {Ip}, {Isani}, {Jackson}, {Keyes}, {Koekemoer}, {Kotak},
  {Le}, {Liska}, {Long}, {Lucey}, {Liu}, {Martin}, {Masci}, {McLean}, {Mindel},
  {Misra}, {Morganson}, {Murphy}, {Obaika}, {Narayan}, {Nieto-Santisteban},
  {Norberg}, {Peacock}, {Pier}, {Postman}, {Primak}, {Rae}, {Rai}, {Riess},
  {Riffeser}, {Rix}, {R{\"o}ser}, {Russel}, {Rutz}, {Schilbach}, {Schultz},
  {Scolnic}, {Strolger}, {Szalay}, {Seitz}, {Small}, {Smith}, {Soderblom},
  {Taylor}, {Thomson}, {Taylor}, {Thakar}, {Thiel}, {Thilker}, {Unger},
  {Urata}, {Valenti}, {Wagner}, {Walder}, {Walter}, {Watters}, {Werner},
  {Wood-Vasey}, \& {Wyse}}]{Chambers2016arXiv161205560C}
{Chambers}, K.~C., {Magnier}, E.~A., {Metcalfe}, N., {et~al.} 2016, arXiv
  e-prints, arXiv:1612.05560

\bibitem[{{Chatterjee} {et~al.}(2017){Chatterjee}, {Law}, {Wharton},
  {Burke-Spolaor}, {Hessels}, {Bower}, {Cordes}, {Tendulkar}, {Bassa},
  {Demorest}, {Butler}, {Seymour}, {Scholz}, {Abruzzo}, {Bogdanov}, {Kaspi},
  {Keimpema}, {Lazio}, {Marcote}, {McLaughlin}, {Paragi}, {Ransom}, {Rupen},
  {Spitler}, \& {van Langevelde}}]{Chatterjee2017Natur.541...58C}
{Chatterjee}, S., {Law}, C.~J., {Wharton}, R.~S., {et~al.} 2017, \nat, 541, 58

\bibitem[{{Chen} {et~al.}(2022){Chen}, {Ravi}, \&
  {Hallinan}}]{Chen2022arXiv220100999C}
{Chen}, G., {Ravi}, V., \& {Hallinan}, G.~W. 2022, arXiv e-prints,
  arXiv:2201.00999

\bibitem[{{Chime/Frb Collaboration} {et~al.}(2020){Chime/Frb Collaboration},
  {Amiri}, {Andersen}, {Bandura}, {Bhardwaj}, {Boyle}, {Brar}, {Chawla},
  {Chen}, {Cliche}, {Cubranic}, {Deng}, {Denman}, {Dobbs}, {Dong}, {Fandino},
  {Fonseca}, {Gaensler}, {Giri}, {Good}, {Halpern}, {Hessels}, {Hill},
  {H{\"o}fer}, {Josephy}, {Kania}, {Karuppusamy}, {Kaspi}, {Keimpema},
  {Kirsten}, {Landecker}, {Lang}, {Leung}, {Li}, {Lin}, {Marcote}, {Masui},
  {McKinven}, {Mena-Parra}, {Merryfield}, {Michilli}, {Milutinovic},
  {Mirhosseini}, {Naidu}, {Newburgh}, {Ng}, {Nimmo}, {Paragi}, {Patel}, {Pen},
  {Pinsonneault-Marotte}, {Pleunis}, {Rafiei-Ravandi}, {Rahman}, {Ransom},
  {Renard}, {Sanghavi}, {Scholz}, {Shaw}, {Shin}, {Siegel}, {Singh}, {Smegal},
  {Smith}, {Stairs}, {Tendulkar}, {Tretyakov}, {Vanderlinde}, {Wang}, {Wang},
  {Wulf}, {Yadav}, \& {Zwaniga}}]{Chime2020Natur.582..351C}
{Chime/Frb Collaboration}, {Amiri}, M., {Andersen}, B.~C., {et~al.} 2020, \nat,
  582, 351

\bibitem[{{Condon} {et~al.}(1998){Condon}, {Cotton}, {Greisen}, {Yin},
  {Perley}, {Taylor}, \& {Broderick}}]{Condon1998AJ....115.1693C}
{Condon}, J.~J., {Cotton}, W.~D., {Greisen}, E.~W., {et~al.} 1998, \aj, 115,
  1693

\bibitem[{{Condon} {et~al.}(1991){Condon}, {Huang}, {Yin}, \&
  {Thuan}}]{Condon1991ApJ...378...65C}
{Condon}, J.~J., {Huang}, Z.~P., {Yin}, Q.~F., \& {Thuan}, T.~X. 1991, \apj,
  378, 65

\bibitem[{{Condon} {et~al.}(2019){Condon}, {Matthews}, \&
  {Broderick}}]{Condon2019ApJ...872..148C}
{Condon}, J.~J., {Matthews}, A.~M., \& {Broderick}, J.~J. 2019, \apj, 872, 148

\bibitem[{{Cook} {et~al.}(2019){Cook}, {Kasliwal}, {Van Sistine}, {Kaplan},
  {Sutter}, {Kupfer}, {Shupe}, {Laher}, {Masci}, {Dale}, {Sesar}, {Brady},
  {Yan}, {Ofek}, {Reitze}, \& {Kulkarni}}]{Cook2019ApJ...880....7C}
{Cook}, D.~O., {Kasliwal}, M.~M., {Van Sistine}, A., {et~al.} 2019, \apj, 880,
  7

\bibitem[{{Cruces} {et~al.}(2021){Cruces}, {Spitler}, {Scholz}, {Lynch},
  {Seymour}, {Hessels}, {Gouiff{\'e}s}, {Hilmarsson}, {Kramer}, \&
  {Munjal}}]{Cruces2021MNRAS.500..448C}
{Cruces}, M., {Spitler}, L.~G., {Scholz}, P., {et~al.} 2021, \mnras, 500, 448

\bibitem[{{Curti} {et~al.}(2020){Curti}, {Mannucci}, {Cresci}, \&
  {Maiolino}}]{Curti2020MNRAS.491..944C}
{Curti}, M., {Mannucci}, F., {Cresci}, G., \& {Maiolino}, R. 2020, \mnras, 491,
  944

\bibitem[{{Davis} {et~al.}(2022){Davis}, {Kaviraj}, {Hardcastle}, {Martin},
  {Jackson}, {Kraljic}, {Malek}, {Peirani}, {Smith}, {Volonteri}, \&
  {Wang}}]{Davis2022MNRAS.511.4109D}
{Davis}, F., {Kaviraj}, S., {Hardcastle}, M.~J., {et~al.} 2022, \mnras, 511,
  4109

\bibitem[{{Deller} \& {Middelberg}(2014)}]{DellerMiddelberg2014AJ....147...14D}
{Deller}, A.~T. \& {Middelberg}, E. 2014, \aj, 147, 14

\bibitem[{{Dong} \& {Hallinan}(2023)}]{Dong2023ApJ...948..119D}
{Dong}, D.~Z. \& {Hallinan}, G. 2023, \apj, 948, 119

\bibitem[{{Eckart} {et~al.}(1986){Eckart}, {Witzel}, {Biermann}, {Johnston},
  {Simon}, {Schalinski}, \& {Kuhr}}]{Eckart1986A&A...168...17E}
{Eckart}, A., {Witzel}, A., {Biermann}, P., {et~al.} 1986, \aap, 168, 17

\bibitem[{{Eftekhari} {et~al.}(2020){Eftekhari}, {Berger}, {Margalit},
  {Metzger}, \& {Williams}}]{Eftekhari2020ApJ...895...98E}
{Eftekhari}, T., {Berger}, E., {Margalit}, B., {Metzger}, B.~D., \& {Williams},
  P.~K.~G. 2020, \apj, 895, 98

\bibitem[{{Flewelling} {et~al.}(2020){Flewelling}, {Magnier}, {Chambers},
  {Heasley}, {Holmberg}, {Huber}, {Sweeney}, {Waters}, {Calamida}, {Casertano},
  {Chen}, {Farrow}, {Hasinger}, {Henderson}, {Long}, {Metcalfe}, {Narayan},
  {Nieto-Santisteban}, {Norberg}, {Rest}, {Saglia}, {Szalay}, {Thakar},
  {Tonry}, {Valenti}, {Werner}, {White}, {Denneau}, {Draper}, {Hodapp},
  {Jedicke}, {Kaiser}, {Kudritzki}, {Price}, {Wainscoat}, {Chastel}, {McLean},
  {Postman}, \& {Shiao}}]{Flewelling2020ApJS..251....7F}
{Flewelling}, H.~A., {Magnier}, E.~A., {Chambers}, K.~C., {et~al.} 2020, \apjs,
  251, 7

\bibitem[{{Gaensler} \& {Slane}(2006)}]{Gaensler2006ARA&A..44...17G}
{Gaensler}, B.~M. \& {Slane}, P.~O. 2006, \araa, 44, 17

\bibitem[{{Ger{\'e}b} {et~al.}(2015){Ger{\'e}b}, {Morganti}, {Oosterloo},
  {Hoppmann}, \& {Staveley-Smith}}]{Gereb2015A&A...580A..43G}
{Ger{\'e}b}, K., {Morganti}, R., {Oosterloo}, T.~A., {Hoppmann}, L., \&
  {Staveley-Smith}, L. 2015, \aap, 580, A43

\bibitem[{{Ginsburg} {et~al.}(2019){Ginsburg}, {Sip{\H{o}}cz}, {Brasseur},
  {Cowperthwaite}, {Craig}, {Deil}, {Guillochon}, {Guzman}, {Liedtke}, {Lian
  Lim}, {Lockhart}, {Mommert}, {Morris}, {Norman}, {Parikh}, {Persson},
  {Robitaille}, {Segovia}, {Singer}, {Tollerud}, {de Val-Borro}, {Valtchanov},
  {Woillez}, {Astroquery Collaboration}, \& {a subset of astropy
  Collaboration}}]{Ginsburg2019AJ....157...98G}
{Ginsburg}, A., {Sip{\H{o}}cz}, B.~M., {Brasseur}, C.~E., {et~al.} 2019, \aj,
  157, 98

\bibitem[{{Gonz{\'a}lez Delgado} {et~al.}(2015){Gonz{\'a}lez Delgado},
  {Garc{\'\i}a-Benito}, {P{\'e}rez}, {Cid Fernandes}, {de Amorim},
  {Cortijo-Ferrero}, {Lacerda}, {L{\'o}pez Fern{\'a}ndez}, {Vale-Asari},
  {S{\'a}nchez}, {Moll{\'a}}, {Ruiz-Lara}, {S{\'a}nchez-Bl{\'a}zquez},
  {Walcher}, {Alves}, {Aguerri}, {Bekerait{\'e}}, {Bland-Hawthorn}, {Galbany},
  {Gallazzi}, {Husemann}, {Iglesias-P{\'a}ramo}, {Kalinova},
  {L{\'o}pez-S{\'a}nchez}, {Marino}, {M{\'a}rquez}, {Masegosa}, {Mast},
  {M{\'e}ndez-Abreu}, {Mendoza}, {del Olmo}, {P{\'e}rez}, {Quirrenbach}, \&
  {Zibetti}}]{Gonzalez2015A&A...581A.103G}
{Gonz{\'a}lez Delgado}, R.~M., {Garc{\'\i}a-Benito}, R., {P{\'e}rez}, E.,
  {et~al.} 2015, \aap, 581, A103

\bibitem[{{Greene} {et~al.}(2020){Greene}, {Strader}, \&
  {Ho}}]{Greene2020ARA&A..58..257G}
{Greene}, J.~E., {Strader}, J., \& {Ho}, L.~C. 2020, \araa, 58, 257

\bibitem[{{G{\"u}ltekin} {et~al.}(2019){G{\"u}ltekin}, {King}, {Cackett},
  {Nyland}, {Miller}, {Di Matteo}, {Markoff}, \&
  {Rupen}}]{Gultekin2019ApJ...871...80G}
{G{\"u}ltekin}, K., {King}, A.~L., {Cackett}, E.~M., {et~al.} 2019, \apj, 871,
  80

\bibitem[{{Guo} {et~al.}(2022){Guo}, {Liu}, {Wang}, {Wang}, {Zhang}, {Ji},
  {Han}, \& {Chen}}]{Guo2022yCat..36670044G}
{Guo}, Y., {Liu}, C., {Wang}, L., {et~al.} 2022, VizieR Online Data Catalog,
  J/A+A/667/A44

\bibitem[{{G{\"u}rkan} {et~al.}(2018){G{\"u}rkan}, {Hardcastle}, {Smith},
  {Best}, {Bourne}, {Calistro-Rivera}, {Heald}, {Jarvis}, {Prandoni},
  {R{\"o}ttgering}, {Sabater}, {Shimwell}, {Tasse}, \&
  {Williams}}]{Gurkan2018MNRAS.475.3010G}
{G{\"u}rkan}, G., {Hardcastle}, M.~J., {Smith}, D.~J.~B., {et~al.} 2018,
  \mnras, 475, 3010

\bibitem[{{Hessels} {et~al.}(2019){Hessels}, {Spitler}, {Seymour}, {Cordes},
  {Michilli}, {Lynch}, {Gourdji}, {Archibald}, {Bassa}, {Bower}, {Chatterjee},
  {Connor}, {Crawford}, {Deneva}, {Gajjar}, {Kaspi}, {Keimpema}, {Law},
  {Marcote}, {McLaughlin}, {Paragi}, {Petroff}, {Ransom}, {Scholz}, {Stappers},
  \& {Tendulkar}}]{Hessels2019ApJ...876L..23H}
{Hessels}, J.~W.~T., {Spitler}, L.~G., {Seymour}, A.~D., {et~al.} 2019, \apjl,
  876, L23

\bibitem[{{Ivezi{\'c}} {et~al.}(2002){Ivezi{\'c}}, {Menou}, {Knapp}, {Strauss},
  {Lupton}, {Vanden Berk}, {Richards}, {Tremonti}, {Weinstein}, {Anderson},
  {Bahcall}, {Becker}, {Bernardi}, {Blanton}, {Eisenstein}, {Fan},
  {Finkbeiner}, {Finlator}, {Frieman}, {Gunn}, {Hall}, {Kim}, {Kinkhabwala},
  {Narayanan}, {Rockosi}, {Schlegel}, {Schneider}, {Strateva}, {SubbaRao},
  {Thakar}, {Voges}, {White}, {Yanny}, {Brinkmann}, {Doi}, {Fukugita},
  {Hennessy}, {Munn}, {Nichol}, \& {York}}]{Ivezic2002AJ....124.2364I}
{Ivezi{\'c}}, {\v{Z}}., {Menou}, K., {Knapp}, G.~R., {et~al.} 2002, \aj, 124,
  2364

\bibitem[{{Jarrett} {et~al.}(2011){Jarrett}, {Cohen}, {Masci}, {Wright},
  {Stern}, {Benford}, {Blain}, {Carey}, {Cutri}, {Eisenhardt}, {Lonsdale},
  {Mainzer}, {Marsh}, {Padgett}, {Petty}, {Ressler}, {Skrutskie}, {Stanford},
  {Surace}, {Tsai}, {Wheelock}, \& {Yan}}]{Jarrett2011ApJ...735..112J}
{Jarrett}, T.~H., {Cohen}, M., {Masci}, F., {et~al.} 2011, \apj, 735, 112

\bibitem[{{Kauffmann} {et~al.}(2003){Kauffmann}, {Heckman}, {Tremonti},
  {Brinchmann}, {Charlot}, {White}, {Ridgway}, {Brinkmann}, {Fukugita}, {Hall},
  {Ivezi{\'c}}, {Richards}, \& {Schneider}}]{Kauffmann2003MNRAS.346.1055K}
{Kauffmann}, G., {Heckman}, T.~M., {Tremonti}, C., {et~al.} 2003, \mnras, 346,
  1055

\bibitem[{{Kewley} {et~al.}(2001){Kewley}, {Dopita}, {Sutherland}, {Heisler},
  \& {Trevena}}]{Kewley2001ApJ...556..121K}
{Kewley}, L.~J., {Dopita}, M.~A., {Sutherland}, R.~S., {Heisler}, C.~A., \&
  {Trevena}, J. 2001, \apj, 556, 121

\bibitem[{{Koss} {et~al.}(2017){Koss}, {Trakhtenbrot}, {Ricci}, {Lamperti},
  {Oh}, {Berney}, {Schawinski}, {Balokovi{\'c}}, {Baronchelli}, {Crenshaw},
  {Fischer}, {Gehrels}, {Harrison}, {Hashimoto}, {Hogg}, {Ichikawa}, {Masetti},
  {Mushotzky}, {Sartori}, {Stern}, {Treister}, {Ueda}, {Veilleux}, \&
  {Winter}}]{Koss2017ApJ...850...74K}
{Koss}, M., {Trakhtenbrot}, B., {Ricci}, C., {et~al.} 2017, \apj, 850, 74

\bibitem[{{Kothes} {et~al.}(2006){Kothes}, {Fedotov}, {Foster}, \&
  {Uyan{\i}ker}}]{Kothes2006A&A...457.1081K}
{Kothes}, R., {Fedotov}, K., {Foster}, T.~J., \& {Uyan{\i}ker}, B. 2006, \aap,
  457, 1081

\bibitem[{{Kron}(1980)}]{Kron1980ApJS...43..305K}
{Kron}, R.~G. 1980, \apjs, 43, 305

\bibitem[{{Lacy} {et~al.}(2020){Lacy}, {Baum}, {Chandler}, {Chatterjee},
  {Clarke}, {Deustua}, {English}, {Farnes}, {Gaensler}, {Gugliucci},
  {Hallinan}, {Kent}, {Kimball}, {Law}, {Lazio}, {Marvil}, {Mao}, {Medlin},
  {Mooley}, {Murphy}, {Myers}, {Osten}, {Richards}, {Rosolowsky}, {Rudnick},
  {Schinzel}, {Sivakoff}, {Sjouwerman}, {Taylor}, {White}, {Wrobel},
  {Andernach}, {Beasley}, {Berger}, {Bhatnager}, {Birkinshaw}, {Bower},
  {Brandt}, {Brown}, {Burke-Spolaor}, {Butler}, {Comerford}, {Demorest}, {Fu},
  {Giacintucci}, {Golap}, {G{\"u}th}, {Hales}, {Hiriart}, {Hodge}, {Horesh},
  {Ivezi{\'c}}, {Jarvis}, {Kamble}, {Kassim}, {Liu}, {Loinard}, {Lyons},
  {Masters}, {Mezcua}, {Moellenbrock}, {Mroczkowski}, {Nyland}, {O'Dea},
  {O'Sullivan}, {Peters}, {Radford}, {Rao}, {Robnett}, {Salcido}, {Shen},
  {Sobotka}, {Witz}, {Vaccari}, {van Weeren}, {Vargas}, {Williams}, \&
  {Yoon}}]{Lacy2020PASP..132c5001L}
{Lacy}, M., {Baum}, S.~A., {Chandler}, C.~J., {et~al.} 2020, \pasp, 132, 035001

\bibitem[{{Law} {et~al.}(2022){Law}, {Connor}, \&
  {Aggarwal}}]{Law2022ApJ...927...55L}
{Law}, C.~J., {Connor}, L., \& {Aggarwal}, K. 2022, \apj, 927, 55

\bibitem[{{Marcote} {et~al.}(2017){Marcote}, {Paragi}, {Hessels}, {Keimpema},
  {van Langevelde}, {Huang}, {Bassa}, {Bogdanov}, {Bower}, {Burke-Spolaor},
  {Butler}, {Campbell}, {Chatterjee}, {Cordes}, {Demorest}, {Garrett}, {Ghosh},
  {Kaspi}, {Law}, {Lazio}, {McLaughlin}, {Ransom}, {Salter}, {Scholz},
  {Seymour}, {Siemion}, {Spitler}, {Tendulkar}, \&
  {Wharton}}]{Marcote2017ApJ...834L...8M}
{Marcote}, B., {Paragi}, Z., {Hessels}, J.~W.~T., {et~al.} 2017, \apjl, 834, L8

\bibitem[{{Margalit} \& {Metzger}(2018)}]{Margalit2018ApJ...868L...4M}
{Margalit}, B. \& {Metzger}, B.~D. 2018, \apjl, 868, L4

\bibitem[{{Martin} {et~al.}(2005){Martin}, {Fanson}, {Schiminovich},
  {Morrissey}, {Friedman}, {Barlow}, {Conrow}, {Grange}, {Jelinsky},
  {Milliard}, {Siegmund}, {Bianchi}, {Byun}, {Donas}, {Forster}, {Heckman},
  {Lee}, {Madore}, {Malina}, {Neff}, {Rich}, {Small}, {Surber}, {Szalay},
  {Welsh}, \& {Wyder}}]{Martin2005ApJ...619L...1M}
{Martin}, D.~C., {Fanson}, J., {Schiminovich}, D., {et~al.} 2005, \apjl, 619,
  L1

\bibitem[{{McConnell} {et~al.}(2020){McConnell}, {Hale}, {Lenc}, {Banfield},
  {Heald}, {Hotan}, {Leung}, {Moss}, {Murphy}, {O'Brien}, {Pritchard}, {Raja},
  {Sadler}, {Stewart}, {Thomson}, {Whiting}, {Allison}, {Amy}, {Anderson},
  {Ball}, {Bannister}, {Bell}, {Bock}, {Bolton}, {Bunton}, {Chippendale},
  {Collier}, {Cooray}, {Cornwell}, {Diamond}, {Edwards}, {Gupta}, {Hayman},
  {Heywood}, {Jackson}, {Koribalski}, {Lee-Waddell}, {McClure-Griffiths}, {Ng},
  {Norris}, {Phillips}, {Reynolds}, {Roxby}, {Schinckel}, {Shields},
  {Tremblay}, {Tzioumis}, {Voronkov}, \&
  {Westmeier}}]{McConnell2020PASA...37...48M}
{McConnell}, D., {Hale}, C.~L., {Lenc}, E., {et~al.} 2020, \pasa, 37, e048

\bibitem[{{McConnell} \& {Ma}(2013)}]{McConnell2013ApJ...764..184M}
{McConnell}, N.~J. \& {Ma}, C.-P. 2013, \apj, 764, 184

\bibitem[{{Michilli} {et~al.}(2018){Michilli}, {Seymour}, {Hessels}, {Spitler},
  {Gajjar}, {Archibald}, {Bower}, {Chatterjee}, {Cordes}, {Gourdji}, {Heald},
  {Kaspi}, {Law}, {Sobey}, {Adams}, {Bassa}, {Bogdanov}, {Brinkman},
  {Demorest}, {Fernandez}, {Hellbourg}, {Lazio}, {Lynch}, {Maddox}, {Marcote},
  {McLaughlin}, {Paragi}, {Ransom}, {Scholz}, {Siemion}, {Tendulkar}, {van
  Rooy}, {Wharton}, \& {Whitlow}}]{Michilli2018Natur.553..182M}
{Michilli}, D., {Seymour}, A., {Hessels}, J.~W.~T., {et~al.} 2018, \nat, 553,
  182

\bibitem[{{Mohan} \& {Rafferty}(2015)}]{Mohan2015ascl.soft02007M}
{Mohan}, N. \& {Rafferty}, D. 2015, {PyBDSF: Python Blob Detection and Source
  Finder}, Astrophysics Source Code Library, record ascl:1502.007

\bibitem[{{Molina} {et~al.}(2021){Molina}, {Reines}, {Greene}, {Darling}, \&
  {Condon}}]{Molina2021ApJ...910....5M}
{Molina}, M., {Reines}, A.~E., {Greene}, J.~E., {Darling}, J., \& {Condon},
  J.~J. 2021, \apj, 910, 5

\bibitem[{{Mondal} {et~al.}(2020){Mondal}, {Bera}, {Chandra}, \&
  {Das}}]{Mondal2020MNRAS.498.3863M}
{Mondal}, S., {Bera}, A., {Chandra}, P., \& {Das}, B. 2020, \mnras, 498, 3863

\bibitem[{{Morabito} {et~al.}(2022){Morabito}, {Jackson}, {Mooney}, {Sweijen},
  {Badole}, {Kukreti}, {Venkattu}, {Groeneveld}, {Kappes}, {Bonnassieux},
  {Drabent}, {Iacobelli}, {Croston}, {Best}, {Bondi}, {Callingham}, {Conway},
  {Deller}, {Hardcastle}, {McKean}, {Miley}, {Moldon}, {R{\"o}ttgering},
  {Tasse}, {Shimwell}, {van Weeren}, {Anderson}, {Asgekar}, {Avruch}, {van
  Bemmel}, {Bentum}, {Bonafede}, {Brouw}, {Butcher}, {Ciardi}, {Corstanje},
  {Coolen}, {Damstra}, {de Gasperin}, {Duscha}, {Eisl{\"o}ffel}, {Engels},
  {Falcke}, {Garrett}, {Griessmeier}, {Gunst}, {van Haarlem}, {Hoeft}, {van der
  Horst}, {J{\"u}tte}, {Kadler}, {Koopmans}, {Krankowski}, {Mann}, {Nelles},
  {Oonk}, {Orru}, {Paas}, {Pandey}, {Pizzo}, {Pandey-Pommier}, {Reich},
  {Rothkaehl}, {Ruiter}, {Schwarz}, {Shulevski}, {Soida}, {Tagger}, {Vocks},
  {Wijers}, {Wijnholds}, {Wucknitz}, {Zarka}, \&
  {Zucca}}]{Morabito2022A&A...658A...1M}
{Morabito}, L.~K., {Jackson}, N.~J., {Mooney}, S., {et~al.} 2022, \aap, 658, A1

\bibitem[{{Murase} {et~al.}(2016){Murase}, {Kashiyama}, \&
  {M{\'e}sz{\'a}ros}}]{Murase2016MNRAS.461.1498M}
{Murase}, K., {Kashiyama}, K., \& {M{\'e}sz{\'a}ros}, P. 2016, \mnras, 461,
  1498

\bibitem[{{Niu} {et~al.}(2022){Niu}, {Aggarwal}, {Li}, {Zhang}, {Chatterjee},
  {Tsai}, {Yu}, {Law}, {Burke-Spolaor}, {Cordes}, {Zhang}, {Ocker}, {Yao},
  {Wan}, {Feng}, {Niino}, {Bochenek}, {Cruces}, {Connor}, {Jiang}, {Dai},
  {Luo}, {Li}, {Miao}, {Niu}, {Anna-Thomas}, {Sydnor}, {Stern}, {Wang}, {Yuan},
  {Yue}, {Zhou}, {Yan}, {Zhu}, \& {Zhang}}]{Niu2022Natur.606..873N}
{Niu}, C.~H., {Aggarwal}, K., {Li}, D., {et~al.} 2022, \nat, 606, 873

\bibitem[{{Nyland} {et~al.}(2020){Nyland}, {Dong}, {Hallinan}, {Patil},
  {Sarbadhicary}, {Lacy}, {Polisensky}, {Kassim}, {Kimball}, \&
  {Peters}}]{Nyland2020AAS...23512901N}
{Nyland}, K., {Dong}, D., {Hallinan}, G., {et~al.} 2020, in American
  Astronomical Society Meeting Abstracts, Vol. 235, American Astronomical
  Society Meeting Abstracts \#235, 129.01

\bibitem[{{Ofek}(2017)}]{Ofek2017ApJ...846...44O}
{Ofek}, E.~O. 2017, \apj, 846, 44

\bibitem[{{Oke} \& {Gunn}(1983)}]{Oke1983ApJ...266..713O}
{Oke}, J.~B. \& {Gunn}, J.~E. 1983, \apj, 266, 713

\bibitem[{{Paragi} {et~al.}(2014){Paragi}, {Frey}, {Kaaret}, {Cseh},
  {Overzier}, \& {Kharb}}]{Paragi2014ApJ...791....2P}
{Paragi}, Z., {Frey}, S., {Kaaret}, P., {et~al.} 2014, \apj, 791, 2

\bibitem[{{Planck Collaboration} {et~al.}(2011){Planck Collaboration},
  {Aatrokoski}, {Ade}, {Aghanim}, {Aller}, {Aller}, {Angelakis}, {Arnaud},
  {Ashdown}, {Aumont}, {Baccigalupi}, {Balbi}, {Banday}, {Barreiro},
  {Bartlett}, {Battaner}, {Benabed}, {Beno{\^\i}t}, {Berdyugin}, {Bernard},
  {Bersanelli}, {Bhatia}, {Bonaldi}, {Bonavera}, {Bond}, {Borrill}, {Bouchet},
  {Bucher}, {Burigana}, {Burrows}, {Cabella}, {Capalbi}, {Cappellini},
  {Cardoso}, {Catalano}, {Cavazzuti}, {Cay{\'o}n}, {Challinor}, {Chamballu},
  {Chary}, {Chiang}, {Christensen}, {Clements}, {Colafrancesco}, {Colombi},
  {Couchot}, {Coulais}, {Cutini}, {Cuttaia}, {Danese}, {Davies}, {Davis}, {de
  Bernardis}, {de Gasperis}, {de Rosa}, {de Zotti}, {Delabrouille}, {Delouis},
  {Dickinson}, {Dole}, {Donzelli}, {Dor{\'e}}, {D{\"o}rl}, {Douspis}, {Dupac},
  {Efstathiou}, {En{\ss}lin}, {Finelli}, {Forni}, {Frailis}, {Franceschi},
  {Fuhrmann}, {Galeotta}, {Ganga}, {Gargano}, {Gasparrini}, {Gehrels}, {Giard},
  {Giardino}, {Giglietto}, {Giommi}, {Giordano}, {Giraud-H{\'e}raud},
  {Gonz{\'a}lez-Nuevo}, {G{\'o}rski}, {Gratton}, {Gregorio}, {Gruppuso},
  {Harrison}, {Henrot-Versill{\'e}}, {Herranz}, {Hildebrandt}, {Hivon},
  {Hobson}, {Holmes}, {Hovest}, {Hoyland}, {Huffenberger}, {Jaffe}, {Juvela},
  {Keih{\"a}nen}, {Keskitalo}, {King}, {Kisner}, {Kneissl}, {Knox},
  {Krichbaum}, {Kurki-Suonio}, {Lagache}, {L{\"a}hteenm{\"a}ki}, {Lamarre},
  {Lasenby}, {Laureijs}, {Lavonen}, {Lawrence}, {Leach}, {Leonardi},
  {Le{\'o}n-Tavares}, {Linden-V{\o}rnle}, {Lindfors}, {L{\'o}pez-Caniego},
  {Lubin}, {Mac{\'\i}as-P{\'e}rez}, {Maffei}, {Maino}, {Mandolesi}, {Mann},
  {Maris}, {Mart{\'\i}nez-Gonz{\'a}lez}, {Masi}, {Massardi}, {Matarrese},
  {Matthai}, {Max-Moerbeck}, {Mazziotta}, {Mazzotta}, {Melchiorri}, {Mendes},
  {Mennella}, {Michelson}, {Mingaliev}, {Mitra}, {Miville-Desch{\^e}nes},
  {Moneti}, {Monte}, {Montier}, {Morgante}, {Mortlock}, {Munshi}, {Murphy},
  {Naselsky}, {Natoli}, {Nestoras}, {Netterfield}, {Nieppola}, {Nilsson},
  {N{\o}rgaard-Nielsen}, {Noviello}, {Novikov}, {Novikov}, {O'Dwyer},
  {Osborne}, {Pajot}, {Partridge}, {Pasian}, {Patanchon}, {Pavlidou},
  {Pearson}, {Perdereau}, {Perotto}, {Perri}, {Perrotta}, {Piacentini}, {Piat},
  {Plaszczynski}, {Platania}, {Pointecouteau}, {Polenta}, {Ponthieu},
  {Poutanen}, {Pr{\'e}zeau}, {Procopio}, {Prunet}, {Puget}, {Rachen},
  {Rain{\`o}}, {Reach}, {Readhead}, {Rebolo}, {Reeves}, {Reinecke}, {Reinthal},
  {Renault}, {Ricciardi}, {Richards}, {Riller}, {Riquelme}, {Ristorcelli},
  {Rocha}, {Rosset}, {Rowan-Robinson}, {Rubi{\~n}o-Mart{\'\i}n}, {Rusholme},
  {Saarinen}, {Sandri}, {Savolainen}, {Scott}, {Seiffert}, {Sievers},
  {Sillanp{\"a}{\"a}}, {Smoot}, {Sotnikova}, {Starck}, {Stevenson}, {Stivoli},
  {Stolyarov}, {Sudiwala}, {Sygnet}, {Takalo}, {Tammi}, {Tauber}, {Terenzi},
  {Thompson}, {Toffolatti}, {Tomasi}, {Tornikoski}, {Torre}, {Tosti},
  {Tramacere}, {Tristram}, {Tuovinen}, {T{\"u}rler}, {Turunen}, {Umana},
  {Ungerechts}, {Valenziano}, {Valtaoja}, {Varis}, {Verrecchia}, {Vielva},
  {Villa}, {Vittorio}, {Wandelt}, {Wu}, {Yvon}, {Zacchei}, {Zensus}, {Zhou}, \&
  {Zonca}}]{Planck2011A&A...536A..15P}
{Planck Collaboration}, {Aatrokoski}, J., {Ade}, P.~A.~R., {et~al.} 2011, \aap,
  536, A15

\bibitem[{{Platts} {et~al.}(2021){Platts}, {Caleb}, {Stappers}, {Main},
  {Weltman}, {Shock}, {Kramer}, {Bezuidenhout}, {Jankowski}, {Morello},
  {Possenti}, {Rajwade}, {Rhodes}, \& {Wu}}]{Platts2021MNRAS.505.3041P}
{Platts}, E., {Caleb}, M., {Stappers}, B.~W., {et~al.} 2021, \mnras, 505, 3041

\bibitem[{{Rajwade} {et~al.}(2020){Rajwade}, {Mickaliger}, {Stappers},
  {Morello}, {Agarwal}, {Bassa}, {Breton}, {Caleb}, {Karastergiou}, {Keane}, \&
  {Lorimer}}]{Rajwade2020MNRAS.495.3551R}
{Rajwade}, K.~M., {Mickaliger}, M.~B., {Stappers}, B.~W., {et~al.} 2020,
  \mnras, 495, 3551

\bibitem[{{Reines} {et~al.}(2020){Reines}, {Condon}, {Darling}, \&
  {Greene}}]{Reines2020ApJ...888...36R}
{Reines}, A.~E., {Condon}, J.~J., {Darling}, J., \& {Greene}, J.~E. 2020, \apj,
  888, 36

\bibitem[{{Reines} {et~al.}(2013){Reines}, {Greene}, \&
  {Geha}}]{Reines2013ApJ...775..116R}
{Reines}, A.~E., {Greene}, J.~E., \& {Geha}, M. 2013, \apj, 775, 116

\bibitem[{{Reines} \& {Volonteri}(2015)}]{Reines2015ApJ...813...82R}
{Reines}, A.~E. \& {Volonteri}, M. 2015, \apj, 813, 82

\bibitem[{{Resmi} {et~al.}(2021){Resmi}, {Vink}, \&
  {Ishwara-Chandra}}]{Resmi2021A&A...655A.102R}
{Resmi}, L., {Vink}, J., \& {Ishwara-Chandra}, C.~H. 2021, \aap, 655, A102

\bibitem[{{Robitaille} \& {Bressert}(2012)}]{Robitaille2012ascl.soft08017R}
{Robitaille}, T. \& {Bressert}, E. 2012, {APLpy: Astronomical Plotting Library
  in Python}, Astrophysics Source Code Library, record ascl:1208.017

\bibitem[{{Sargent} {et~al.}(2022){Sargent}, {Johnson}, {Reines}, {Secrest},
  {van der Horst}, {Cigan}, {Darling}, \&
  {Greene}}]{Sargent2022ApJ...933..160S}
{Sargent}, A.~J., {Johnson}, M.~C., {Reines}, A.~E., {et~al.} 2022, \apj, 933,
  160

\bibitem[{{Schechter}(1976)}]{1976ApJ...203..297S}
{Schechter}, P. 1976, \apj, 203, 297

\bibitem[{{Shimwell} {et~al.}(2022){Shimwell}, {Hardcastle}, {Tasse}, {Best},
  {R{\"o}ttgering}, {Williams}, {Botteon}, {Drabent}, {Mechev}, {Shulevski},
  {van Weeren}, {Bester}, {Br{\"u}ggen}, {Brunetti}, {Callingham}, {Chy{\.z}y},
  {Conway}, {Dijkema}, {Duncan}, {de Gasperin}, {Hale}, {Haverkorn}, {Hugo},
  {Jackson}, {Mevius}, {Miley}, {Morabito}, {Morganti}, {Offringa}, {Oonk},
  {Rafferty}, {Sabater}, {Smith}, {Schwarz}, {Smirnov}, {O'Sullivan},
  {Vedantham}, {White}, {Albert}, {Alegre}, {Asabere}, {Bacon}, {Bonafede},
  {Bonnassieux}, {Brienza}, {Bilicki}, {Bonato}, {Calistro Rivera}, {Cassano},
  {Cochrane}, {Croston}, {Cuciti}, {Dallacasa}, {Danezi}, {Dettmar}, {Di
  Gennaro}, {Edler}, {En{\ss}lin}, {Emig}, {Franzen}, {Garc{\'\i}a-Vergara},
  {Grange}, {G{\"u}rkan}, {Hajduk}, {Heald}, {Heesen}, {Hoang}, {Hoeft},
  {Horellou}, {Iacobelli}, {Jamrozy}, {Jeli{\'c}}, {Kondapally}, {Kukreti},
  {Kunert-Bajraszewska}, {Magliocchetti}, {Mahatma}, {Ma{\l}ek}, {Mandal},
  {Massaro}, {Meyer-Zhao}, {Mingo}, {Mostert}, {Nair}, {Nakoneczny},
  {Nikiel-Wroczy{\'n}ski}, {Orr{\'u}}, {Pajdosz-{\'S}mierciak}, {Pasini},
  {Prandoni}, {van Piggelen}, {Rajpurohit}, {Retana-Montenegro}, {Riseley},
  {Rowlinson}, {Saxena}, {Schrijvers}, {Sweijen}, {Siewert}, {Timmerman},
  {Vaccari}, {Vink}, {West}, {Wo{\l}owska}, {Zhang}, \&
  {Zheng}}]{Shimwell2022A&A...659A...1S}
{Shimwell}, T.~W., {Hardcastle}, M.~J., {Tasse}, C., {et~al.} 2022, \aap, 659,
  A1

\bibitem[{{Smith} {et~al.}(2021){Smith}, {Haskell}, {G{\"u}rkan}, {Best},
  {Hardcastle}, {Kondapally}, {Williams}, {Duncan}, {Cochrane}, {McCheyne},
  {R{\"o}ttgering}, {Sabater}, {Shimwell}, {Tasse}, {Bonato}, {Bondi},
  {Jarvis}, {Leslie}, {Prandoni}, \& {Wang}}]{Smith2021A&A...648A...6S}
{Smith}, D.~J.~B., {Haskell}, P., {G{\"u}rkan}, G., {et~al.} 2021, \aap, 648,
  A6

\bibitem[{{Spitler} {et~al.}(2016){Spitler}, {Scholz}, {Hessels}, {Bogdanov},
  {Brazier}, {Camilo}, {Chatterjee}, {Cordes}, {Crawford}, {Deneva}, {Ferdman},
  {Freire}, {Kaspi}, {Lazarus}, {Lynch}, {Madsen}, {McLaughlin}, {Patel},
  {Ransom}, {Seymour}, {Stairs}, {Stappers}, {van Leeuwen}, \&
  {Zhu}}]{Spitler2016Natur.531..202S}
{Spitler}, L.~G., {Scholz}, P., {Hessels}, J.~W.~T., {et~al.} 2016, \nat, 531,
  202

\bibitem[{{Sridhar} \& {Metzger}(2022)}]{Sridhar2022ApJ...937....5S}
{Sridhar}, N. \& {Metzger}, B.~D. 2022, \apj, 937, 5

\bibitem[{{Tasse} {et~al.}(2018){Tasse}, {Hugo}, {Mirmont}, {Smirnov},
  {Atemkeng}, {Bester}, {Hardcastle}, {Lakhoo}, {Perkins}, \&
  {Shimwell}}]{Tasse2018A&A...611A..87T}
{Tasse}, C., {Hugo}, B., {Mirmont}, M., {et~al.} 2018, \aap, 611, A87

\bibitem[{{Taylor}(2011)}]{Taylor2011ascl.soft01010T}
{Taylor}, M. 2011, {TOPCAT: Tool for OPerations on Catalogues And Tables},
  Astrophysics Source Code Library, record ascl:1101.010

\bibitem[{{Tendulkar} {et~al.}(2017){Tendulkar}, {Bassa}, {Cordes}, {Bower},
  {Law}, {Chatterjee}, {Adams}, {Bogdanov}, {Burke-Spolaor}, {Butler},
  {Demorest}, {Hessels}, {Kaspi}, {Lazio}, {Maddox}, {Marcote}, {McLaughlin},
  {Paragi}, {Ransom}, {Scholz}, {Seymour}, {Spitler}, {van Langevelde}, \&
  {Wharton}}]{Tendulkar2017ApJ...834L...7T}
{Tendulkar}, S.~P., {Bassa}, C.~G., {Cordes}, J.~M., {et~al.} 2017, \apjl, 834,
  L7

\bibitem[{{Tendulkar} {et~al.}(2021){Tendulkar}, {Gil de Paz}, {Kirichenko},
  {Hessels}, {Bhardwaj}, {{\'A}vila}, {Bassa}, {Chawla}, {Fonseca}, {Kaspi},
  {Keimpema}, {Kirsten}, {Lazio}, {Marcote}, {Masui}, {Nimmo}, {Paragi},
  {Rahman}, {Pay{\'a}}, {Scholz}, \& {Stairs}}]{Tendulkar2021ApJ...908L..12T}
{Tendulkar}, S.~P., {Gil de Paz}, A., {Kirichenko}, A.~Y., {et~al.} 2021,
  \apjl, 908, L12

\bibitem[{{Tremonti} {et~al.}(2004){Tremonti}, {Heckman}, {Kauffmann},
  {Brinchmann}, {Charlot}, {White}, {Seibert}, {Peng}, {Schlegel}, {Uomoto},
  {Fukugita}, \& {Brinkmann}}]{Tremonti2004ApJ...613..898T}
{Tremonti}, C.~A., {Heckman}, T.~M., {Kauffmann}, G., {et~al.} 2004, \apj, 613,
  898

\bibitem[{{Truebenbach} \& {Darling}(2017)}]{Truebenbach2017MNRAS.468..196T}
{Truebenbach}, A.~E. \& {Darling}, J. 2017, \mnras, 468, 196

\bibitem[{{Venkattu} {et~al.}(2023){Venkattu}, {Lundqvist}, {P{\'e}rez-Torres},
  {Morabito}, {Mold{\'o}n}, {Conway}, {Chandra}, \&
  {Tasse}}]{Venkattu2023arXiv230702365V}
{Venkattu}, D., {Lundqvist}, P., {P{\'e}rez-Torres}, M., {et~al.} 2023, arXiv
  e-prints, arXiv:2307.02365

\bibitem[{{V{\'e}ron-Cetty} \& {V{\'e}ron}(2010)}]{Veron2010A&A...518A..10V}
{V{\'e}ron-Cetty}, M.~P. \& {V{\'e}ron}, P. 2010, \aap, 518, A10

\bibitem[{Vohl(2023)}]{vohl_d_2023_7986693}
Vohl, D. 2023, matchmaker, Zenodo. DOI:10.5281/zenodo.7986693

\bibitem[{{Vohl} {et~al.}(2023){Vohl}, {Vedantham}, {Hessels}, \&
  {Bassa}}]{Vohl2023arXiv230311967V}
{Vohl}, D., {Vedantham}, H. .~K., {Hessels}, J.~W.~T., \& {Bassa}, C.~G. 2023,
  arXiv e-prints, arXiv:2303.11967

\bibitem[{{Wright} {et~al.}(2010){Wright}, {Eisenhardt}, {Mainzer}, {Ressler},
  {Cutri}, {Jarrett}, {Kirkpatrick}, {Padgett}, {McMillan}, {Skrutskie},
  {Stanford}, {Cohen}, {Walker}, {Mather}, {Leisawitz}, {Gautier}, {McLean},
  {Benford}, {Lonsdale}, {Blain}, {Mendez}, {Irace}, {Duval}, {Liu}, {Royer},
  {Heinrichsen}, {Howard}, {Shannon}, {Kendall}, {Walsh}, {Larsen}, {Cardon},
  {Schick}, {Schwalm}, {Abid}, {Fabinsky}, {Naes}, \&
  {Tsai}}]{Wright2010AJ....140.1868W}
{Wright}, E.~L., {Eisenhardt}, P. R.~M., {Mainzer}, A.~K., {et~al.} 2010, \aj,
  140, 1868

\bibitem[{{Yang} {et~al.}(2020){Yang}, {Gurvits}, {Paragi}, {Frey}, {Conway},
  {Liu}, \& {Cui}}]{Yang2020MNRAS.495L..71Y}
{Yang}, J., {Gurvits}, L.~I., {Paragi}, Z., {et~al.} 2020, \mnras, 495, L71

\bibitem[{{Zaja{\v{c}}ek} {et~al.}(2019){Zaja{\v{c}}ek}, {Busch},
  {Valencia-S.}, {Eckart}, {Britzen}, {Fuhrmann}, {Schneeloch}, {Fazeli},
  {Harrington}, \& {Zensus}}]{Zajacek2019A&A...630A..83Z}
{Zaja{\v{c}}ek}, M., {Busch}, G., {Valencia-S.}, M., {et~al.} 2019, \aap, 630,
  A83

\bibitem[{{Zhao} \& {Wang}(2021)}]{Zhao2021ApJ...923L..17Z}
{Zhao}, Z.~Y. \& {Wang}, F.~Y. 2021, \apjl, 923, L17

\end{thebibliography}
\bibliographystyle{aa}

%-------------------------------------------------------------------

\begin{appendix}

\section{The obtained L--SFR distribution}
\label{app:l_sfr}

\begin{figure*}
    \centering
    \includegraphics[width=17cm]{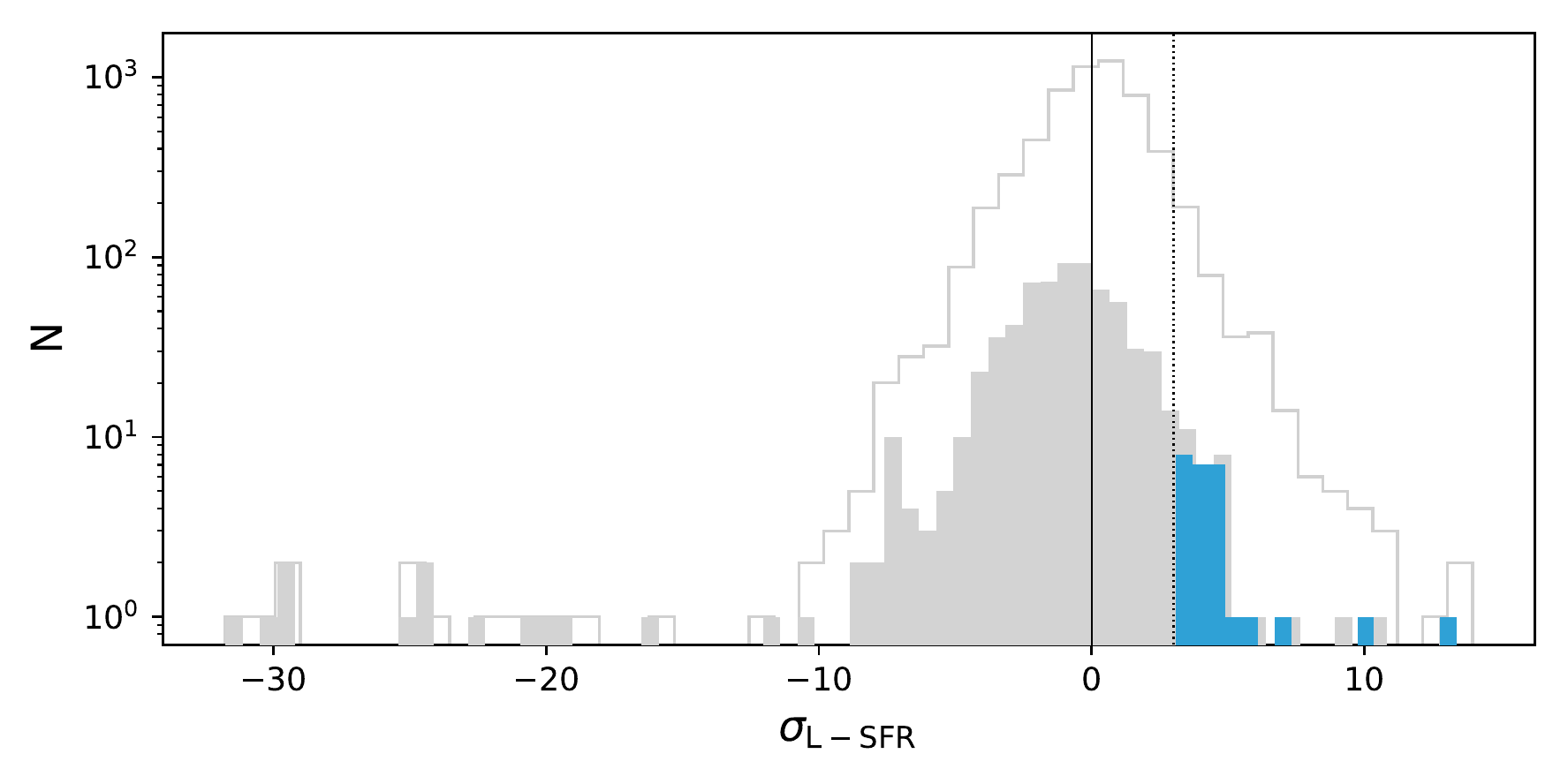} 
    \caption{
    Matched compact radio objects along the L--SFR plane. 
    Gray-filled bins indicate cases matched to dwarf galaxies. 
    Gray border bins indicate cases matched to galaxies at all masses available in CLU.
    Blue bins indicate OCRs.
    Very low-luminosity cases encountered in Figure \ref{fig:selection} are further investigated in Figure \ref{fig:high_sigmas_vs_redshift_offset_dwarfs}. 
    }%
    \label{fig:l_sfr_plane}%
\end{figure*}

The zero point on the L--SFR relation from \citet{Gurkan2018MNRAS.475.3010G} was obtained on more massive ($\sim10^{8.5}$--$10^11\,M_\odot$) galaxies than our dwarf sample. 
In figure \ref{fig:l_sfr_plane}, we show that the zero point corresponds to the peak of the distribution when matching compact radio sources to all galaxy masses available within CLU using a $6\arcsec$ matching radius. 
The peak of the distribution of matched dwarf galaxies falls just below the zero point.
Objects below $-10\sigma$ (14 dwarf galaxies) have redshifts below 0.002, except two extra cases at higher masses that  have redshifts of 0.003 and 0.027, respectively.

\begin{figure*}
    \centering
    \includegraphics[width=17cm]{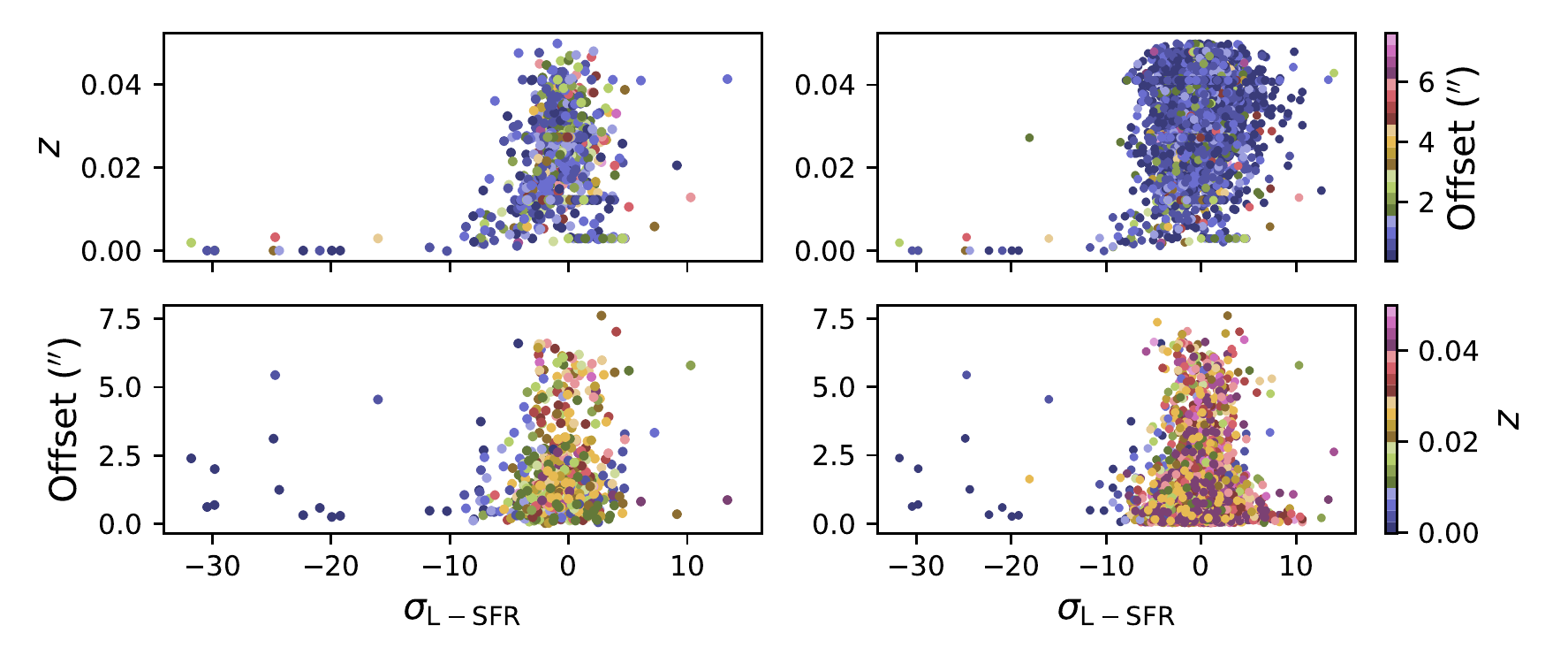}
    \caption{Compact radio objects matched to dwarf galaxies (left) and all galaxies (right) along the L--SFR plane as a function of redshift and offset. 
    Very low-luminosity cases encountered in Figure \ref{fig:selection} all have redshifts below 0.002.}%
    \label{fig:high_sigmas_vs_redshift_offset_dwarfs}%
\end{figure*}

\section{2MASX~J12594007+2751177 and SDSS~J143037.09+352052.8}
\label{sec:appendix1}
The dwarf galaxy 2MASX~J12594007+2751177 was shown to be of very low surface density in PS1 (Figure \ref{fig:family_plot_continued}). To better highlight the environment in which this galaxy resides, Figure \ref{fig:HSC} shows the same field as observed by PS1 in r filter (right) as shown previously, and a JVO Subaru/Suprime-Cam~\citep{Aihara2019PASJ...71..114A} composite image (left) using all filters---where 2MASX~J12594007+2751177 is apparent. 
Green markers (x, +) show the location of the LoTSS and CLU detections, respectively.
The Subaru Suprime-Cam image strengthens the hypothesis that 2MASX~J12594007+2751177 and ILTJ125944.53+275800.9 are unrelated.

\begin{figure*}
    \centering
    \includegraphics[width=17cm]{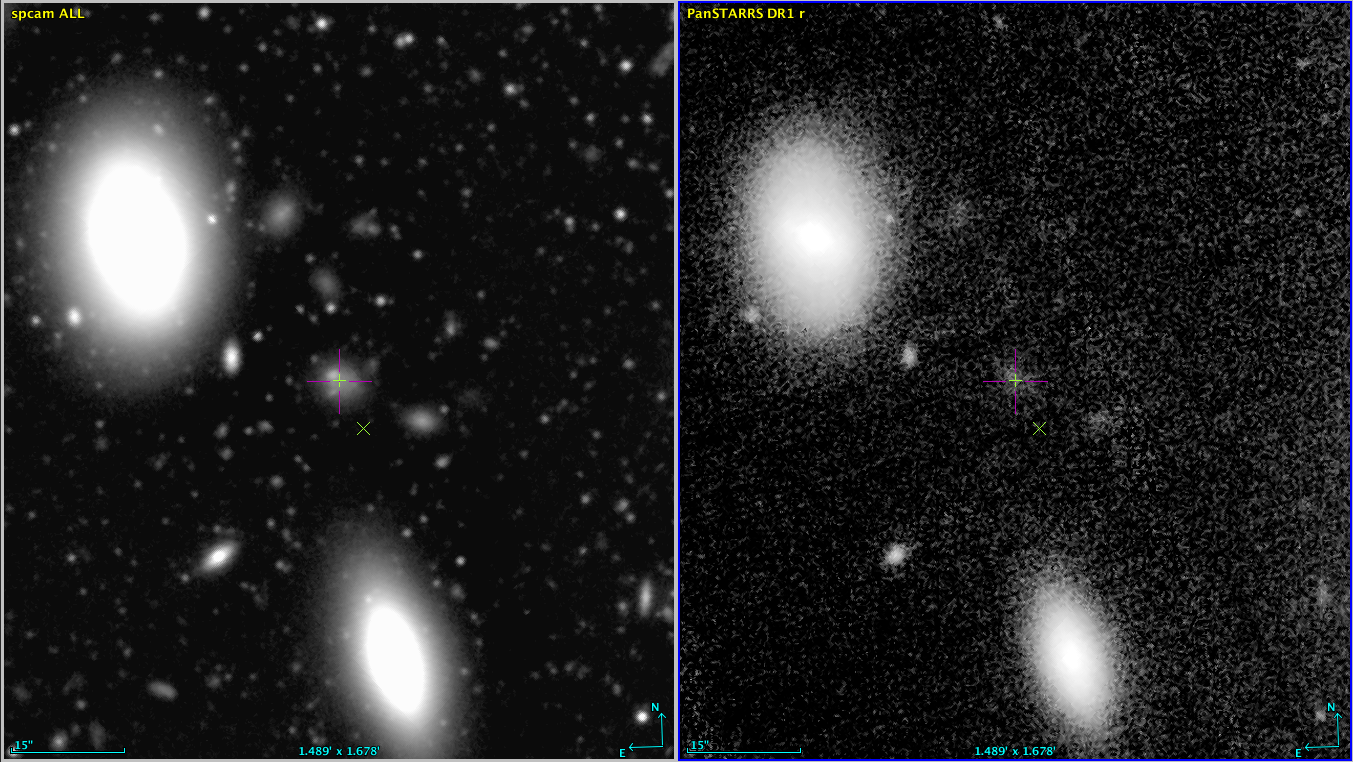} 
    \caption{Environment around 2MASX~J12594007+2751177 as observed by PS1 in r filter (right, Figure \ref{fig:family_plot_continued}), and JVO Subaru/Suprime-Cam composite image (left) using all filters---where 2MASX~J12594007+2751177 (pink markers) is apparent.
    Green markers ($\times$, +) show the location of the LoTSS and CLU detections, respectively.
    Figure generated with Aladin Desktop~\citep{Bonnarel2000A&AS..143...33B}.
    }%
    \label{fig:HSC}%
\end{figure*}

\begin{figure*}
    \centering
    \includegraphics[width=17cm]{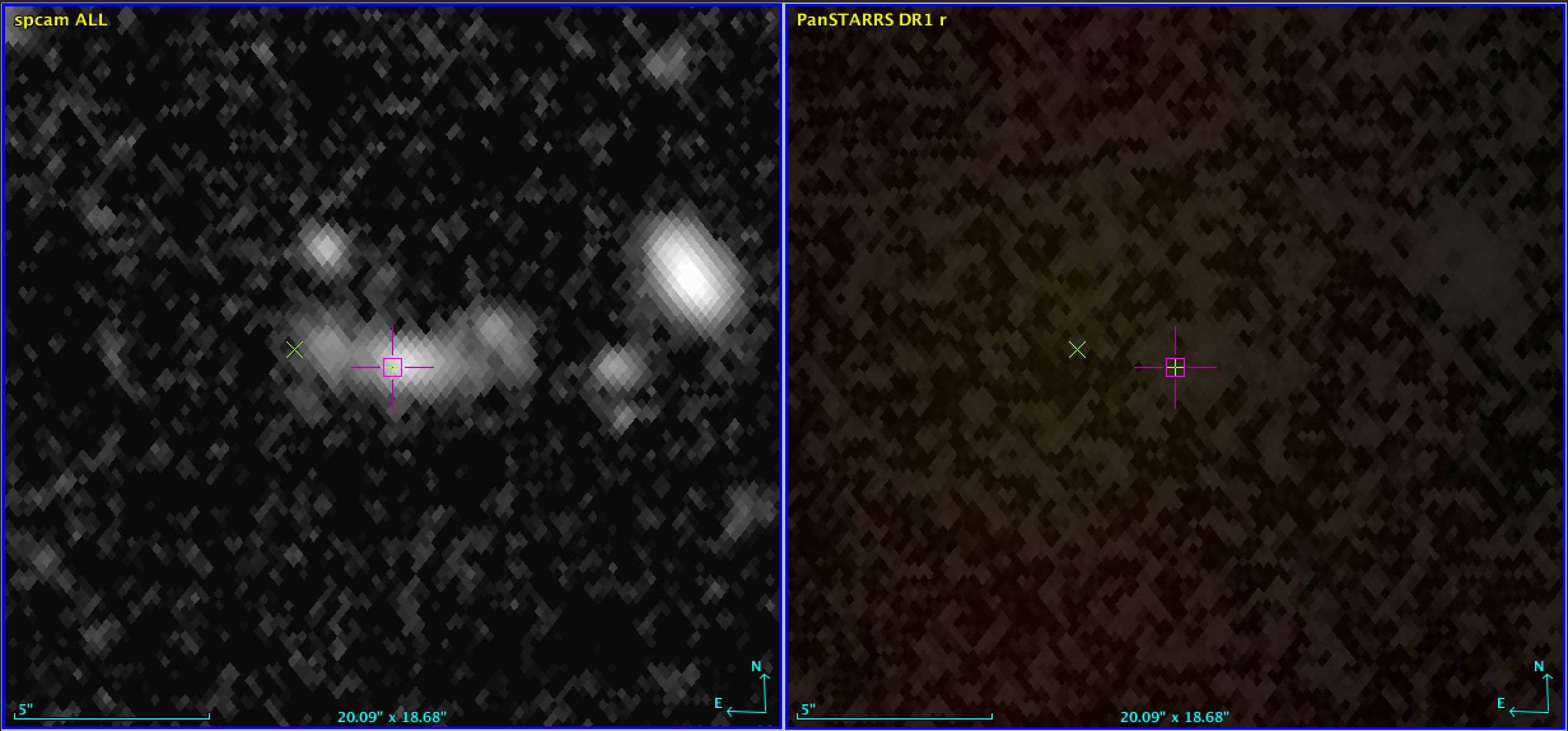} 
    \caption{Environment around SDSS~J143037.09+352052.8 as observed by PS1 in r filter (right, Figure \ref{fig:family_plot_continued}), and JVO Subaru/Suprime-Cam composite image (left) using all filters---where SDSS~J143037.09+352052.8 (pink markers) is apparent.
    Green markers ($\times$, +) show the location of the LoTSS and CLU detections, respectively.
    }%
    \label{fig:agn}%
\end{figure*}

Finally, the AGN candidate ILT~143037.30+352053.3 was matched to the host galaxy SDSS~J143037.09+352052.8. 
SDSS~J143037.09+352052.8 is not resolved in PS1 r filter shown in Figure \ref{fig:family_plot_continued}. 
We found a JVO Subaru/Suprime-Cam image for this field, which indicates that the radio source is within the optical footprint of the host, which we show in Figure \ref{fig:agn}.

\section{ACO~1656}
\label{sec:appendix2}

We note in \S\ref{sec:candidates} that four of our selected compact radio sources fall within galaxies that are members of the cluster of galaxies ACO~1656. These are galaxies and respective matched radio sources (Table~\ref{table:candidates}) are 2MASS J13002220+2814499 (ILTJ130022.42+281451.7), 
SDSS J125944.76+275807.1 (ILTJ125944.53+275800.9), 
2MASX J12594007+2751177 (ILTJ125940.18 +275123.5), and 
SSTSL2 J125915.27+274604.1 (ILTJ125915.34+274604.2). 
We show the position of these galaxies along with the central coordinate of ACO~1656 in Figure \ref{fig:HSC}, which displays a PS1 composite image of the z and g filters, as plotted in Aladin Desktop~\citep{Bonnarel2000A&AS..143...33B}.

\begin{figure*}
    \centering
    \includegraphics[width=17cm]{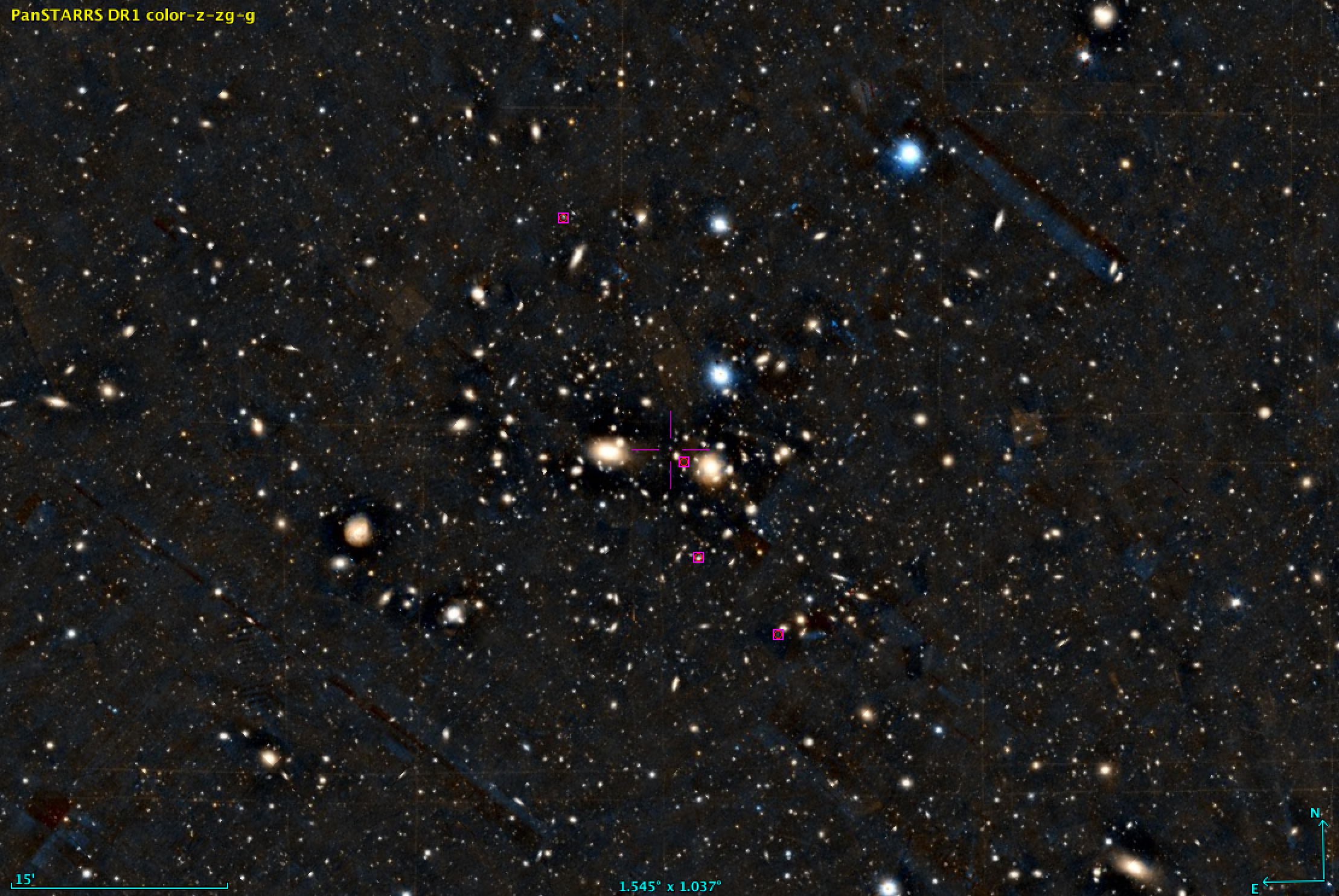} 
    \caption{Cluster of galaxies ACO~1656 (pink markers), and four dwarf galaxies matched to compact radio sources (pink squares).}%
    \label{fig:HSC}%
\end{figure*}

\section{Host properties}
\label{app:host_properties}

In Table \ref{table:candidates}, a number of physical properties are listed, including luminosity of the compact radio source, as well as redshift and star formation rate of the host galaxy. 
Here, we also present the distribution relative to stellar mass and specific star formation rate (sSFR, the star formation rate normalized by the stellar mass) in Figure \ref{fig:mstar_vs_sfr_ssfr}. 
Similarly, Figure \ref{fig:mstar_vs_sfr_ssfr} shows SFR and sSFR in relation to stellar mass. 
Here, we note that none of our candidates reach the same level of SFR as the host galaxies from \RI~\citep{Tendulkar2017ApJ...834L...7T} and \RItwin~\citep{Niu2022Natur.606..873N}, with \RItwin\  being within range. The  host galaxy of 
\RI\ stands out with respect of its sSFR compared to the rest of the data points. 
The host galaxies of the IMBH candidates studied by  \citeauthor{Reines2020ApJ...888...36R}  fall within the SFR and sSFR ranges of our candidates. 
We note that two out of the three galaxies matched to CLU in our study correspond to unresolved radio sources (J0909+5655, J1136+2643) in \citet{Sargent2022ApJ...933..160S}.
We performed two-sample KS tests between the respective candidates distributions to that of the background sample (``compact radio sources-dwarf galaxy'' matches below 3$\sigma$ on the L--SFR relation) for stellar mass, SFR, and sSFR that yield $p$-values of 0.2, 0.8, and 0.4 respectively, and are therefore consistent with being drawn from the same population.

\begin{figure*}
    \centering
    \includegraphics[width=17cm]{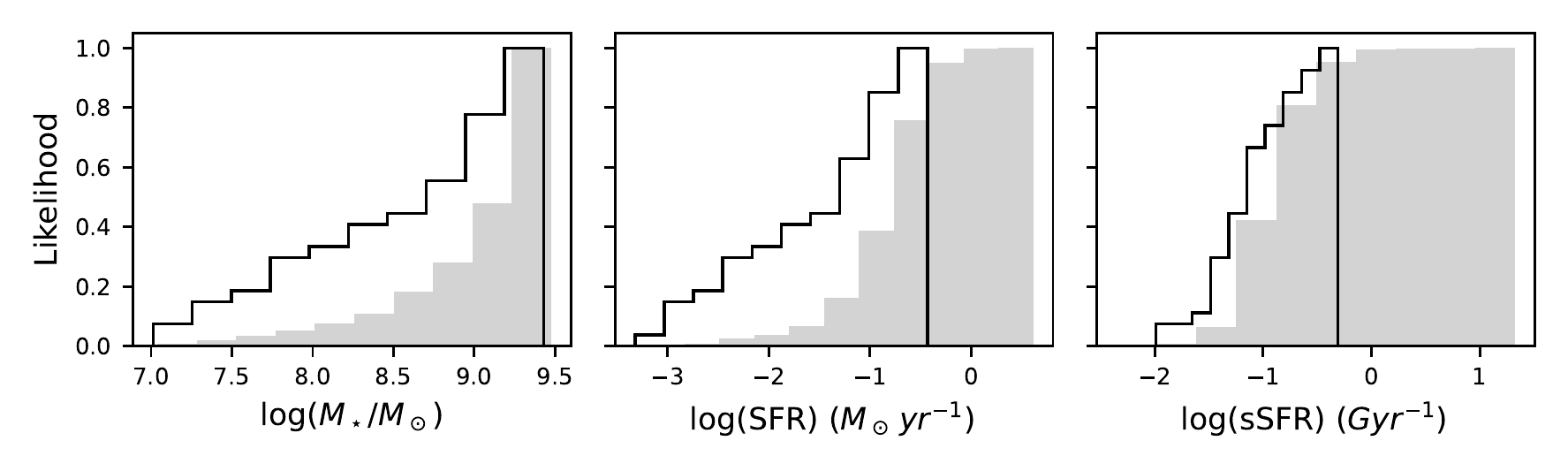} 
    \caption{Distribution of host stellar mass (left panel), SFR (middle panel), and sSFR (right panel) for candidates listed in Table \ref{table:candidates}.
    Gray bars indicate values for all ``compact radio source-dwarf galaxy'' matches below 3$\sigma$ on the L--SFR relation.
    }%
    \label{fig:mstar_sfr_ssfr}%
\end{figure*}

\begin{figure*}
    \centering
    \includegraphics[width=17cm]{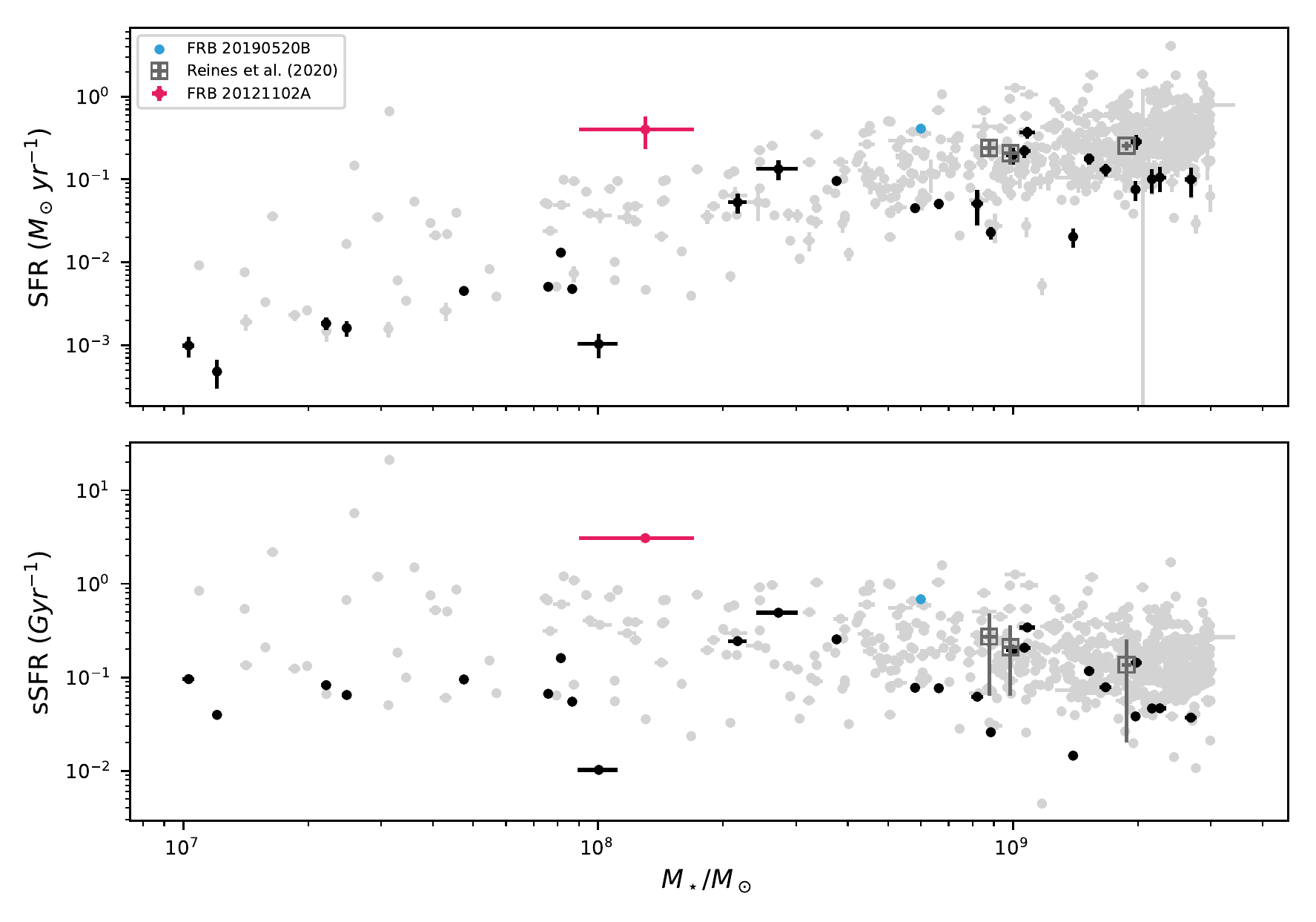} 
    \caption{Host stellar mass as a function of SFR and sSFR for our candidates. We also indicate values for \RI, \RItwin, and \citet{Reines2020ApJ...888...36R} sources from Figure \ref{fig:selection}.
    Gray markers indicate all ``compact radio source-dwarf galaxy'' matches below 3$\sigma$ on the L--SFR relation.
    }%
    \label{fig:mstar_vs_sfr_ssfr}%
\end{figure*}

\section{Radio luminosity versus ${\sc OI}/{\rm H\upalpha}$}

\begin{figure}
    \centering
    \resizebox{\hsize}{!}{\includegraphics{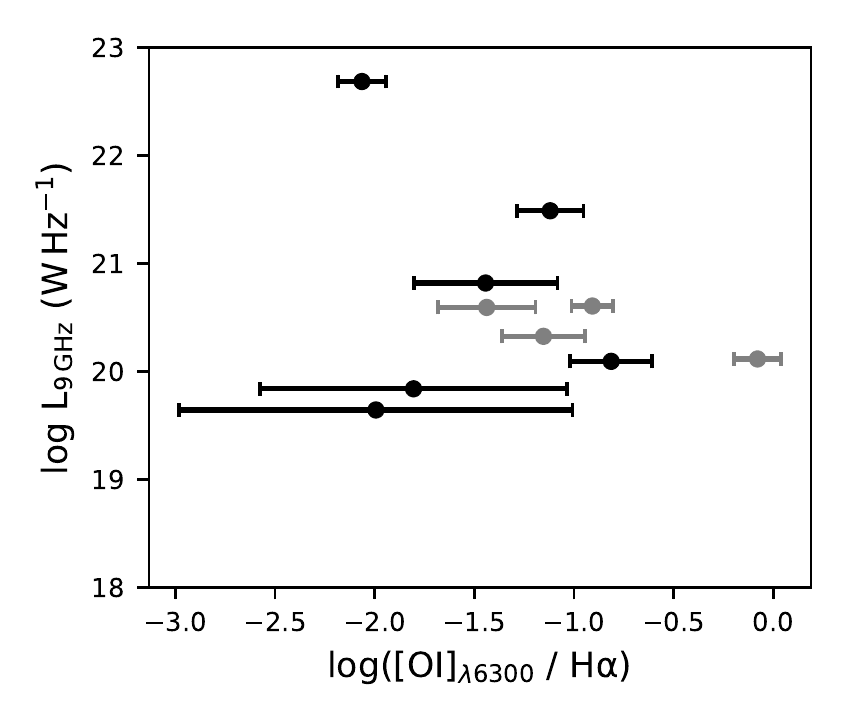}}
    \caption{
    Black markers indicate values where the luminosity was scaled to 9\,GHz with spectral indices obtained in \S\ref{subsec:spectral}, while gray markers were scaled using the canonical $\alpha=-0.7$ for synchrotron spectra of optically thin radio sources. 
    }%
    \label{fig:L_OI_Halpha}%
\end{figure}

\citet{Reines2020ApJ...888...36R} showed that the relation between the emission line ratio $[{\sc OI}]_{\lambda3600}/{\rm H\upalpha}$ and the luminosity at 9\,GHz could be used to separate radio emission from radio AGN and star formation. 
We used the spectral indices fitted in \S\ref{subsec:spectral} (and the canonical $\alpha= -0.7$ otherwise) to scale the luminosity of our sources to 9\,GHz for sources with spectral line measurements available in SDSS. 
We show the results in Figure \ref{fig:L_OI_Halpha}. 
Most of our sources fill the gap region found in Figure 10 of \citet{Reines2020ApJ...888...36R} that divides radio emission consistent with star formation and radio AGN. 
We note that, despite the fact that scaling flux from 144\,MHz to 9\,GHz remains an approximation, typically 10\% of the flux being resolved out at higher frequencies would only marginally affect the luminosity of sources plotted here.

As mentioned in Appendix \ref{app:host_properties}, galaxies J0909+5655 and J1136+2643 were matched to CLU, with both radio source being unresolved by the VLBA \citep{Sargent2022ApJ...933..160S}.  
The remaining match (J1220+3020, $L_{\rm 144\,MHz}\sim5\times10^{21}\,{\rm W\,Hz^{-1}}$) is one of Reines' sources with optical spectroscopic signature (obtained with targeted observation at the location of the radio source with GMOS-N/IFU) consistent with accreting IMBH, via the AGN coronal line [Fe~X] and enhanced [O~I] emission coincident with the radio source~\citep{Molina2021ApJ...910....5M, Sargent2022ApJ...933..160S}.
We do not detect these sources in LoTSS, despite two of them being within the DR2 footprint.

The two VLBA detections at 144\,MHz-scaled luminosity from \citet{Reines2020ApJ...888...36R} using spectral indices by \citet{Eftekhari2020ApJ...895...98E} yield $\sim1.39\times10^{22}$ and $\sim2.47\times10^{22}\,{\rm W}\,{\rm Hz^{-1}}$. 
It is therefore plausible that a fraction of the radio luminosity reported for our candidates could be attributed to an unresolved compact component. 
However, our scaling from 9\,GHz to 144\,MHz should be seen as an informed approximation, and radio luminosity values presented in Figure \ref{fig:L_OI_Halpha} should be regarded as order-of-magnitude estimates.

\section{Spatial coverage of ancillary surveys}
\label{app:coverage}

\begin{figure*}
    \centering
    \includegraphics[trim={0 0.5in 0 0.5in},clip,width=17cm]{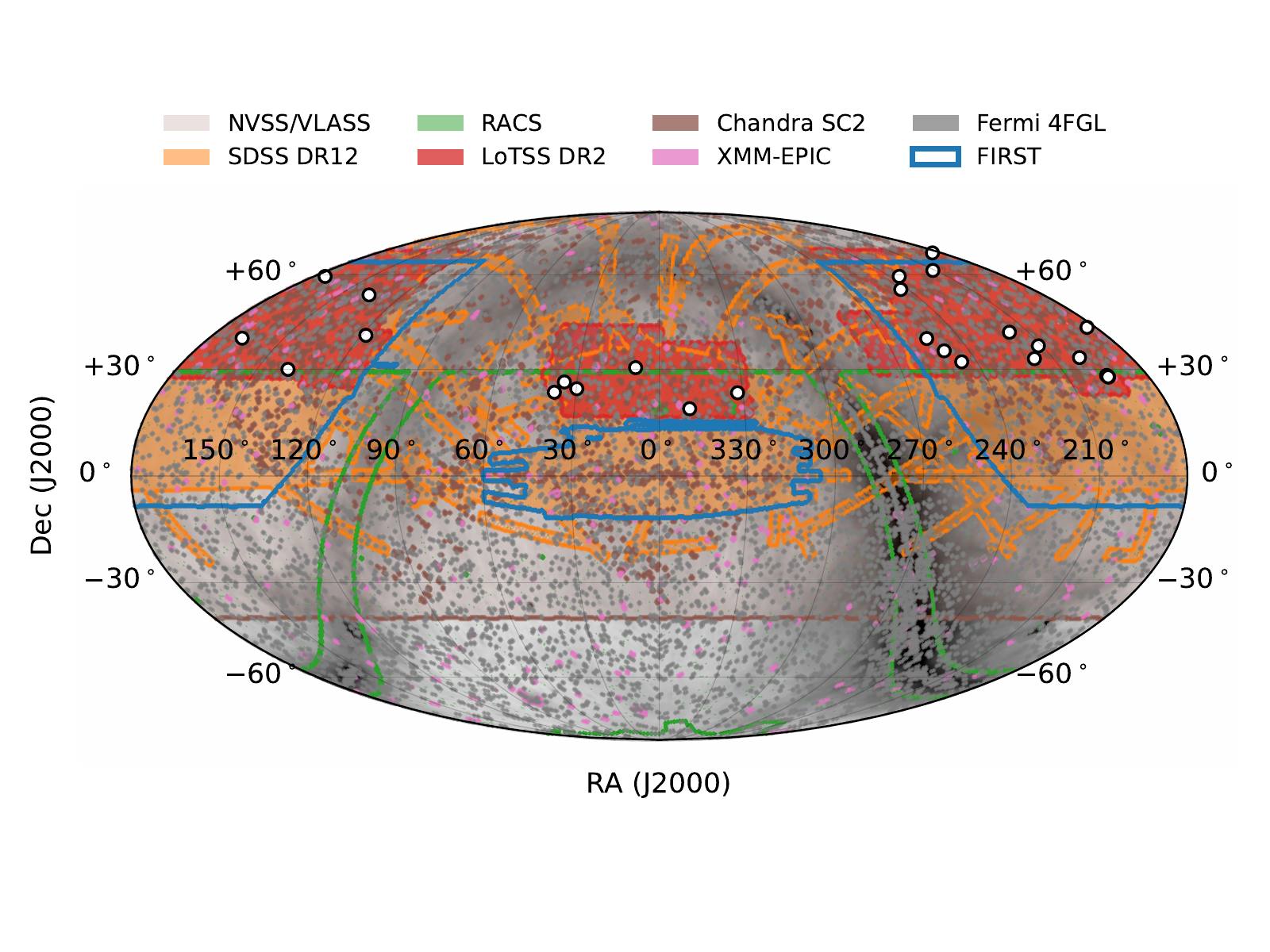} 
    \caption{Coverage maps of surveys searched in \S\ref{sec:candidates}.
    Source candidates discussed in this paper are shown as white circles.
    }%
    \label{fig:mocs}%
\end{figure*}

In this section, we evaluate the spatial coverage of radio surveys used in \S\ref{subsec:spectral}. Figure \ref{fig:mocs} shows multiorder coverage (MOC) maps for each of these surveys, seen as subdivided cells. 
Each survey is color coded and described in the figure caption. 
In addition, we overplot our candidates using yellow `+' markers. 
We note that seven sources have not been observed by FIRST and nine have been observed by RACS (Figure \ref{fig:spectral_indices}). 
All other sources could have been observed by the remaining surveys.
Finally, we also show coverage for SDSS DR12 (\S\ref{subsec:bpt}),  as well as Chandra SC2, XMM-Newton EPIC, and Fermi 4FGL (\S\ref{subsec:xray}). 

\end{appendix}

\end{document}